\documentclass[floatfix,superscriptaddress,numerical,amsmath,amssymb, pre, aps, reprint,showpacs]{revtex4-2}
\pdfoutput=1
\usepackage{packages}
\makeatletter
\AtBeginDocument{\let\LS@rot\@undefined}
\makeatother

\begin{document}

\title{Effects of homophily and heterophily
on preferred-degree networks: mean-field analysis and overwhelming transition
}
\author{Xiang Li}
\affiliation{Department of Applied Mathematics, School of Mathematics, University of
Leeds, Leeds LS2 9JT, United Kingdom}
\author{Mauro Mobilia}
\affiliation{Department of Applied Mathematics, School of Mathematics, University of
Leeds, Leeds LS2 9JT, United Kingdom}

\author{Alastair M. Rucklidge}
\affiliation{Department of Applied Mathematics, School of Mathematics, University of
Leeds, Leeds LS2 9JT, United Kingdom}

\author{R.K.P. Zia}
\affiliation{Center for Soft Matter and Biological Physics, Department of
Physics, Virginia Polytechnic Institute \& State University, Blacksburg, Virginia
24061, USA}
\affiliation{Physics Department, University of Houston, Houston, Texas 77204, USA
}

\begin{abstract}
We investigate the long-time properties of a dynamic, out-of-equilibrium network of individuals holding one of two opinions in a population consisting of two communities of different sizes. Here, while the agents' opinions are fixed, they have a preferred degree which leads them to endlessly create and delete links. Our evolving network is shaped by homophily/heterophily, a form of social interaction by which individuals tend to establish links with others having similar/dissimilar opinions. Using Monte Carlo simulations and a detailed mean-field analysis, we investigate how the sizes of the communities and the degree of homophily/heterophily affect the network structure. In particular, we show that when the network is subject to enough heterophily, an ``overwhelming transition'' occurs: individuals of the smaller community are overwhelmed by links from  the larger group, and their mean degree greatly exceeds the  preferred degree. This and related phenomena are characterized by the network's total and joint degree distributions, as well as the fraction of links across both communities and that of agents having fewer edges than the  preferred degree. We use our mean-field theory to discuss the network's polarization when the group sizes and level of homophily vary. 

\end{abstract}

\maketitle
\section{Introduction}
The relevance of simple mathematical models in describing collective social
phenomena has a long history~\cite{asch1955opinions,asch1956studies,schelling1980strategy}, and the importance
of relating ``micro-level" to
``macro-level" phenomena is well established~\cite{baronchelli2018emergence,latane1981psychology,axelrod1997dissemination,mcpherson2001birds,yavacs2014impact}. 
Yet, it is only recently that connections between simple models of
social dynamics and those traditionally used in statistical mechanics, such
as Ising-like spin models, have been systematically exploited in ``sociophysics'' or ``opinion dynamics''~\cite{Castellano-rev,Galam-book,Sen-book,mobilia2015nonlinear,castellano2009nonlinear,mobilia2007role}. 
As a result, there is an intense field of research dedicated to the study of
social networks by means of models and tools borrowed from statistical physics~\cite{Albert-rev,Dorogovtsev-book,newman2018networks,mellor2017heterogeneous}. In particular, 
various dynamical processes have been studied 
on complex networks whose structure is random but static, see, {\it e.g.},
~\cite{Antal-2006,sood2008voter,Baxter-2008,Blythe-2010,Castellano-2010,moretti2013mean,Szolnoki-2014,Sabsovich-2017}, while in other models agents and links co-evolve~\cite{holme2006nonequilibrium,evans2007exact,vazquez2008analytical,vazquez2008generic,lindquist2009network,durrett2012graph,henry2011emergence}.

Naturally, collective phenomena such as 
phase transitions and polarization that emerge from agent interactions 
have received great attention. 
An important example of social interaction is {\it homophily}, which is the tendency for nodes to create links with similar ones~\cite{mcpherson2001birds,centola2011experimental,centola2007complex,del2017modeling,centola2007homophily,newman2018networks},
while the tendency to establish ties with different others is referred to as {\it heterophily}~\cite{xie2016,ramazi2018,barranco2019,yokomatsu2021}.
Homophily is 
commonly seen in political parties~\cite{pariser2011filter,Iyengar2012,barbera2015tweeting,barbera2015,bakshy2015exposure,del2016spreading,wang2020public}, and the  effects of this form of ``assortative mixing'' 
on network dynamics have been investigated in
sociological~\cite{yavacs2014impact,shalizi2011homophily,mcpherson1987homophily,zeltzer2020gender,yavacs2014impact,gargiulo2017role,centola2007complex,centola2011experimental,centola2007homophily} and interdisciplinary physics studies~\cite{Boguna-2004,wong2006spatial,karimi2018homophily,kimura2008coevolutionary,papadopoulos2012popularity,asikainen2020cumulative,krapivsky2021divergence}.
In this literature, homophily is often modeled by assuming a  
biased probability of creating an edge or of rewiring an existing link (edge weighting),
and notably features in growing~\cite{papadopoulos2012popularity,karimi2018homophily,gargiulo2017role,overgoor2019choosing} and  nodal attribute network models~\cite{Boguna-2004,wong2006spatial,gorski2020}, and 
is often considered together with other ``structural balance'' 
processes that mitigate  tensions between connected agents~\cite{heider58,gorski2020,asikainen2020cumulative}. 

While homophily appears to be ubiquitous in social networks with many  examples of
``birds of a feather flock together'' behaviors,  heterophily appears to be more 
elusive, with some empirical evidence in team formation processes~\cite{johnson2009}, and in professional cooperation networks~\cite{xie2016,barranco2019}.
Quite interestingly however, 
in a two-community growing network according to  preferential attachment, it has recently been found that heterophily
is responsible for the increase of the average degree of the agents of the smaller group~\cite{karimi2018homophily}.


Here we consider an evolving network model in which links 
fluctuate continuously as the result of the homophilic/heterophilic interactions between
individuals of two communities, holding 
one of two different opinions that remains fixed, who try to satisfy a prescribed {\it preferred degree}~\cite{liu2013modeling,liu2014modeling,bassler2015networks}. 

The objective of this work is to understand in some detail
how homophily and heterophily affect the stationary state of the network
and its emerging properties. While some of these aspects are 
considered in Ref.~\cite{pre2021} for the special case of opinion groups of the 
same size, here we focus on the general case of communities of arbitrary sizes, showing that this leads to surprising results at a price of a considerably more challenging analysis.
%
%
%
Our main contribution is the detailed characterization of 
the ``overwhelming transition'' arising under enough heterophily 
in communities of different sizes.  Remarkably, as observed in Ref.~\cite{pre2021}, the network then undergoes a transition separating 
 a  phase where it is  homogeneous
and an ``overwhelming phase'' in which 
agents of the smaller community are overwhelmed by links from those of the larger group, with degrees greatly exceeding the  preferred degree. Here, we unveil the properties of the  overwhelming phase (and ordinary regime) notably
in terms of the total and joint degree distributions
by devising suitable mean-field theories and  Monte Carlo simulations.

The remainder of the paper is organized as follows: the formulation of the
model and quantities of interest are introduced in the next section.  
In Sec.~III, we  summarize the main properties of the symmetric system reported in 
Ref.~\cite{pre2021}. Our main findings 
regarding the ordinary and overwhelming phases 
are presented in Secs.~IV and V
centred respectively on simulation results  and mean-field analysis. Our conclusions are presented in Sec.~VI. 
Further details are provided in the Supplementary Material (attached at the end of the document; or on \url{https://stacks. iop.org/JSTAT/2022/013402/mmedia}).

\section{Model formulation and quantities of interest}
Our model consists of $N$ agents (nodes) with a varying number of
connections (links) between them, forming a fluctuating network. Each agent $j=1, \cdots, N$ is endowed with one of two possible states (``opinions''), $\sigma_j =\pm 1$, and a preferred number of links $\kappa$. A fraction $n_{\pm}$ of the agents is in opinion state $\pm 1$, 
so that the network consists of a  number $N_{\pm}=Nn_{\pm}$ of agents holding opinion $\pm 1$. 
 Using
the terminology of Ising-like models~\cite{krapivsky2010kinetic}, $(N_+,N_-)$ can be replaced by  $(N,m)$, where $m\equiv \left(
N_{+}-N_{-}\right)/N=n_+-n_-$ is an intensive quantity  called ``magnetization".

The basis of our model is a preferred degree network (PDN)~\cite{bassler2015networks,liu2013modeling,liu2014modeling,liu2012extraordinary} in which, at
discrete time steps, different individuals (nodes) are chosen to add or cut
links to others depending on whether its degree is less or greater than  $\kappa$. The former are referred to as ``adders'' and the latter as ``cutters''.
So as to
avoid frozen networks,  $\kappa$ is chosen to be a half integer.
Here, the  PDN dynamics is supplemented by two ingredients: (i) the network consists of two communities of differing opinions and sizes; (ii) there are social interactions among agents embodied by the notion of \textit{homophily}.
 Controlled by a parameter $J\in \left[-1,1\right]$, homophily models the behavior of individuals who prefer to
``make friends" and establish links with those holding the
same opinion (positive homophily or simply homophily, $J>0$) or those with opposite opinions
(negative homophily or simply heterophily, $J<0$), see Fig.~S1 and Sec.~S1 in the supplementary material.

By combining Monte Carlo and mean-field techniques, we here investigate how 
 the network
topology respond to parameters $(\kappa, J, N, m)$ in its {\it out-of-equilibrium} stationary state (see Sec.~S2 of the supplementary material),
and focus on the unexpected phenomena arising when $m\neq 0$, briefly reported in~\cite{pre2021}.

\subsection{Model update rules}

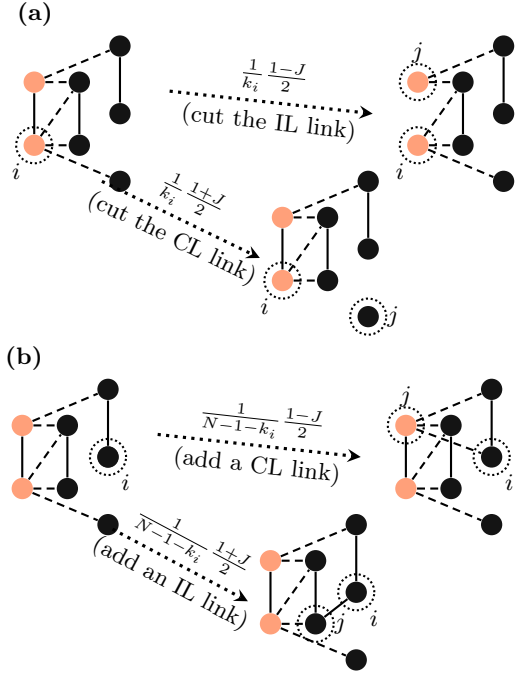
\begin{figure}[tph]
\begin{tikzpicture}[>=stealth,inner sep=1mm,scale=0.6]
\definecolor{bluec}{RGB}{20,20,20}
\definecolor{redc}{RGB}{255,160,122}
\tikzset{redn/.style={>=stealth,circle,fill=redc}}
\tikzset{bluen/.style={>=stealth,circle,fill=bluec}}

\begin{scope}[name=s1]
\node (1) at (-0.5,-0.3) [redn] {};
 \node[label=240:\diagfont$i$] (2) at (-0.5,-1.7) [redn] {};
 \draw[densely dotted,\linewid] (-0.5,-1.7) circle (0.4cm);
 \node (3) at (0.5,-0.3) [bluen] {};
 \node (4) at (0.5,-1.7) [bluen] {};
 \node (5) at (1.4,0.5) [bluen] {};
 \node (6) at (1.4,-1) [bluen] {};
 \node (7) at (1.4,-2.5) [bluen] {};
 \draw[\linewid] (1) -- (2);
 \draw[\linewid] (3) -- (4);
 \draw[\linewid] (5) -- (6);
 \draw[densely dashed,\linewid] (1) -- (5);
 \draw[densely dashed,\linewid] (1) -- (3);
 \draw[densely dashed,\linewid] (2) -- (3);
 \draw[densely dashed,\linewid] (2) -- (4);
 \draw[densely dashed,\linewid] (2) -- (7);
\end{scope}

\begin{scope}[shift={(5,-3)}]
\node (1) at (0,-0.3) [redn] {};
 \node[label=240:\diagfont$i$] (2) at (0,-1.7) [redn] {};
 \draw[densely dotted,\linewid] (0,-1.7) circle (0.4cm);
 \node (3) at (1,-0.3) [bluen] {};
 \node (4) at (1,-1.7) [bluen] {};
 \node (5) at (1.9,0.5) [bluen] {};
 \node (6) at (1.9,-1) [bluen] {};
  \node[label=0:\diagfont$j$] (7) at (1.9,-2.5) [bluen] {};
 \draw[densely dotted,\linewid] (1.9,-2.5) circle (0.4cm);
 \draw[\linewid] (1) -- (2);
 \draw[\linewid] (3) -- (4);
 \draw[\linewid] (5) -- (6);
 \draw[densely dashed,\linewid] (1) -- (5);
 \draw[densely dashed,\linewid] (1) -- (3);
 \draw[densely dashed,\linewid] (2) -- (3);
 \draw[densely dashed,\linewid] (2) -- (4);
\end{scope}

\draw[->,very thick,dotted] (1,-2.5)  --node[midway,above,sloped] {\diagfont$\frac{1}{k_i}\frac{1+J}{2}$} node[midway,below,sloped] {\small (cut the CL link)} (4.5,-4.2);

\begin{scope}[shift={(8,0)}]
 \node[label=90:\diagfont$j$] (1) at (0,-0.3) [redn] {};
 \draw[densely dotted,\linewid] (0,-0.3) circle (0.4cm);
 \node[label=240:\diagfont$i$] (2) at (0,-1.7) [redn] {};
 \draw[densely dotted,\linewid] (0,-1.7) circle (0.4cm);
 \node (3) at (1,-0.3) [bluen] {};
 \node (4) at (1,-1.7) [bluen] {};
 \node (5) at (1.9,0.5) [bluen] {};
 \node (6) at (1.9,-1) [bluen] {};
 \node (7) at (1.9,-2.5) [bluen] {};

 \draw[\linewid] (3) -- (4);
 \draw[\linewid] (5) -- (6);
 \draw[densely dashed,\linewid] (1) -- (5);
 \draw[densely dashed,\linewid] (1) -- (3);
 \draw[densely dashed,\linewid] (2) -- (3);
 \draw[densely dashed,\linewid] (2) -- (4);
 \draw[densely dashed,\linewid] (2) -- (7);
\end{scope}
\draw[->,very thick,dotted] (2.5,-0.5)  --node[midway,above,sloped] {\diagfont$\frac{1}{k_i}\frac{1-J}{2}$} node[midway,below,sloped] {\small (cut the IL link)} (7,-1);


\begin{scope}[scale=1]
\draw (0,1.2) node[right=-18pt] {\textbf{(a)}};
\end{scope}
 
\end{tikzpicture} \\
\begin{tikzpicture}[>=stealth,inner sep=1mm,scale=0.6]
\definecolor{bluec}{RGB}{20,20,20}
\definecolor{redc}{RGB}{255,160,122}
\tikzset{redn/.style={>=stealth,circle,fill=redc}}
\tikzset{bluen/.style={>=stealth,circle,fill=bluec}}

\begin{scope}[name=s1]
\node (1) at (-0.5,-0.3) [redn] {};
\node (2) at (-0.5,-1.7) [redn] {};
 \node (3) at (0.5,-0.3) [bluen] {};
 \node (4) at (0.5,-1.7) [bluen] {};
 \node (5) at (1.4,0.5) [bluen] {};
 \node (6) at (1.4,-1) [bluen] {};
   \node[label=300:\diagfont$i$] (6) at (1.4,-1) [bluen] {};
 \draw[densely dotted,\linewid] (1.4,-1) circle (0.4cm);
 \node (7) at (1.4,-2.5) [bluen] {};
 \draw[\linewid] (1) -- (2);
 \draw[\linewid] (3) -- (4);
 \draw[\linewid] (5) -- (6);
 \draw[densely dashed,\linewid] (1) -- (5);
 \draw[densely dashed,\linewid] (1) -- (3);
 \draw[densely dashed,\linewid] (2) -- (3);
 \draw[densely dashed,\linewid] (2) -- (4);
 \draw[densely dashed,\linewid] (2) -- (7);
\end{scope}

\begin{scope}[shift={(5,-3)}]
\node (1) at (0,-0.3) [redn] {};
\node (2) at (0,-1.7) [redn] {};
 \node (3) at (1,-0.3) [bluen] {};
  \node[label=0:\diagfont$j$] (4) at (1,-1.7) [bluen] {};
 \draw[densely dotted,\linewid] (1,-1.7) circle (0.4cm);
 \node (5) at (1.9,0.5) [bluen] {};
  \node[label=300:\diagfont$i$] (6) at (1.9,-1) [bluen] {};
 \draw[densely dotted,\linewid] (1.9,-1) circle (0.4cm);
 \node (7) at (1.9,-2.5) [bluen] {};
 \draw[\linewid] (1) -- (2);
 \draw[\linewid] (3) -- (4);
 \draw[\linewid] (5) -- (6);
 \draw[\linewid] (4) -- (6);
 \draw[densely dashed,\linewid] (1) -- (5);
 \draw[densely dashed,\linewid] (1) -- (3);
 \draw[densely dashed,\linewid] (2) -- (3);
 \draw[densely dashed,\linewid] (2) -- (4);
 \draw[densely dashed,\linewid] (2) -- (7);
\end{scope}

\draw[->,very thick,dotted] (1.5,-2.6)  --node[midway,above,sloped] {\diagfont$\frac{1}{N-1-k_i}\frac{1+J}{2}$} node[midway,below,sloped] {\small (add an IL link)} (4.5,-4.2);

\begin{scope}[shift={(8,0)}]
 \node[label=90:\diagfont$j$] (1) at (0,-0.3) [redn] {};
 \draw[densely dotted,\linewid] (0,-0.3) circle (0.4cm);
 \node (2) at (0,-1.7) [redn] {};
 \node (3) at (1,-0.3) [bluen] {};
 \node (4) at (1,-1.7) [bluen] {};
 \node (5) at (1.9,0.5) [bluen] {};
   \node[label=300:\diagfont$i$] (6) at (1.9,-1) [bluen] {};
 \draw[densely dotted,\linewid] (1.9,-1) circle (0.4cm);
 \node (7) at (1.9,-2.5) [bluen] {};
 \draw[\linewid] (1) -- (2);
 \draw[\linewid] (3) -- (4);
 \draw[\linewid] (5) -- (6);
 \draw[densely dashed,\linewid] (1) -- (5);
 \draw[densely dashed,\linewid] (1) -- (3);
 \draw[densely dashed,\linewid] (1) -- (6);
 \draw[densely dashed,\linewid] (2) -- (3);
 \draw[densely dashed,\linewid] (2) -- (4);
 \draw[densely dashed,\linewid] (2) -- (7);
\end{scope}
\draw[->,very thick,dotted] (2.5,-0.5)  --node[midway,above,sloped] {\diagfont$\frac{1}{N-1-k_i}\frac{1-J}{2}$} node[midway,below,sloped] {\small (add a CL link)} (7,-1);


\begin{scope}[scale=1]
\draw (0,1.2) node[right=-18pt] {\textbf{(b)}};
\end{scope}
 
\end{tikzpicture} 
\caption{Illustration of the model for $\kappa=2.5$. Dark nodes represent $+1$ voters (majority group) and light nodes are $-1$ voters (minority group). (a) Cutting process of an $ij$ link when $k_i>\kappa$. (b) Adding of an $ij$ link when $k_i<\kappa$. Dashed: cross links (CLs) between agents of different groups; solid:  
internal links (ILs) between voters of the same group.
The probabilities of cutting  an IL and adding a CL are $\check{J}=(1-J)/2$ and $\hat{J}=(1+J)/2$, respectively.
Similarly, the respective probabilities of cutting a CL and adding an IL are 
$\hat{J}$ and $\check{J}$, see text.
}
\label{fig:Fig2}
\end{figure}

The dynamic rules of our model are conveniently stated 
by assuming discrete time steps, $t=1,2,\cdots$.  
These rules are
illustrated in Fig.~\ref{fig:Fig2}. 
At each $t$, an agent $i\ (i = 1, \cdots , N )$ of degree $k_{i}\in \left[0,N-1\right] $ is randomly chosen. For convenience, we will refer to
nodes connected to $i$ as its ``neighbors"
and those not connected as ``non-neighbors". Then, after $i$ is chosen,
\begin{compactitem}
\item If $k_i>\kappa,$ choose a neighbor $j$ and delete the link
\begin{compactitem}
    \item with probability $\check{J}$ if $\sigma_i=\sigma_j$, or
    \item with probability $\hat{J}$ if $\sigma_i\ne\sigma_j$,
\end{compactitem}
\item If $k_i<\kappa,$ choose one of the non-neighbors $j$ and add a link
\begin{compactitem}
    \item with probability $\hat{J}$ if $\sigma_i = \sigma_j$, or
    \item with probability $\check{J}$ if $\sigma_i \ne \sigma_j$.
\end{compactitem}
\end{compactitem}
$\hat{J}$ and $\check{J}$ are defined by
\begin{equation*}
\hat{J}\equiv \left(1+J\right) /2 \quad \text{and}\quad \check{J}\equiv
\left(1-J\right) /2.
\end{equation*}
$J$ thus plays a role similar to the nearest neighbor
interaction in the ordinary Ising model (spin alignment). 
Note that for $J=0$, the 
distinction between the communities  is only nominal and as
the opinions are irrelevant for the dynamics, 
and the system becomes similar to the PDN models of
 Refs.~\cite{liu2012extraordinary,liu2013modeling},
see Sec.~S1 in the supplementary material.
%

\subsection{Quantities of interest}
To study the behavior of this evolving network in various regimes of
parameter space, we focus on a few quantities of interest, summarized in the table of Sec.~S3 of the supplementary material. The most
common of these is the degree distribution (DD) $p_\sigma(k)$
associated with agents in community $\sigma$, and
from these $p$'s, we can extract the average degrees $\mu_{\sigma}\equiv \sum_{k}k~p_{\sigma}\left(k\right)$ for each community $\sigma=\pm$,
as well as the variances $V_{\sigma}\equiv \sum_k k^{2}p_{\sigma}-\mu_{\sigma}^{2}$. 
We also distinguish the number of links an agent has to neighbors
with opinion $\tau\in \{-,+\}$, and denote these by $\ell_{\tau}$ (where 
$k=\ell_{+}+\ell_{-}$). With $\ell_{\tau}$, we compile the \textit{joint} degree distributions (JDD)~\cite{bassler2015networks} for the two communities: $P_{\sigma}\left(\ell_{+},\ell_{-}\right)$, and  
from these  can compute the averages: 
\begin{equation}
\label{eq:barell}
\left(\bar{\ell}_{\pm}\right)_{\sigma}\equiv \sum_{\ell_{+},\ell_{-}}\ell_{\pm}~P_{\sigma}\left(\ell_{+},\ell_{-}\right),
\end{equation}%
which can be regarded as the ``centers of
mass" (CM) of the two JDDs, see Fig.~9 of Ref.~\cite{pre2021}. The average degree in community $\sigma$ is thus
$\mu_{\sigma}=\left(\bar{\ell}_{+}\right)_{\sigma}+\left(\bar{\ell}_{-}\right)
_{\sigma}.$ 
%
%
Another convenient characteristic is the conditional distribution of cross-links:
 \begin{equation}
\label{eq:q}
q_{\sigma}\left(w\,|\,k\right) \equiv \frac{P_{\sigma}\left(\ell_{+},\ell_{-}\right)}{p_{\sigma}\left(k\right)};\ w=\ell_{-\sigma},\ \ell_++\ell_-=k,%
\end{equation}
which gives the probability for a node in the group $\sigma$ to have $w$ cross-links (CLs), {\it provided} it has total degree $k$.

We also study the (fluctuating) total number of connections, which is a global 
quantity denoted by $L=L_{\odot}+L_{\times}$, where $L_{\odot}$ is the overall number of ILs and $L_{\times}$ is the total number of CLs. Denoting by $L_{\sigma \tau}$ the links between agents opinions $\sigma$ and $\tau$, we have $L_{\odot}=L_{++}+L_{--}$ and $L_{\times}=L_{+-}=L_{-+}$, and hence
\begin{equation*}
L=L_{\odot}+L_{\times}\equiv L_{++}+L_{--}+L_{+-}.
\end{equation*}%
Since we  focus on the steady-state
averages of these quantities, we simplify notation by writing $L$ in lieu of $\left\langle L\right\rangle $, etc. (The same below with $\alpha$ and $\rho$.) Clearly, these averages are related to $\left(\bar{\ell}_{\tau}\right)_{\sigma}$: 
\begin{equation}\label{L=Nkappa}
2L_{\sigma \sigma}=N_{\sigma}\left(\bar{\ell}_{\sigma}\right)_{\sigma
};~~L_{\times}=N_{+}\left(\bar{\ell}_{-}\right)_{+}=N_{-}\left(\bar{\ell}%
_{+}\right)_{-}.
\end{equation}%

A natural way to describe polarization (extent of division between the communities) is to start with the ratio of CLs to the total number of links \cite{durrett2012graph}
\begin{equation}\label{rhodef}
\rho \equiv L_{\times}/L,
\end{equation}%
and then a measure of polarization is given by
\begin{equation}\label{Lambda-def}
\Lambda \equiv 1-2\rho =\left(L_{\odot}-L_{\times}\right) /L,
\end{equation}%
so that $\Lambda \left(J=\pm 1\right) =\pm 1$. For asymmetric systems, however, the ratios for the two communities are distinct:
\begin{equation}\label{rhosigma}
\rho_{\sigma}\equiv \frac{\left(\bar{\ell}_{-\sigma}\right)_{\sigma}}{%
\mu_{\sigma}}=\frac{L_{\times}}{L_{\times}+2L_{\sigma \sigma}}. 
\end{equation}%
Further, as will be discussed in Section~IV.E below, $\Lambda$ suffers from some deficiencies. Instead, let us introduce an alternative measure of polarization, $\Pi$, which relies on the (normalized) {\it difference} between the two CMs,
\begin{equation*}
\delta_{x}\equiv \frac{(\bar{\ell}_{+})_{+}-(\bar{\ell}_{+})_{-}}{(\bar{\ell%
}_{+})_{+}+(\bar{\ell}_{+})_{-}},~~\delta_{y}\equiv \frac{(\bar{\ell}%
_{-})_{-}-(\bar{\ell}_{-})_{+}}{(\bar{\ell}_{-})_{-}+(\bar{\ell}_{-})_{+}}.
\end{equation*}%
Specifically, we define
\begin{equation}\label{Pi-def}
\Pi \equiv \frac{\delta_{x}+\delta_{y}}{2}\in \left[ -1,1\right].
\end{equation}%
%

In order to account for the (fluctuating) number of nodes to add/cut links, we
denote by $N^{a}$ and $N^{c}$ the number of  ``adders" and ``cutters", respectively. Further we denote by $N^{\beta}_{\sigma}$ with $\sigma \in \{+,-\}$ and $\beta\in \{a,c\}$, the number of agents who are adders ($\beta=a$) or cutters ($\beta=c$) and hold opinion $\sigma$. Hence $N^{a}+N^{c}=N$ and $N_{\sigma}=N_{\sigma}^{a}+N_{\sigma}^{c}$, and the 
associated fractions are
\begin{equation}\label{sum-ns}
n_{\sigma}^{\beta}\equiv N_{\sigma}^{\beta}/N;~~n_{\sigma}=\Sigma_{\beta}n_{\sigma}^{\beta};~~n^{\beta}=\Sigma_{\sigma}n_{\sigma}^{\beta}. 
\end{equation}%
Clearly, $\Sigma_{\sigma ,\beta}n_{\sigma}^{\beta}=1$. 
The fraction of adders, denoted by
\begin{equation}\label{alpha}
\alpha \equiv n^{a}=N^{a}/N, 
\end{equation}%
plays an important role. It is useful to define the fraction of adders 
in each group $\sigma$ by
\begin{equation}
\alpha_{\sigma}\equiv N_{\sigma}^{a}/N_{\sigma}=n_{\sigma}^{a}/n_{\sigma},
\end{equation}
with $n_{+}\alpha_{+}+n_{-}\alpha_{-}=\alpha$ and $\alpha_{\pm}=\alpha$ when $m=0$.
%

\section{Symmetric Case, $m=0$: Summary of Results}
For the sake of reference and completeness,  we summarize the findings  on the symmetric case $m=0$ reported  in Ref.~\cite{pre2021}. 
We showed that by solving the balance equations for $L_{\times}$
and $L_{\odot}$ in the context of a mean-field approximation, the fractions of adders and CLs
in the steady state of the symmetric case are
\begin{eqnarray}\label{alpha*}
\alpha =\frac{1-J^{2}}{2},   \quad
\rho =\frac{1}{2}-\frac{J}{1+J^{2}}, 
\end{eqnarray}%
which implies that the
response to homophily in a clear unique mean-field expression of the polarization: 
\begin{equation}
\label{eq:polmzero}
\Lambda =\Pi = 1-2\rho=\frac{2J}{1+J^{2}}.
\end{equation}
We also studied the stationary degree distributions $p\left( k\right) $ and $q\left(w|k\right)$. For the former, we found
\begin{equation}
p\left( k
\begin{array}{c}
> \\ 
<%
\end{array}
\kappa \right) =\left( \frac{1\pm J^{2}}{3\pm J^{2}}\right) ^{\tilde{k}};~~%
\text{with }\tilde{k}=%
\begin{array}{c}
k-\left\lceil \kappa \right\rceil \\ 
\left\lfloor \kappa \right\rfloor -k%
\end{array},
\label{eqn:finalpk}
\end{equation}%
from which we obtain the degree average and 
its variance:%
\begin{equation}\label{eqn:avar}
\mu =\mu_{\sigma}=\kappa +3J^{2}/2,~~V=V_{\sigma}=\left( 7+J^{4}\right) /4.
\end{equation}%
We also consider the JDD, $P_{\sigma}\left(\ell _{+},\ell _{-}\right)$,
with $P_{+}\left(x, y\right) =P_{-}\left( y, x\right)$ which,  due to symmetry, gives the probability for either a $+$ or $-$   node to have $x$ ILs and $y$ CLs. The JDD is obtained from \eqref{eq:q}, where the conditional 
distribution of cross-links is approximated by a binomial distribution, i.e. $q\simeq q_{{\rm bin}}\left( w|k\right) =\binom{k}{w}\rho^{w}\left( 1-\rho \right) ^{k-w}$
 ~\cite{pre2021}. Hence, with \eqref{eqn:finalpk} and \eqref{alpha*},
 the product 
\begin{equation}\label{product}
P_{\sigma }\left( \frac{k+u}{2},\frac{k-u}{2}\right) \simeq p\left( k\right)
q_{{\rm bin}}\left( \left( k-\sigma u\right) /2|k\right)  
\end{equation}
is a suitable approximation of the JDD for our purposes.
As explained in Sec.~IV.D, for the generic $m\neq 0$ case with $|mJ|\ll 1$,
the form \eqref{product} can be used to get a three-dimensional impression of the network's properties (see Fig.~\ref{fig:Fig8}).

\section{Asymmetric Case, $m\neq0$: Simulation Results}
In our simulations, $m$ varies 
from $-0.04$ to $-0.6$, and
we will see that $m=-0.04$ and $m=-0.6$ represent two very different regimes, i.e., the ordinary phase and the overwhelming phase, 
the system  
displays properties quite distinct from those summarized in Sec.~III, see below.

We initiate the system with no initial links, run for typically over $10^{2}$ MCS, and verify that quantities like $\alpha_{\sigma}$ and $\rho_{\sigma}$ have settled into steady values. One MCS (Monte Carlo step) is $N$ updates.
Thereafter, simulation runs are carried on for up to another $10^{6}$ MCS, during which we
measured various quantities every MCS. For simplicity,
we show mainly the data associated with $N=1000$ and $\kappa =60.5$, so that $1 \ll \kappa \ll N_{\pm}$. The
data collected for other values give some impression of finite-size effects,
which can be serious when $N$ and $\kappa$ are lowered by an order of
magnitude.  In addition, we have carried out 10 runs with
different random number generators in a handful of cases, to get a better
idea of statistical errors. In all cases tested, the scatter for global
quantities like $\alpha$ and $\rho$ is much smaller than the size of the
symbols (i.e., at most one part in a thousand).

\subsection{Fraction of adders and cross-community links}
\begin{figure}[thp]
\centering
\begin{subfigure}{0.35\textwidth}
         \centering
         \includegraphics[width=\textwidth]{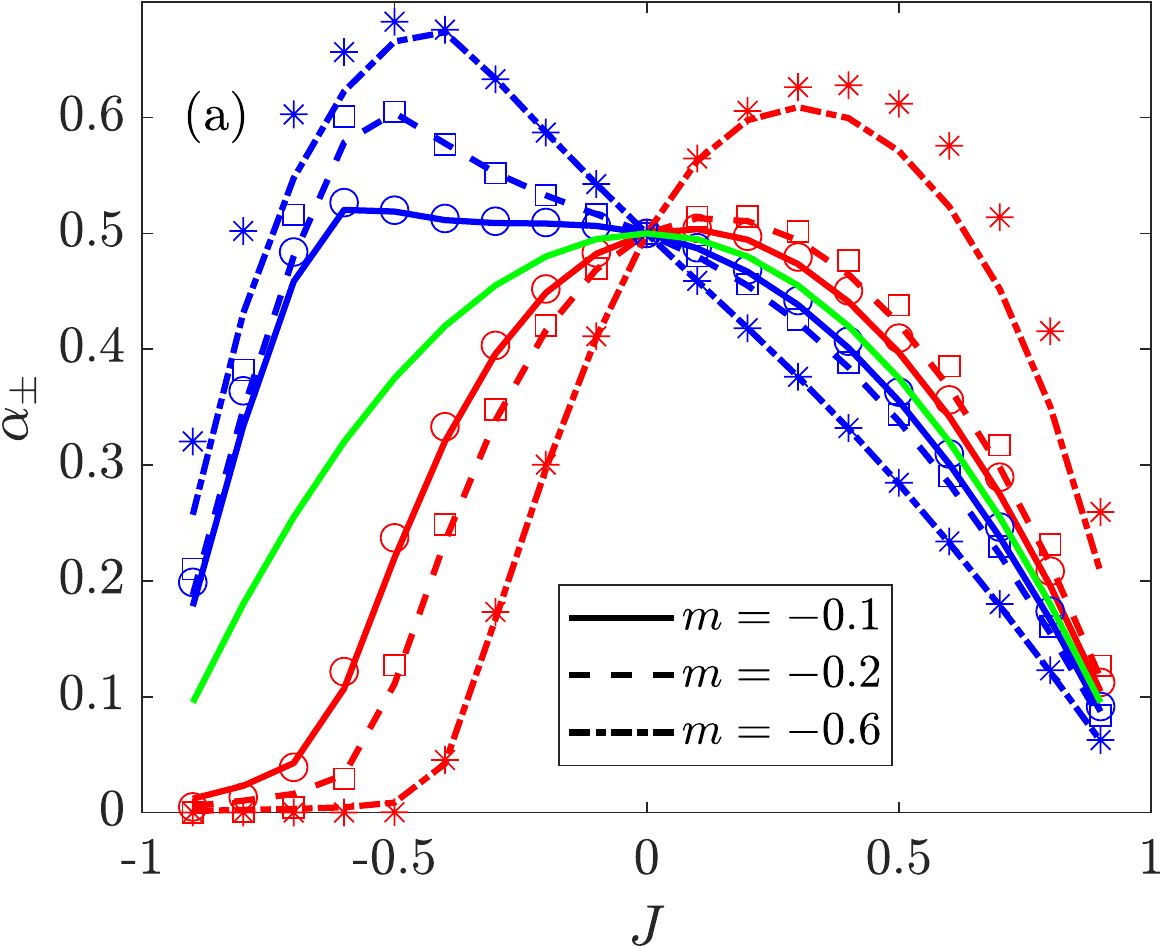}
     \end{subfigure}
     \\
\begin{subfigure}{0.35\textwidth}
         \centering
         \includegraphics[width=\textwidth]{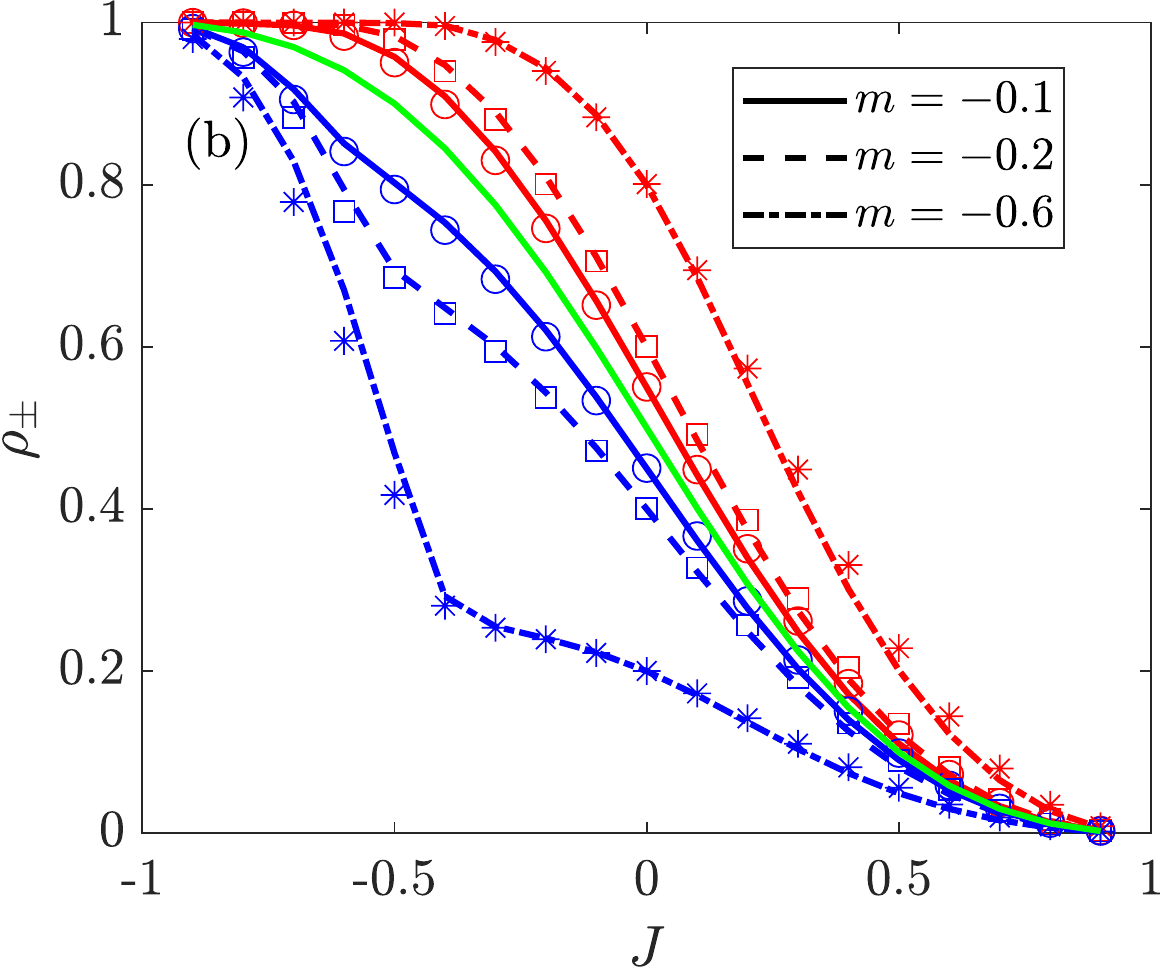}
     \end{subfigure}
\caption{$\alpha_{\pm}$ and $\rho_{\pm}$ \textit{vs.} $J$ in the asymmetric case with different values of $m$
with $N=1000$ and $\kappa=60.5$. Red lines and markers: agents of opinion $+1$ (minority); 
blue lines and markers: agents of opinion $-1$ (majority).
Symbols $\circ$, $\square$ and $\star$ are associated with the simulation results when $m=-0.1, -0.2$ and $-0.6$, respectively. 
Blue and red lines are mean-field predictions obtained by solving  (\ref{eqn:Na}), (\ref{eqn:CLs}), (\ref{SecondEqn}) and~(\ref{mu}), see Sec.~V. For comparison, green lines show the predictions~(\ref{alpha*}) for $m=0$.}
\label{fig:Fig3}
\end{figure}

As Fig.~\ref{fig:Fig3}(a) illustrates, the asymmetric case is of great difference from the symmetric case
even for the smallest $m$, except that 
asymmetry has no effect on the fraction of adders 
at $J=0$. The effect of homophily appears to be \textit{opposite} for the two $\alpha$'s: when $J>0$ and $m<0$, $\alpha_+$ is greater than $\alpha$ in the symmetric case, while $\alpha_-$ is less than that $\alpha$ in the case $m=0$.
On the heterophily side ($J<0$), we find richer phenomena, namely, the presence
of a ``kink" in the $\alpha_{-}\left(J\right) $ curves at larger $-J$, accompanied by $\alpha_{+}$ becoming
vanishingly small. 
This phenomenon is most clearly seen in the $m=-0.6$
data, shown in Fig.~\ref{fig:Fig3}(a), when $-J$ is larger than $\sim 0.5$. The transition to this overwhelmed state is accompanied by rapid changes
in the slopes  of $\alpha_{-}\left( J\right) $. Such kinks can
be seen even for low asymmetry systems (e.g., $m\simeq -0.1$), provided
the strength of heterophily is sufficiently large, see Sec.~IV.C. As we probe deeper into
this regime (i.e., larger $-J$), we find $\alpha_{-}\left( J\right) $
turning downwards, which is consistent with $\alpha$ vanishing in the
limits $\left\vert J\right\vert =1$.

For the fraction $\rho_{\sigma}(J)$ of CLs associated in the group holding
opinion $\sigma$, we have $\rho_{\pm}\left( 0\right) =n_{\mp}=(1\mp m)/2$ and find a monotonic decrease with $J$
in Fig.~\ref{fig:Fig3}(b). 
Similar to the
behavior of $\alpha_{\pm}$, the curves of $\rho_{\pm}$ \textit{vs.} $J$ also deviate from
the symmetric case  in opposite directions. We find that the minority fraction, $\rho_{+}$ approaches $1$, just as the majority $\rho_{-}$ develops
a kink, as $J$ is decreased towards $-1$. As may be expected, these features occur at about the same value of $J\approx J_c$ as those in $\alpha_{\pm}\left( J\right)$. Beyond the critical $J$,
all $\rho_{\pm} $ approach $1$ rapidly when $J<J_c$, a property which can also be
understood intuitively: heterophily is so strong in this regime that the
entire population is locked into establishing CLs. 


\subsection{Mean degrees and associated variances}
\begin{figure}[!t]
\centering
\begin{subfigure}{0.23\textwidth}
         \centering
         \includegraphics[width=\textwidth]{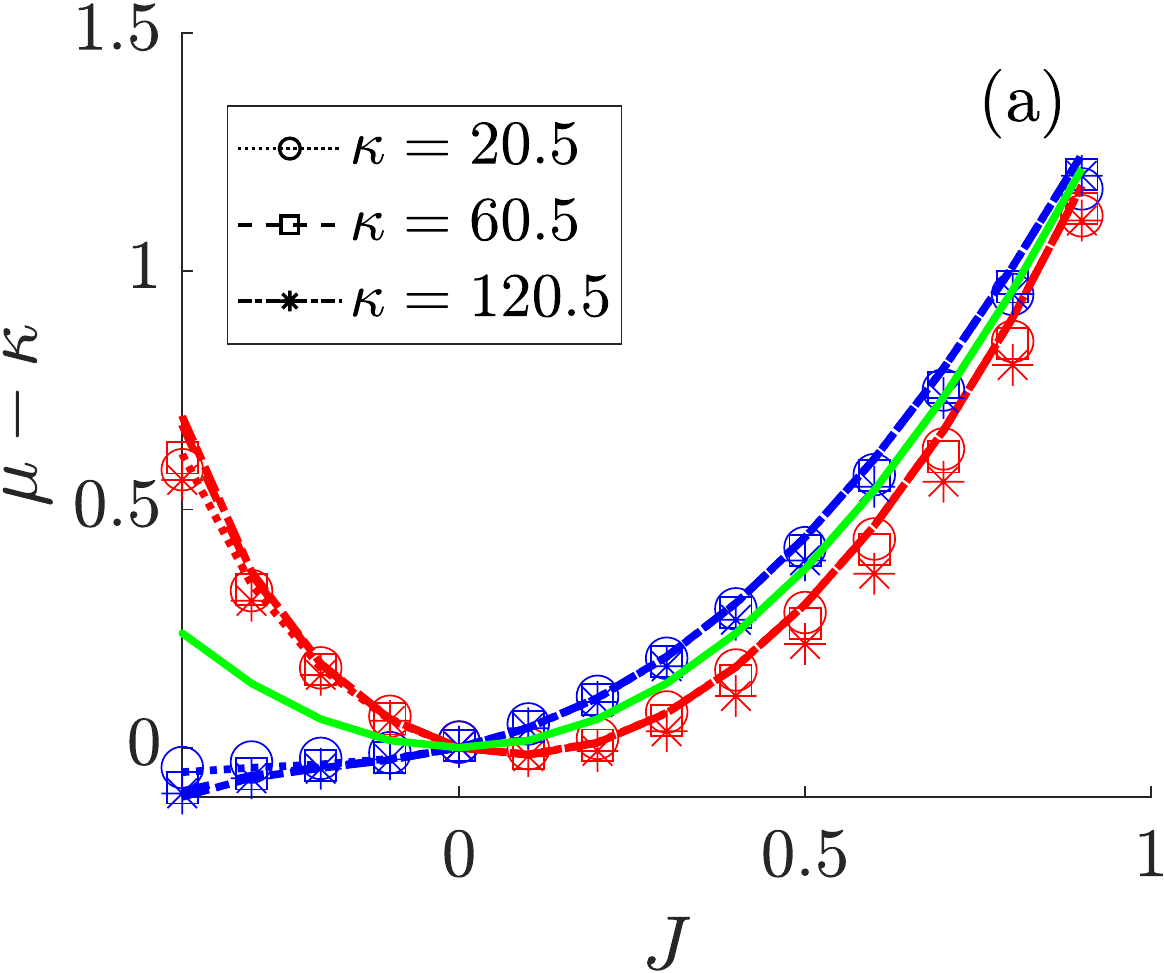}
     \end{subfigure}
\begin{subfigure}{0.23\textwidth}
         \centering
         \includegraphics[width=\textwidth]{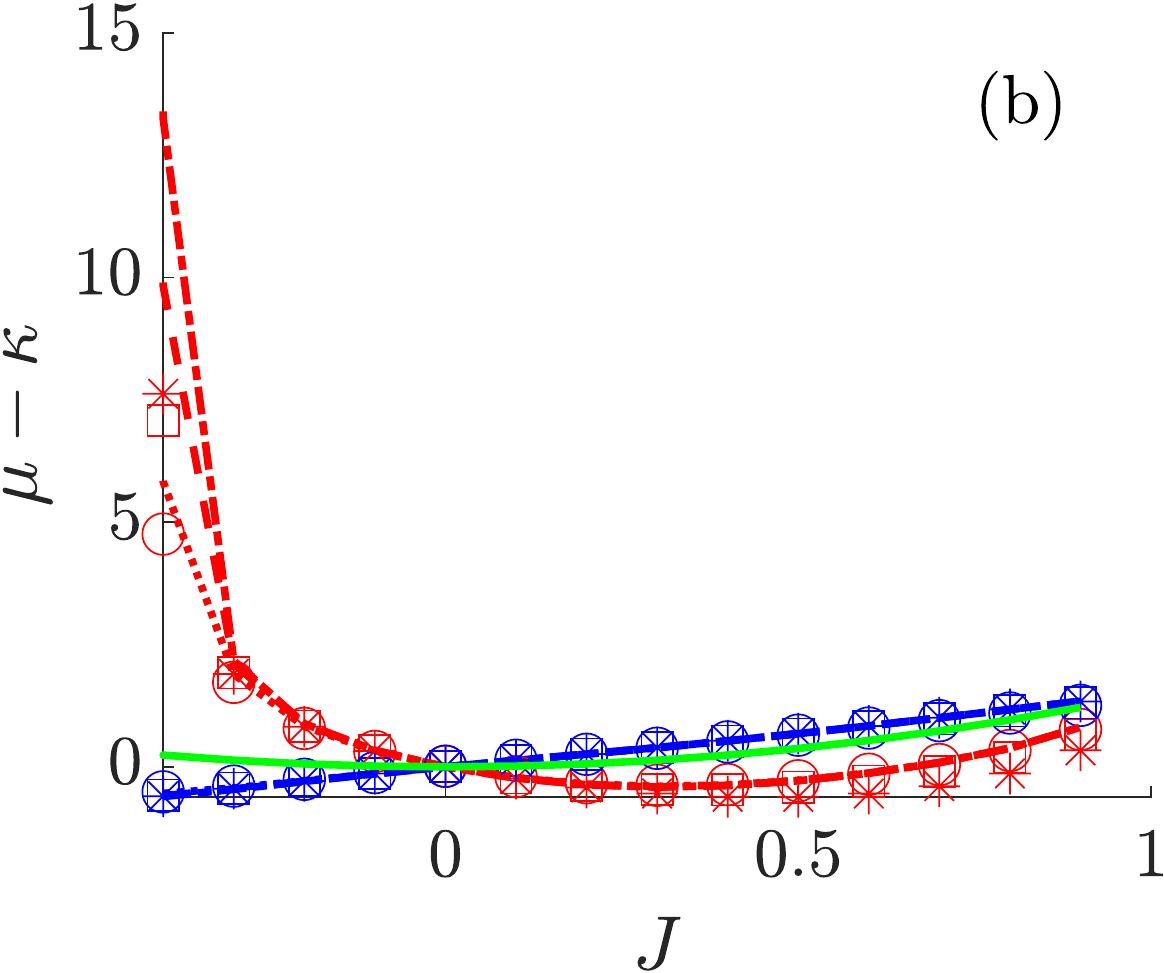}
     \end{subfigure}
     \\
     \begin{subfigure}{0.23\textwidth}
         \centering
         \includegraphics[width=\textwidth]{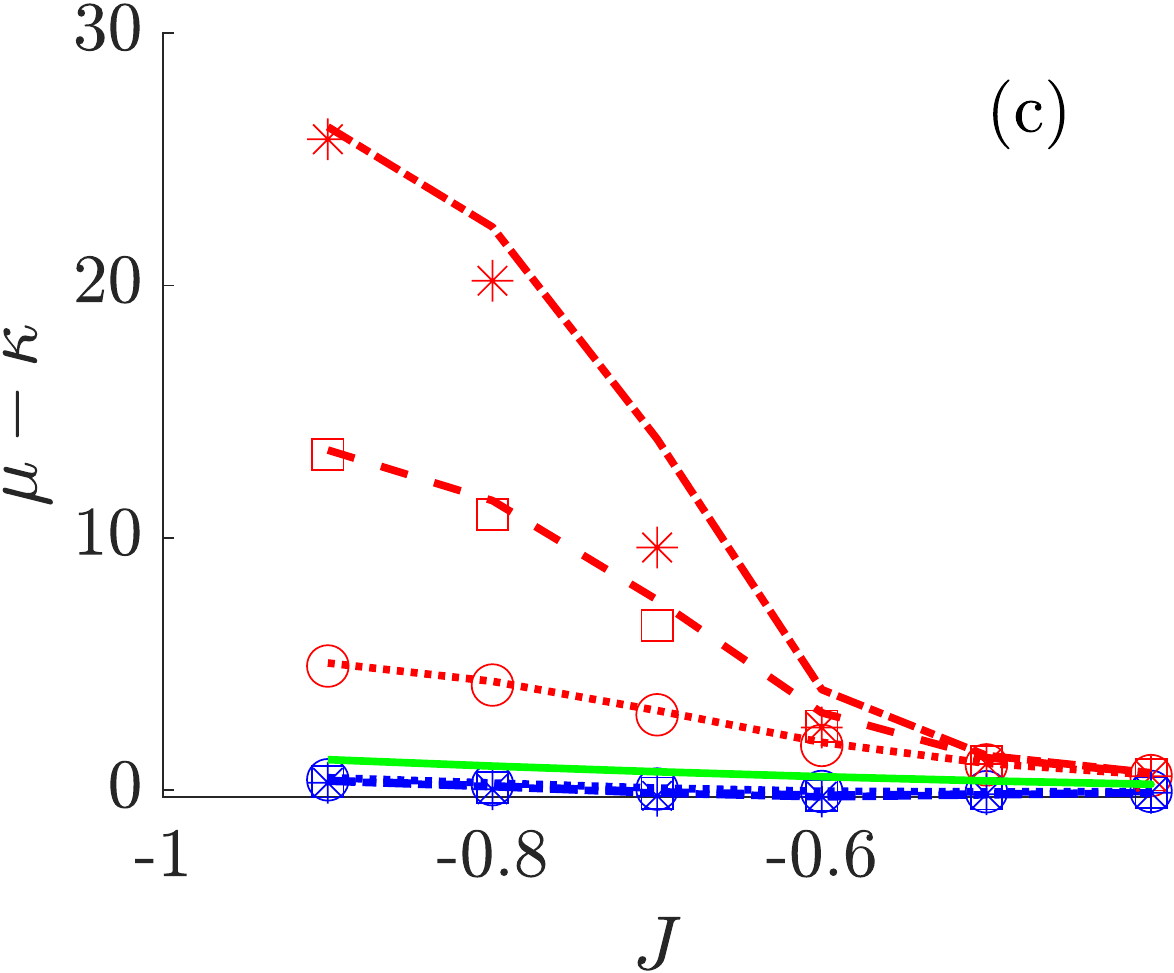}
     \end{subfigure}
\begin{subfigure}{0.23\textwidth}
         \centering
         \includegraphics[width=\textwidth]{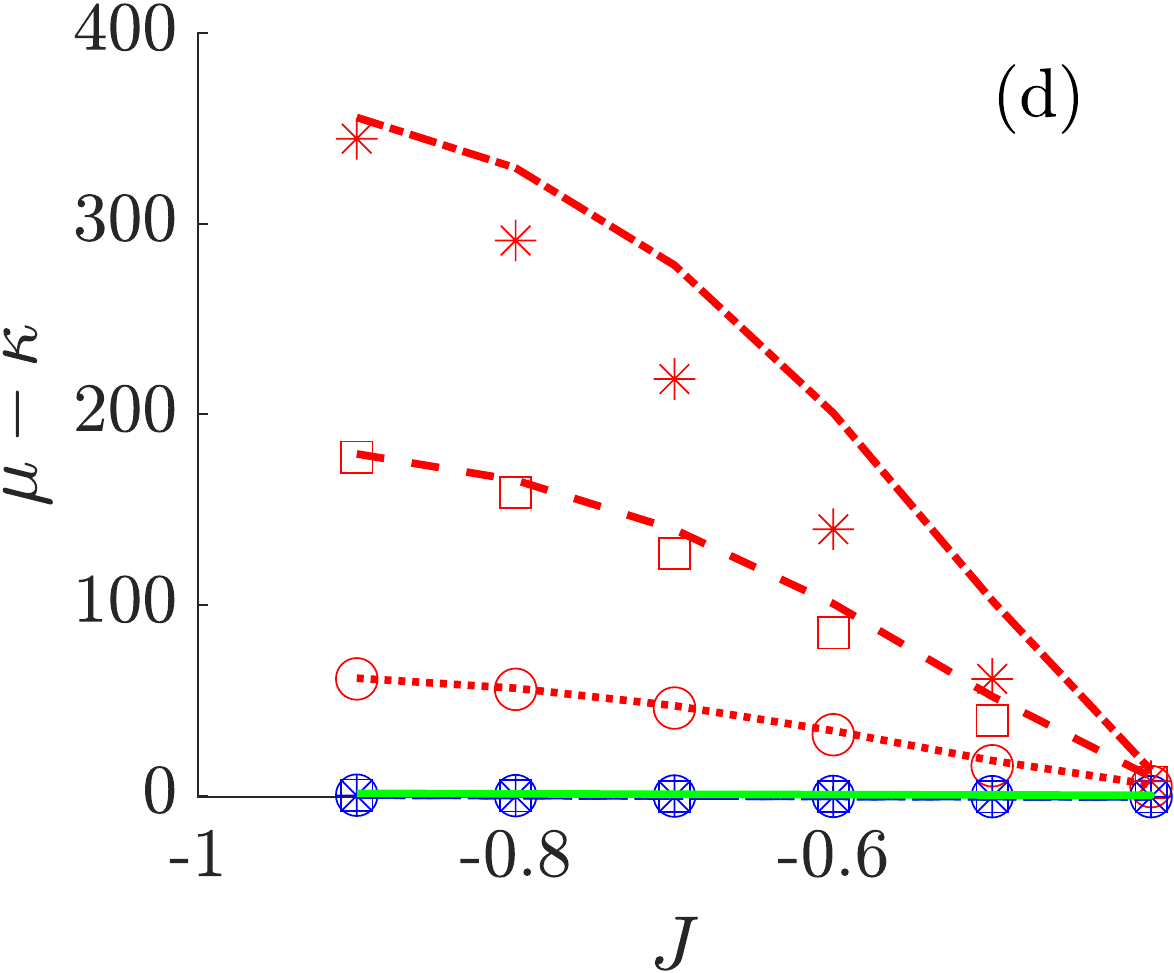}
     \end{subfigure}
     \\
\begin{subfigure}{0.23\textwidth}
         \centering
         \includegraphics[width=\textwidth]{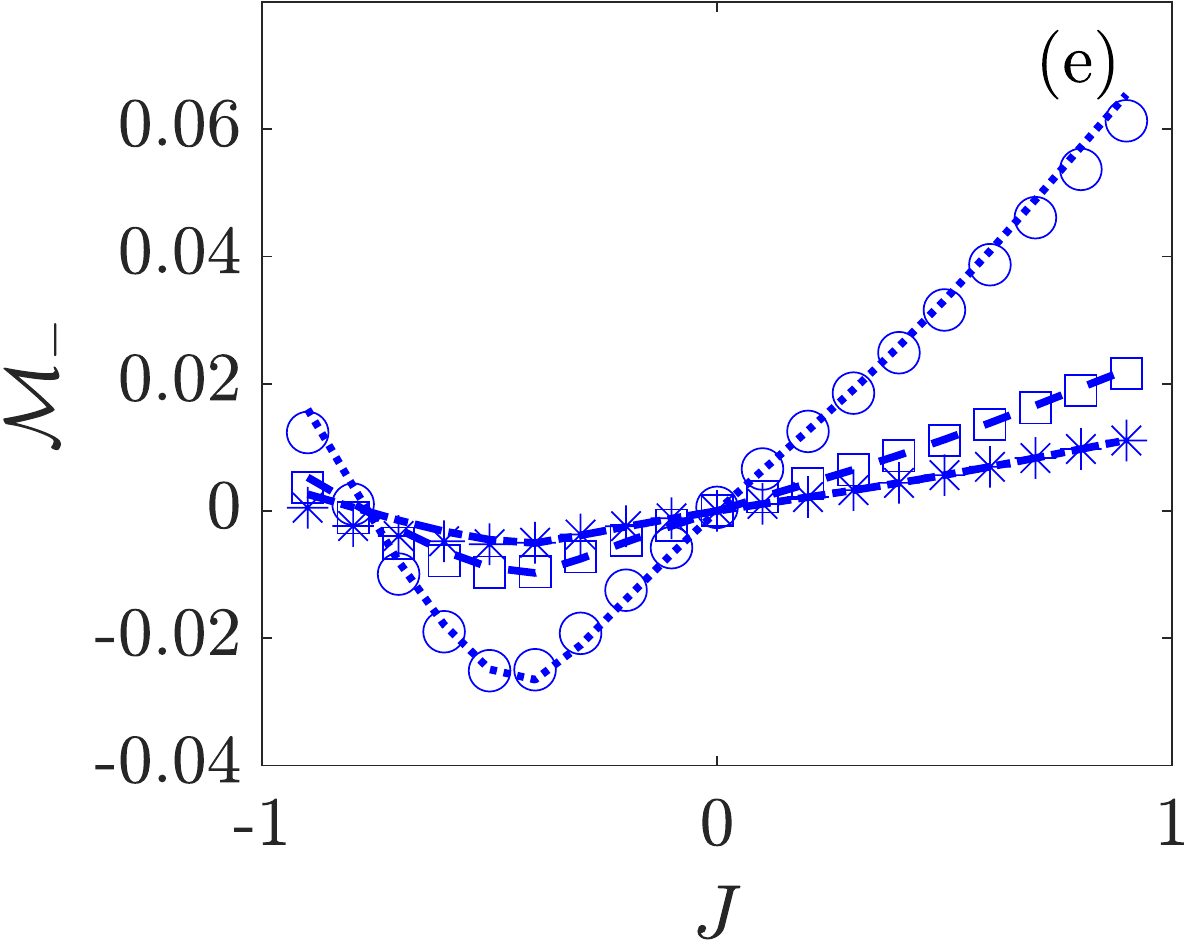}
     \end{subfigure}
\begin{subfigure}{0.23\textwidth}
         \centering
         \includegraphics[width=\textwidth]{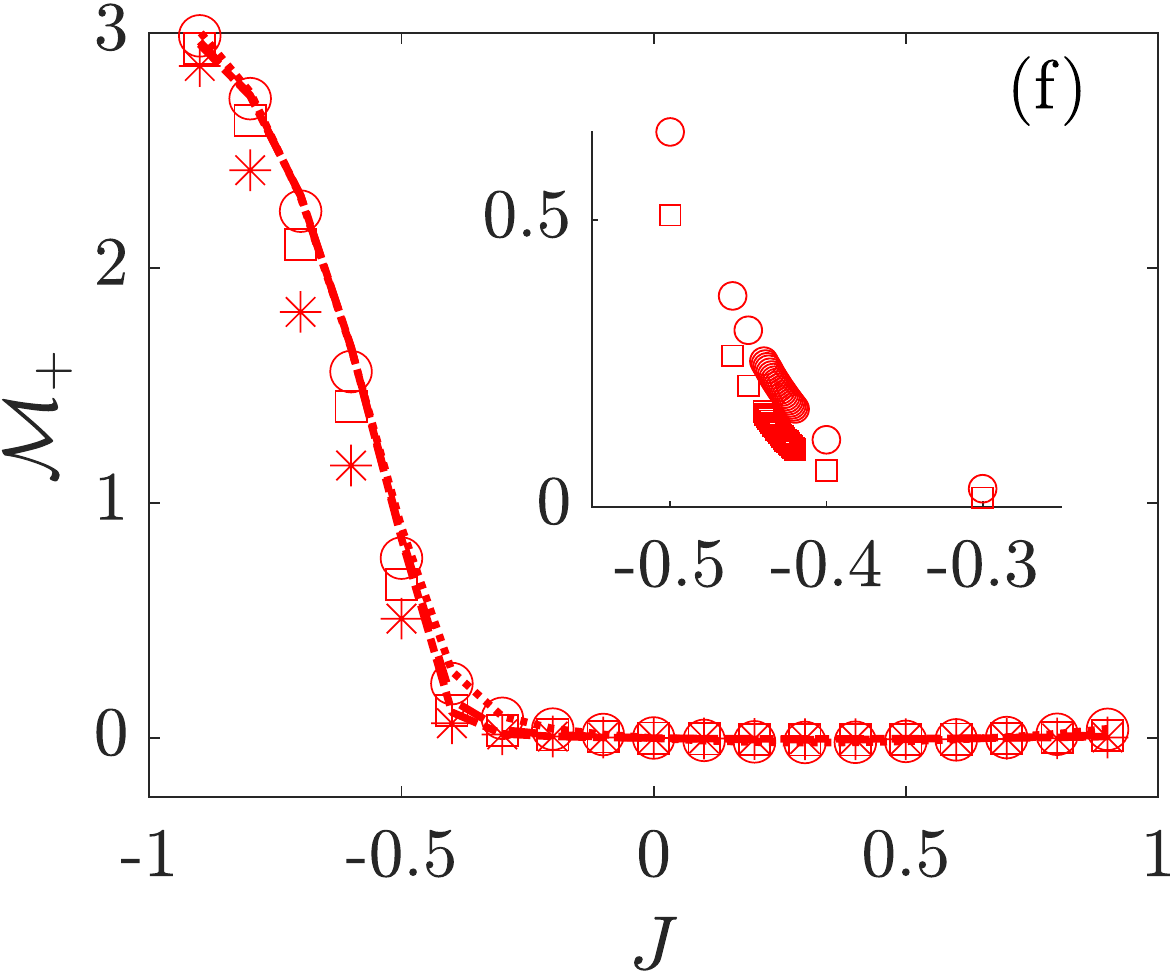}
     \end{subfigure}
     \\
     \begin{subfigure}{0.23\textwidth}
         \centering
         \includegraphics[width=\textwidth]{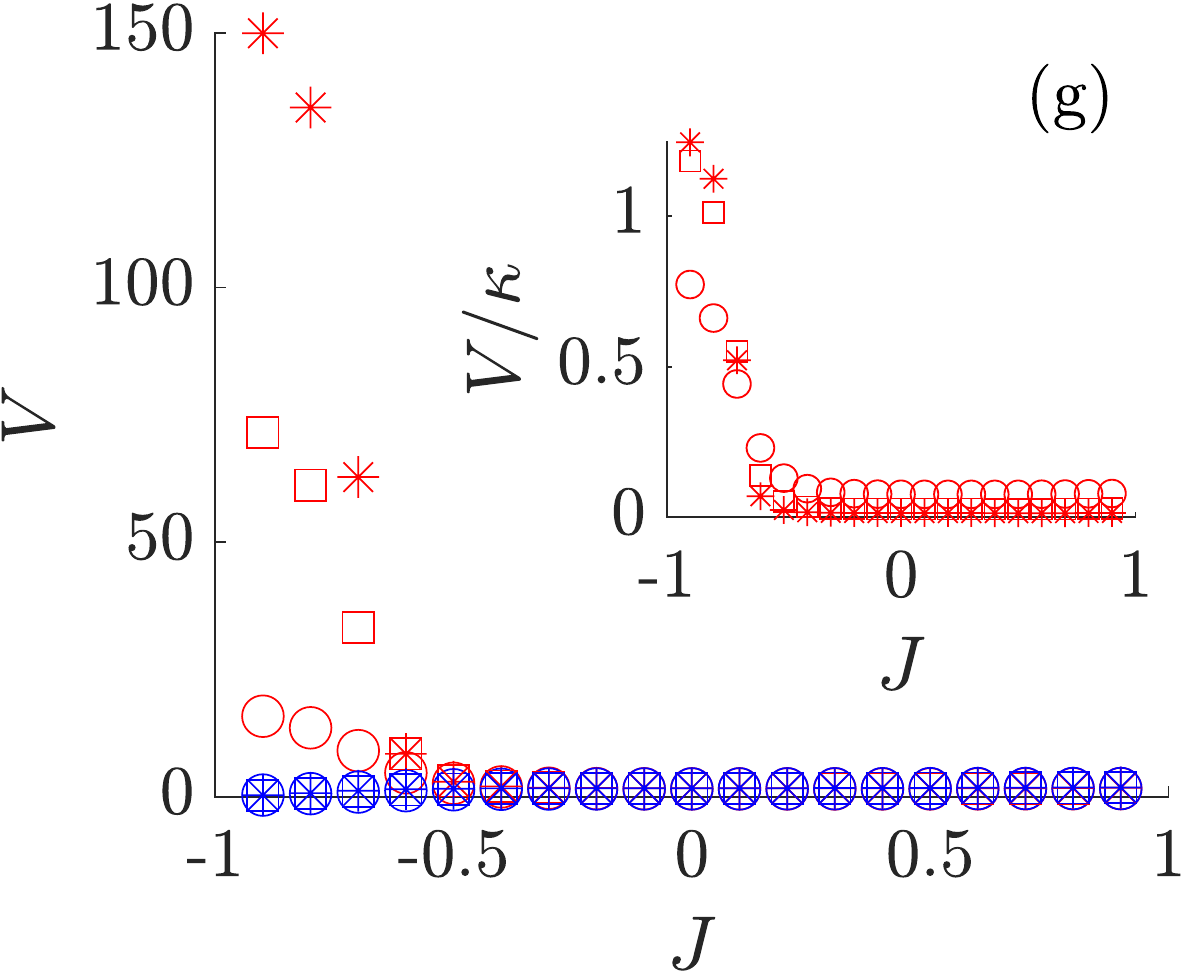}
     \end{subfigure}
     \begin{subfigure}{0.23\textwidth}
         \centering
         \includegraphics[width=\textwidth]{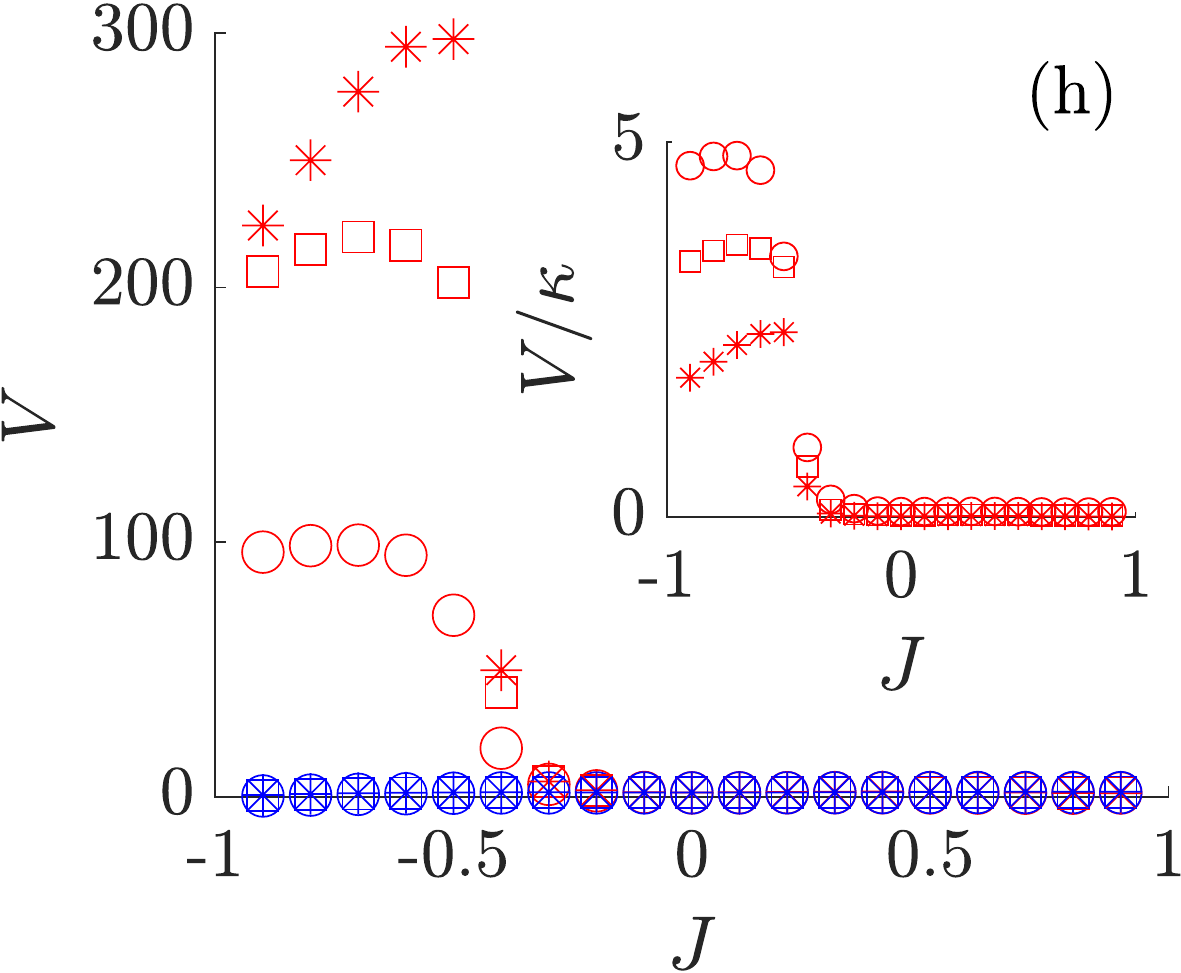}
     \end{subfigure}
\caption{In all panels and insets, markers and lines denote  $\kappa=20.5$ ($\circ$, dotted), $60.5$ ($\square$, dashed), and  $120.5$ ($\ast$, dash-dotted). Red and blue respectively refer to the 
minority ($+$) and majority ($-$) group. Symbols are from simulation data with $N=1000$. In panels (a,c,g), $m=-0.1$ while $m=-0.6$ in (b,d,e,f,h).
In (a)-(d), the green lines show the predictions~(\ref{eqn:avar}) for $m=0$. 
(a,b):  $\mu -\kappa$ \textit{vs.} $J$,  
under low (a) and high (b) asymmetry. Lines are solutions Eqs.~(\ref{eqn:Na}), (\ref{eqn:CLs}), (\ref{SecondEqn}) and~(\ref{mu}) of Sec.~V. The extent of the $J$ axis does not attain  
high heterophily. 
(c,d):  $\mu-\kappa$ \textit{vs.} $J$ for  $J\in[-1,-0.4]$ (significant heterophily). 
(e,f): ${\cal M}_{\mp}$ \textit{vs.} $J$ for the majority (e) and minority (f).
Inset of (f): Blow up of  ${\cal M}_{+}$ \textit{vs.} $J$
 about  transition region (more data points). 
(g,h): DD variances $V_-$ (blue) and $V_+$ (red) \textit{vs.} $J$
under low (g) and high (h) asymmetry.
Insets: $V_+/\kappa$ \textit{vs.} $J$, see text. 
}
  \label{fig:Fig4}
\end{figure}
We now consider the mean degrees ($\mu_{\sigma}$) and 
associated variances ($V_{\sigma}$)
of each community $\sigma=\pm$. 
As in the case of $\alpha_{\pm}$
and $\rho_{\pm}$, the two $\mathcal{\mu}$'s deviate in opposite ways as we
increase $\left\vert J\right\vert$, see Fig.~\ref{fig:Fig4}(a,b). As the data for the
larger $\kappa$'s show, the differences $\mu_{\pm}-\kappa$ in this
regime converge on values which are ${\cal O}\left( 1\right)$, which indicates that the communities are not interacting much. 
In stark contrast, interactions across the communities affect the network 
dramatically under larger heterophily, see  Fig.~\ref{fig:Fig4}(c,d):  while $\mu_{-}$ remains relatively close to $\kappa$, $\mu_{+}-\kappa$ in the minority group 
is strongly $\kappa$ dependent, with a pronounced effect for
large asymmetry, see Fig.~\ref{fig:Fig4}(d) [note the scale of the $\mu-\kappa$ axis]. In this regime, the minority starts being  ``overwhelmed" once $-J$ rises beyond the aforementioned $-J_c$. Indeed,  the average degree of minority agents can be {\it enhanced} by a   factor $\mathcal{E}\equiv \mu_{+}/\kappa\simeq \check{J} N_-/N_+$ which can 
be much larger than unity~\footnote{This can be derived by noting that $\mu_-\simeq \kappa$, so that $L_{\times}\approx N_-\kappa \check{J}$.
 By estimating $\mu_+\simeq L_{\times}/N_+=N_-\kappa \check{J}/N_+=\mathcal{E}\kappa$, we have $\mathcal{E}=\mu_+/\kappa\simeq \check{J} N_-/N_+$. }.
Thus, instead of the difference $\mu_{\sigma} -\kappa $, we
plot (for large asymmetry, $m=-0.6$)%
\begin{equation*}
\mathcal{M_{\sigma}}\equiv \frac{\mu_{\sigma} -\kappa}{\kappa}
\end{equation*}%
\textit{vs.} $J$ in Fig.~\ref{fig:Fig4}(e,f). From simulation 
results for $\mathcal{M}_{-}$ in Fig.~\ref{fig:Fig4}(e), we 
conclude that $\mathcal{M_{-}}\to 0$ as $\kappa \rightarrow\infty$, 
consistently with $\mu_{-}-\kappa = {\cal O}\left(1\right)$.
In  Fig.~\ref{fig:Fig4}(f), we plot $\mathcal{M}_+=\mathcal{E}-1$ vs $J$
and, from the data collapse of the simulation results (red symbols), we conjecture that $\mathcal{M}_{+}$ converges to some definite,
non-trivial thermodynamic limit. 
The behavior of $\mathcal{M}_{+}$ is hence reminiscent of the Ising magnetization, becoming non-zero below a
critical temperature. 
In our simulation results, $\mathcal{M}_{+}$ appears to execute a smooth crossover through the transition region
(inset of Fig.~\ref{fig:Fig4}(f)).
The DD variances  paint a similar picture, with 
$V_{\sigma}={\cal O}\left( 1\right) $ in the ordinary regime and 
$V_+={\cal O}\left(\kappa \right)$ 
in the overwhelmed state, see Fig.~\ref{fig:Fig4}(g,h). The overall behavior is
qualitatively clear, but the dependence on the parameters
is complex as illustrated in Fig.~\ref{fig:Fig4}(h) 
 [note the scale of the axis]. In
the inset of Fig.~\ref{fig:Fig4}(h),  $V_{+}/\kappa$
scales, to some extent, with $\kappa$ but without converging as $\kappa
\rightarrow \infty$, which suggests the need for a further study 
of finite-size effects to draw conclusions
about the thermodynamic limit.

\subsection{Overwhelming transition region characteristics}

\begin{figure}[htp]
\centering
\begin{subfigure}{0.23\textwidth}
         \centering
         \includegraphics[width=\textwidth]{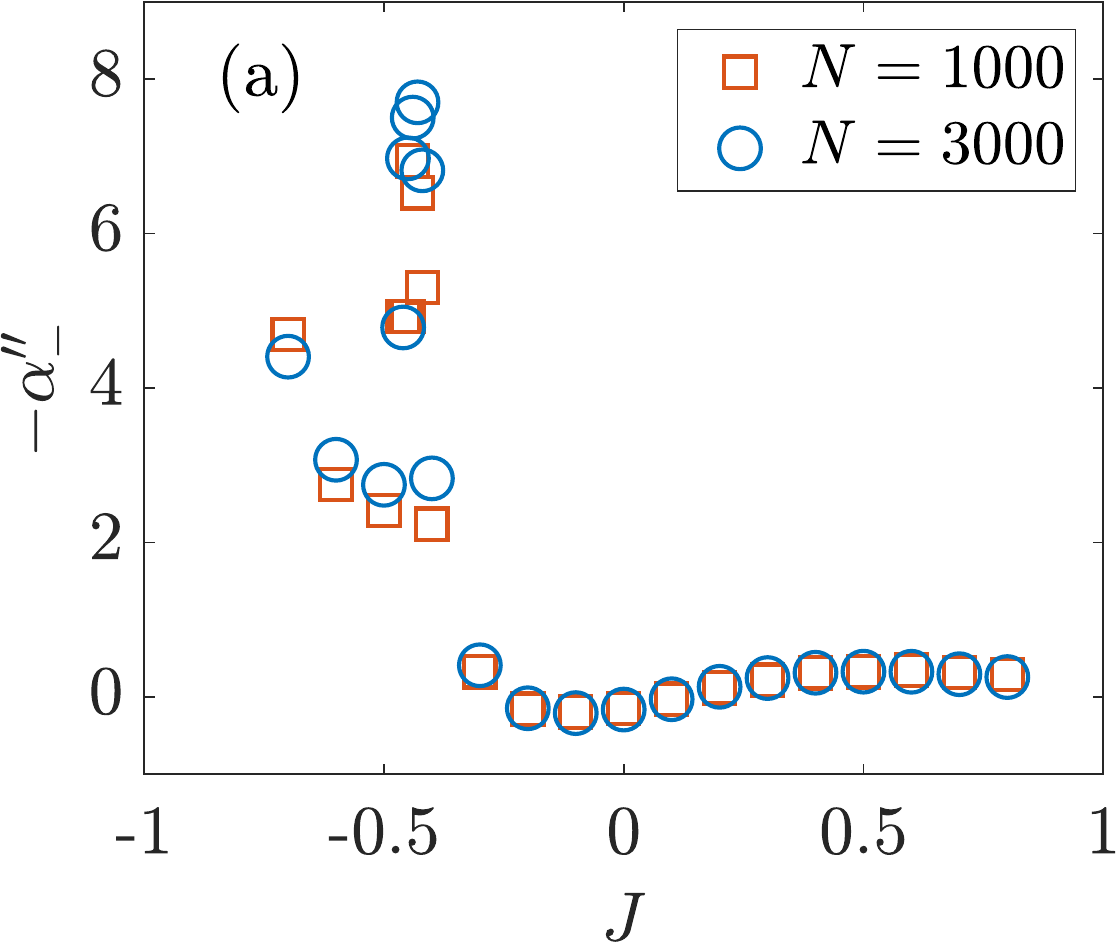}
     \end{subfigure}
 \begin{subfigure}{0.23\textwidth}
         \centering
         \includegraphics[width=\textwidth]{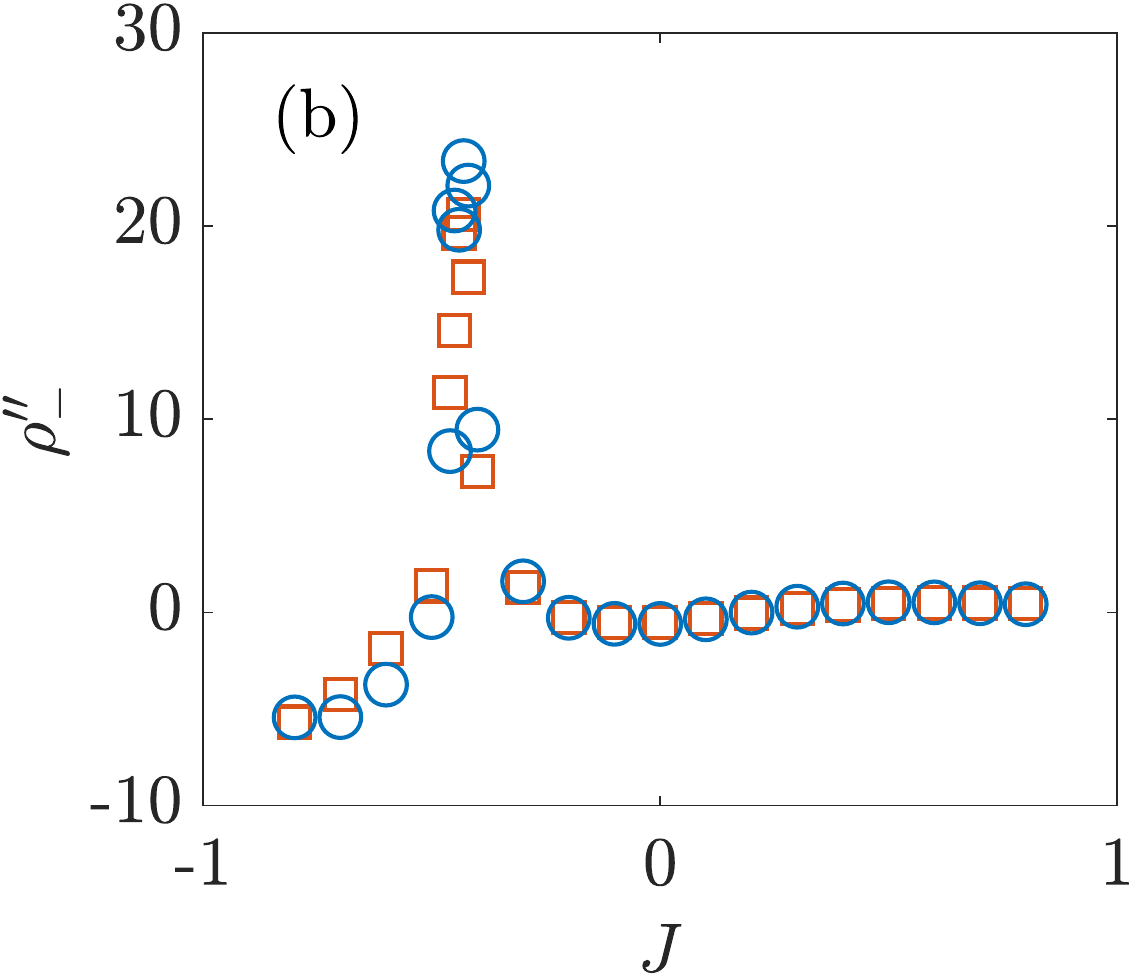}
     \end{subfigure}
     \\
     \begin{subfigure}{0.23\textwidth}
         \centering
         \includegraphics[width=\textwidth]{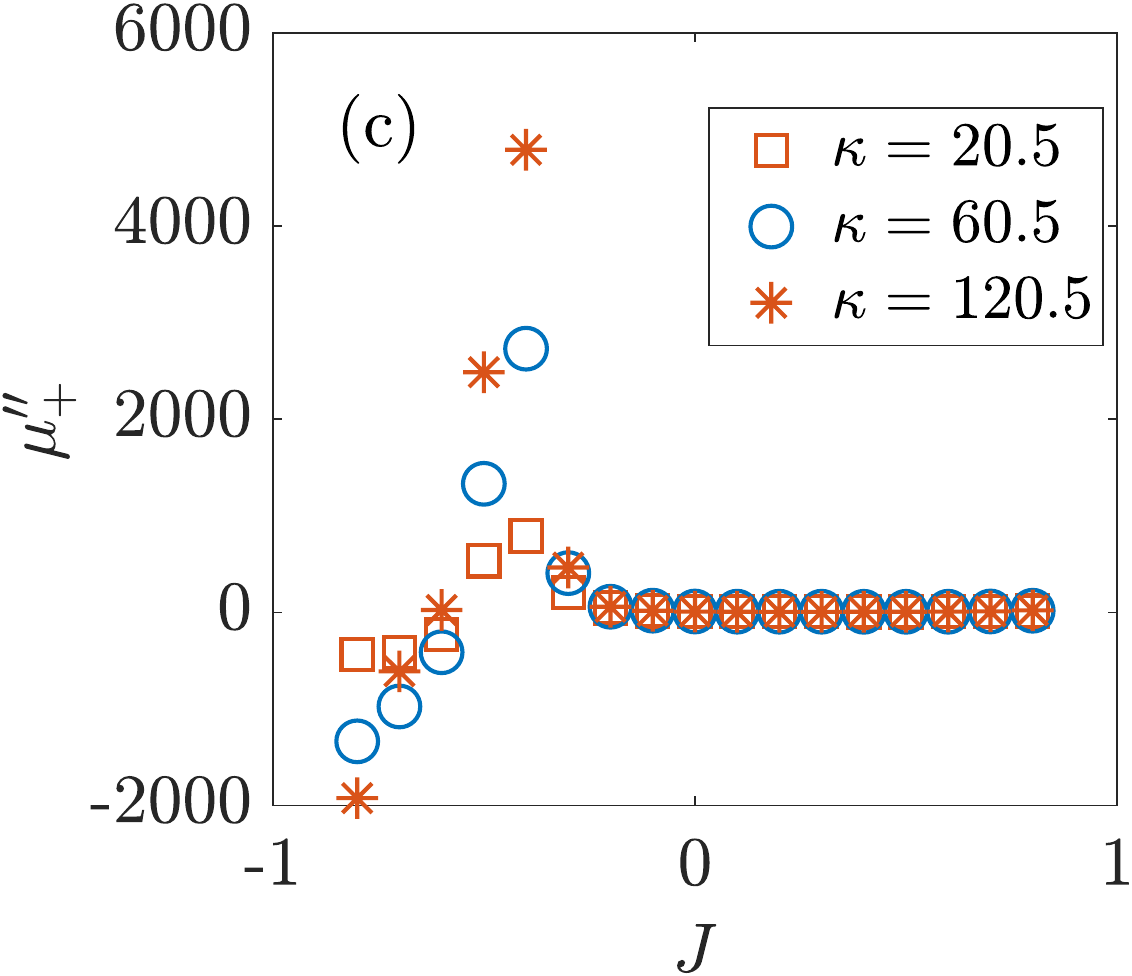}
     \end{subfigure}
 \begin{subfigure}{0.23\textwidth}
         \centering
         \includegraphics[width=\textwidth]{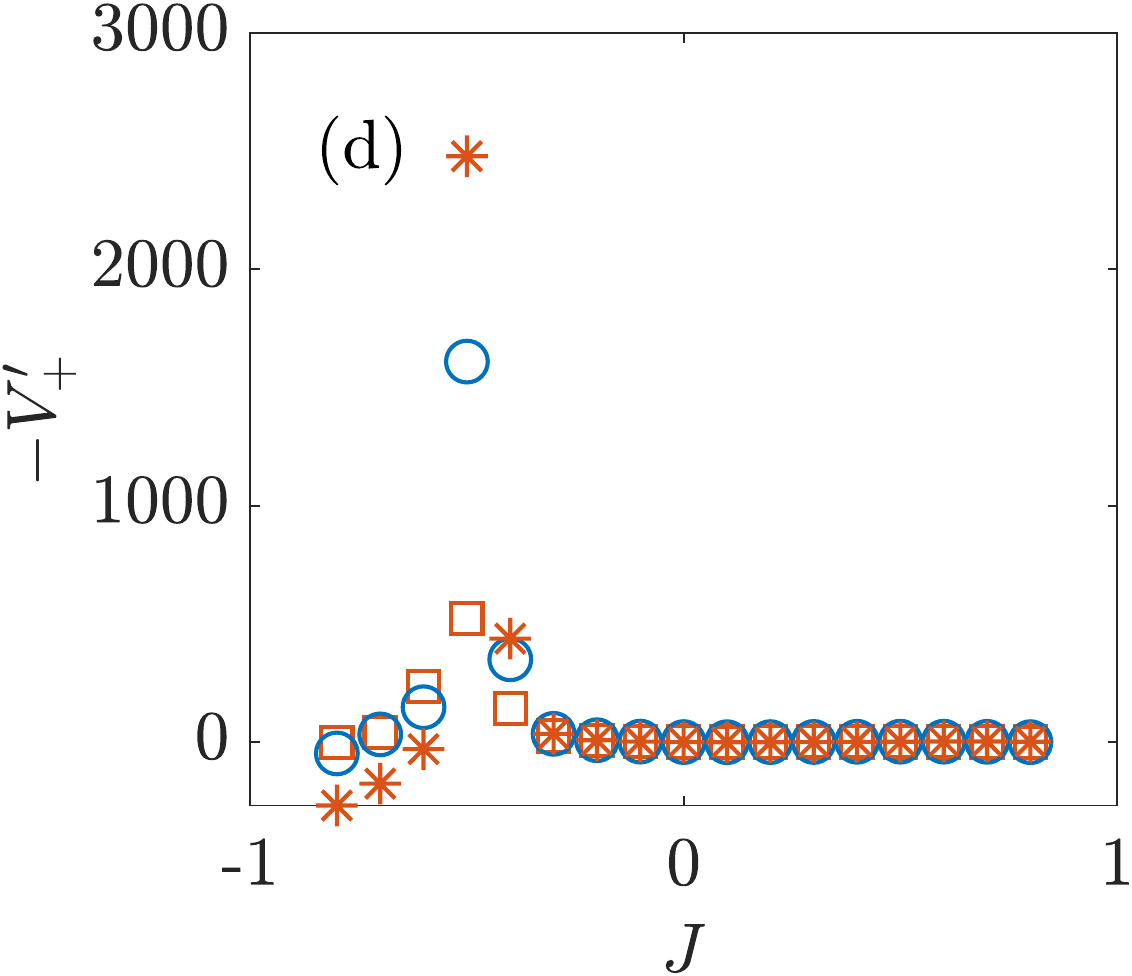}
     \end{subfigure}
\caption{
Highlights of the transition region for systems with $m=-0.6$ (where $()'\equiv d/dJ$):
(a) $-\alpha_-''$ \textit{vs.} $J$; 
(b) $\rho_-''$ vs.$ J$; 
(c) $\mu_+''$ \textit{vs.} $J$; (d) $-V_+'$ \textit{vs.} $J$. All results indicate a sharp peak  around $J\approx -0.42$. 
Other parameters are: $N=1000$ ($\square$), $N=3000$ ($\circ$) and $\kappa=60.5$
in (a,b); $N=1000$, $\kappa = 20.5$ ($\square$), $\kappa =60.5$ ($\circ$), and $\kappa =120.5$ ($\ast$) in (c,d). 
}
  \label{fig:Fig5}
\end{figure}

\begin{figure}[!t]
\centering
         \includegraphics[width=0.35\textwidth]{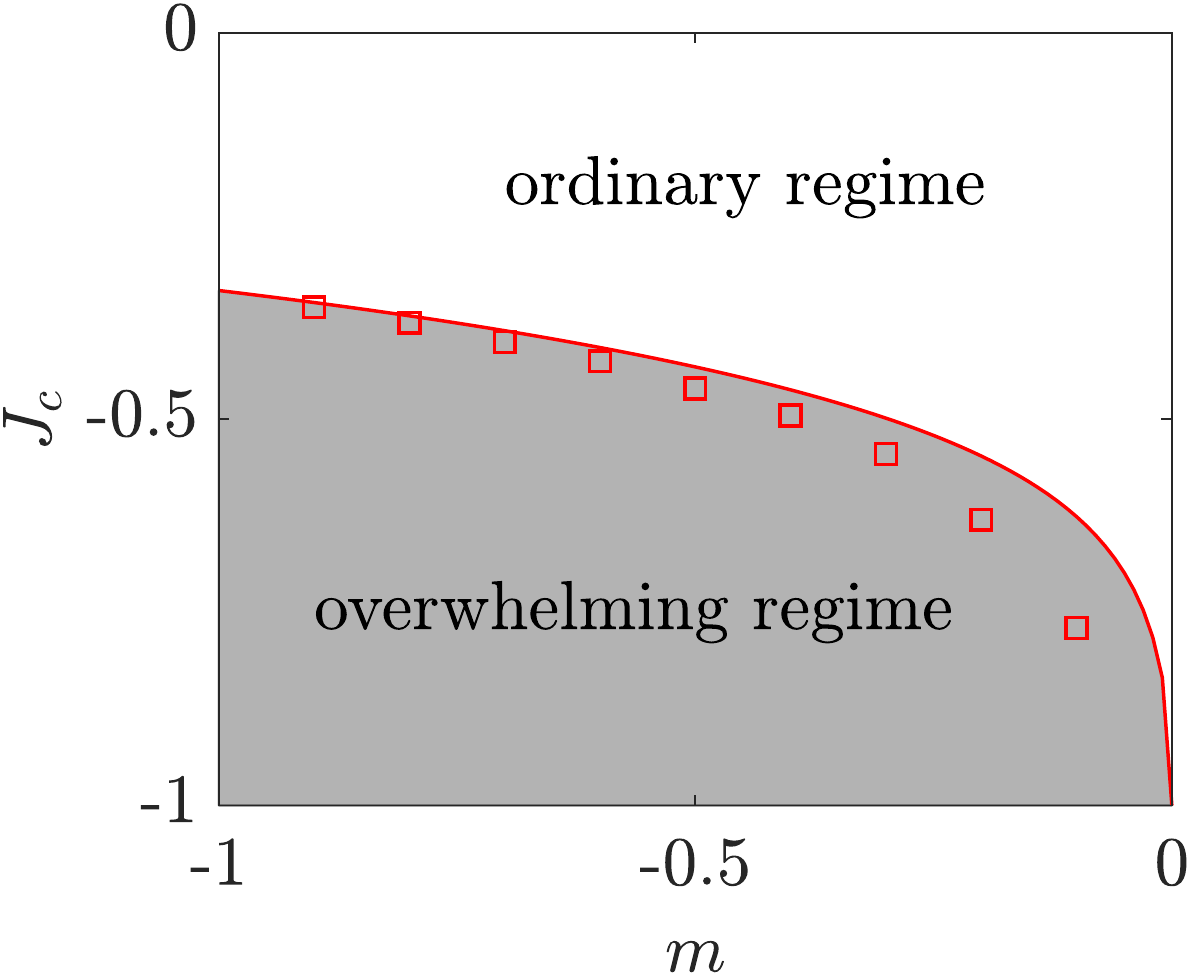}
\caption{$J_c$ \textit{vs.} $m$ (with $N_+<N_-$, i.e. $m<0$) showing the transition line separating the ordinary (white) and overwhelming phases (shaded).
Markers show  $(m,J_c)$ for which simulation data  satisfying $p(\ceil{\kappa})\simeq p(\ceil{\kappa}+1)$. Here,  $N=1000$ and $\kappa=60.5$; while the line is from  Eq.~(\ref{transitionEq}). Note that a mirror diagram with $m \to -m$ is obtained in systems when $N_+>N_-$ ($m>0$).}
\label{fig:Fig6}
\end{figure}

Here, we highlight the transition region
between the ordinary and overwhelming phases 
by providing a perspective of the simulation data based on the derivatives of $\alpha_{-}$, $\rho_{-}$, $\mu_{+}$, $V_{+}$ in Fig.~\ref{fig:Fig5}, see Sec.~S4 in the supplementary material. 
We report the discrete
derivatives 
of these quantities, and 
all results clearly indicate the existence of a sharp peak in the vicinity of $J\approx -0.42$. These peaks correspond to 
the ``kinks'' shown in Fig.~\ref{fig:Fig3}, at which the transition between ordinary and overwhelming phases occurs (``overwhelming transition''). 
The critical point $J_c$ at which this transition occurs depends on $m$,
and  $(m, J_c(m))$ is where $p(\ceil{\kappa})\simeq p(\ceil{\kappa}+1)$ at stationarity, see 
below and Fig.~\ref{fig:Fig7}(d). 
Clearly, $J_c$ is monotonically decreasing when $m$ is negative, implying  that the larger $|m|$, the less heterophily is needed to enter the overwhelming regime. When $m=0$, the system falls into the ordinary regime for any value of $J$~\cite{pre2021}. Remarkably, for any non-zero value of $m$ (non-vanishing level of asymmetry), there is always a critical  level of heterophily $|J_c|$, with $J_c<0$, above which the system is in the overwhelming regime. The features of the transition line $(m, J_c(m))$
are well captured by the mean-field prediction given by Eq.~(\ref{transitionEq}), see  Sec.~V.D and  Fig.~\ref{fig:Fig6}.


\subsection{Degree distributions and joint degree distributions}
\begin{figure}[!t]
\centering
\begin{subfigure}{0.23\textwidth}
         \centering
         \includegraphics[width=\textwidth]{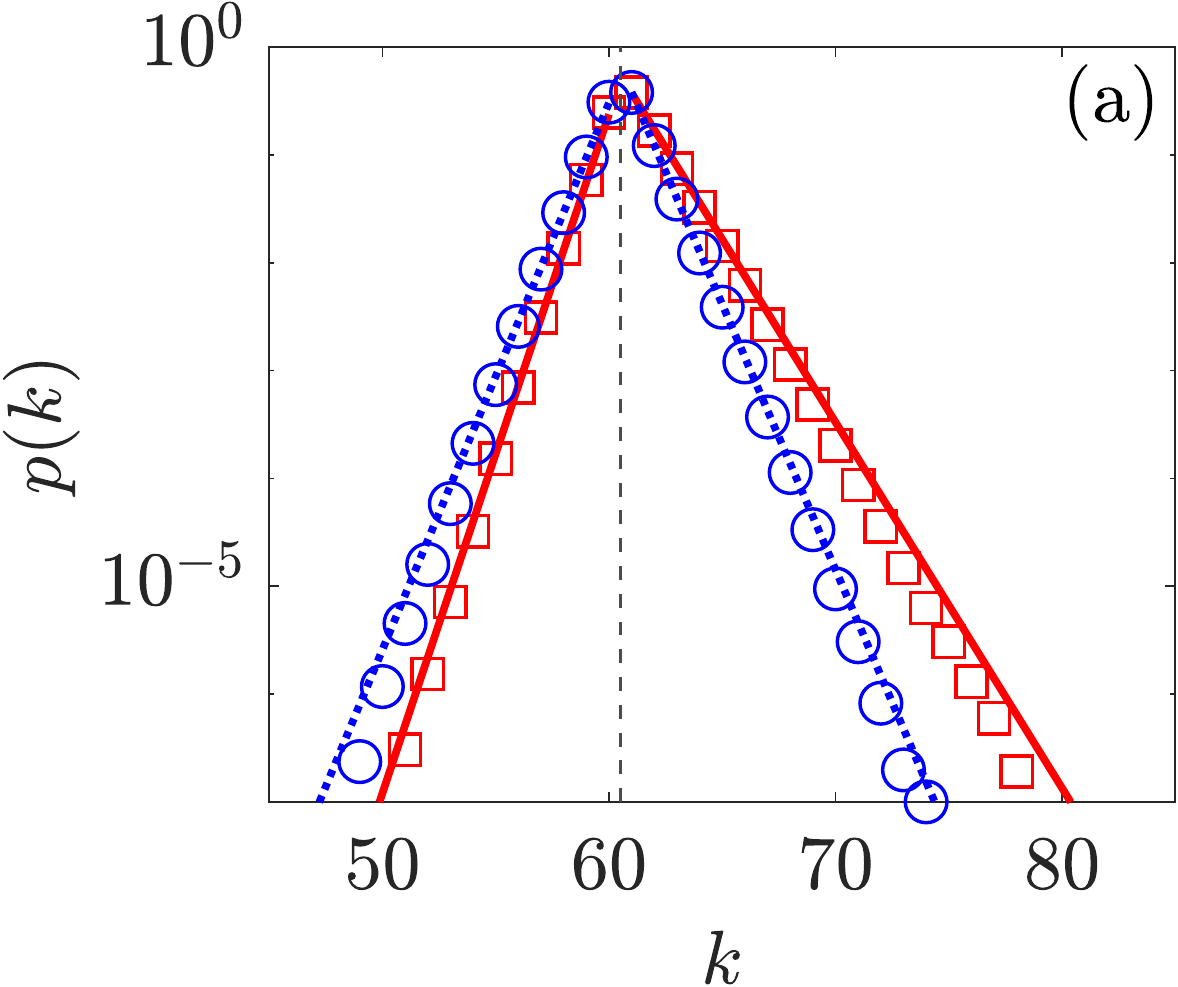}
     \end{subfigure}
\begin{subfigure}{0.23\textwidth}
         \centering
         \includegraphics[width=\textwidth]{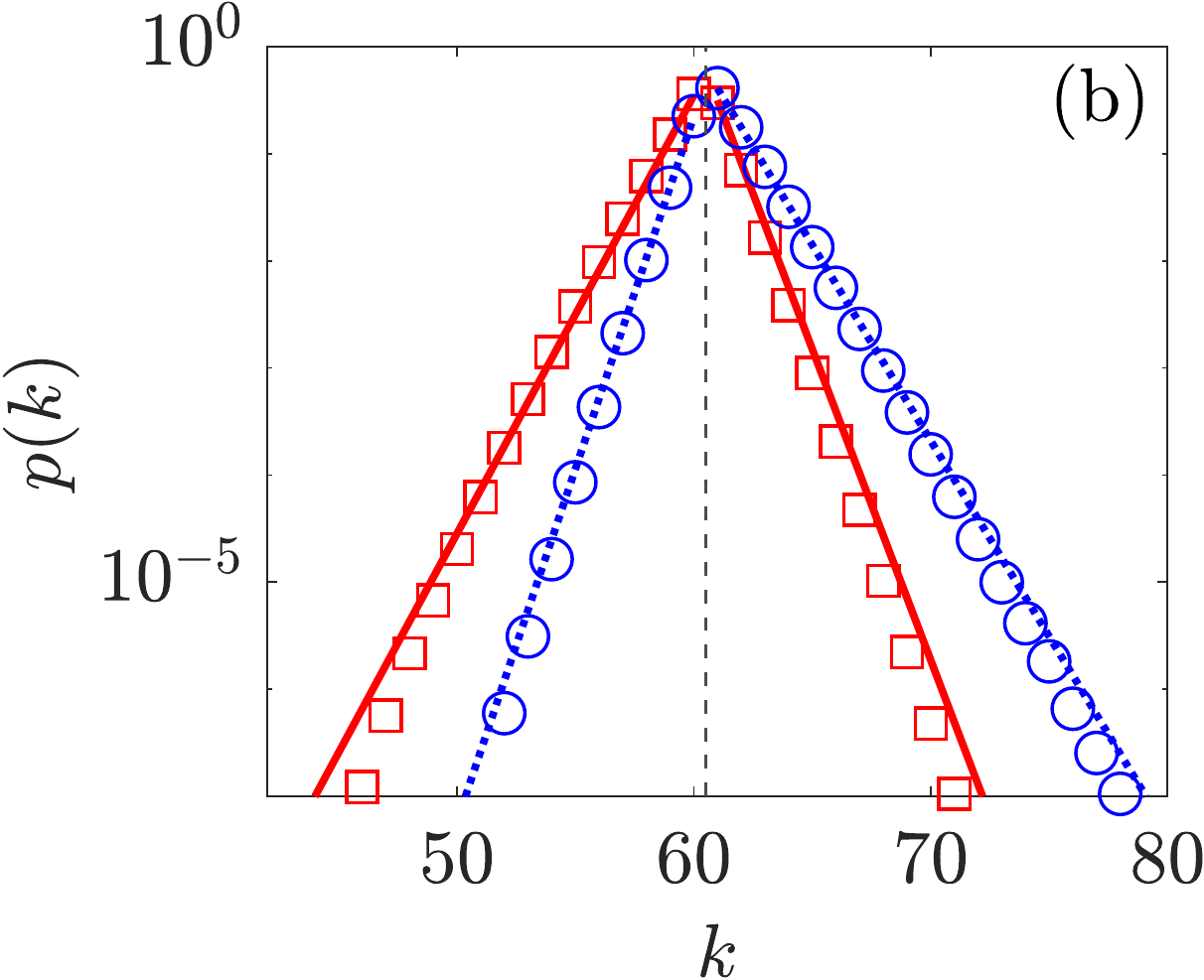}
     \end{subfigure}
     \\
\begin{subfigure}{0.23\textwidth}
         \centering
         \includegraphics[width=\textwidth]{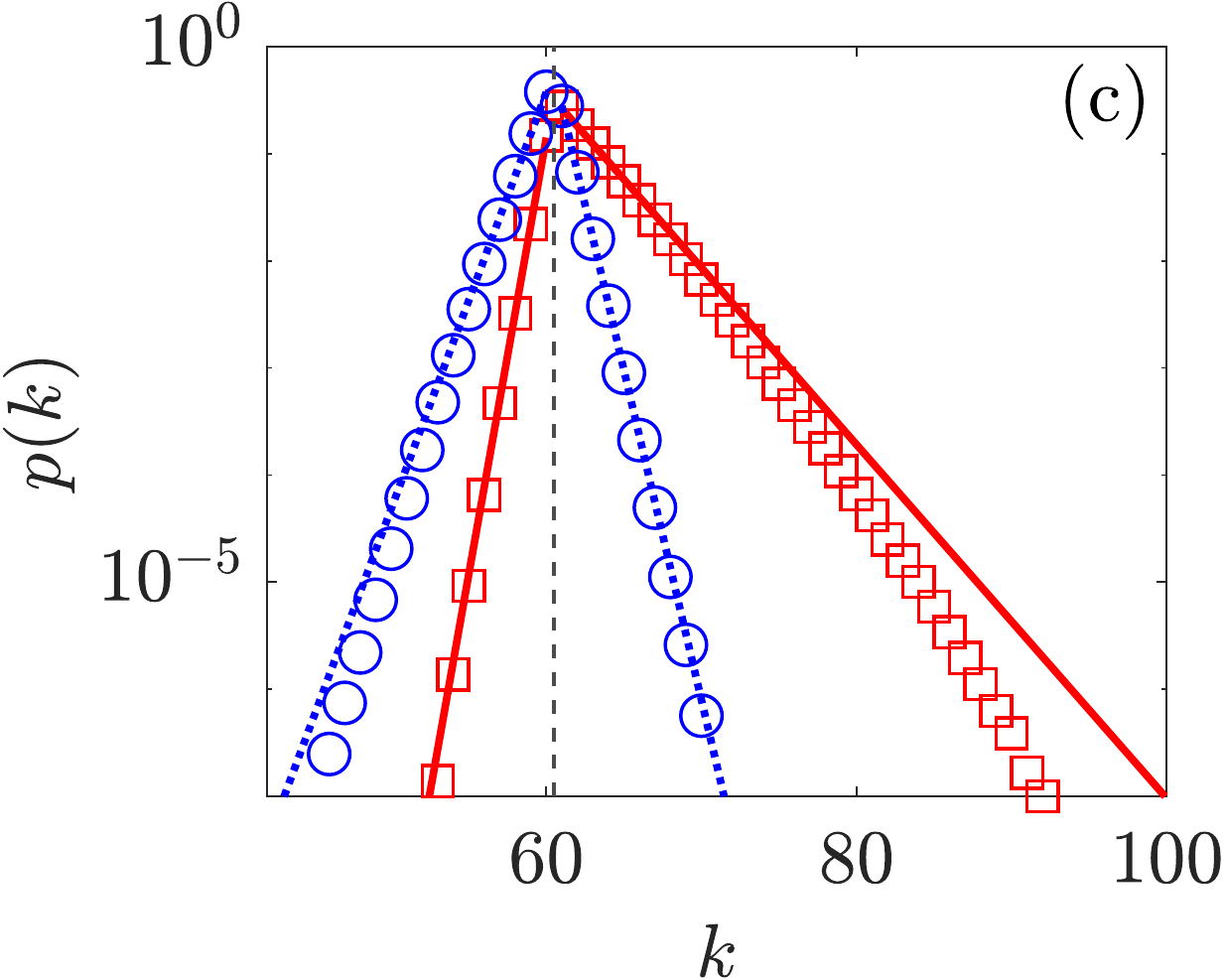}
     \end{subfigure}
 \begin{subfigure}{0.23\textwidth}
         \centering
         \includegraphics[width=\textwidth]{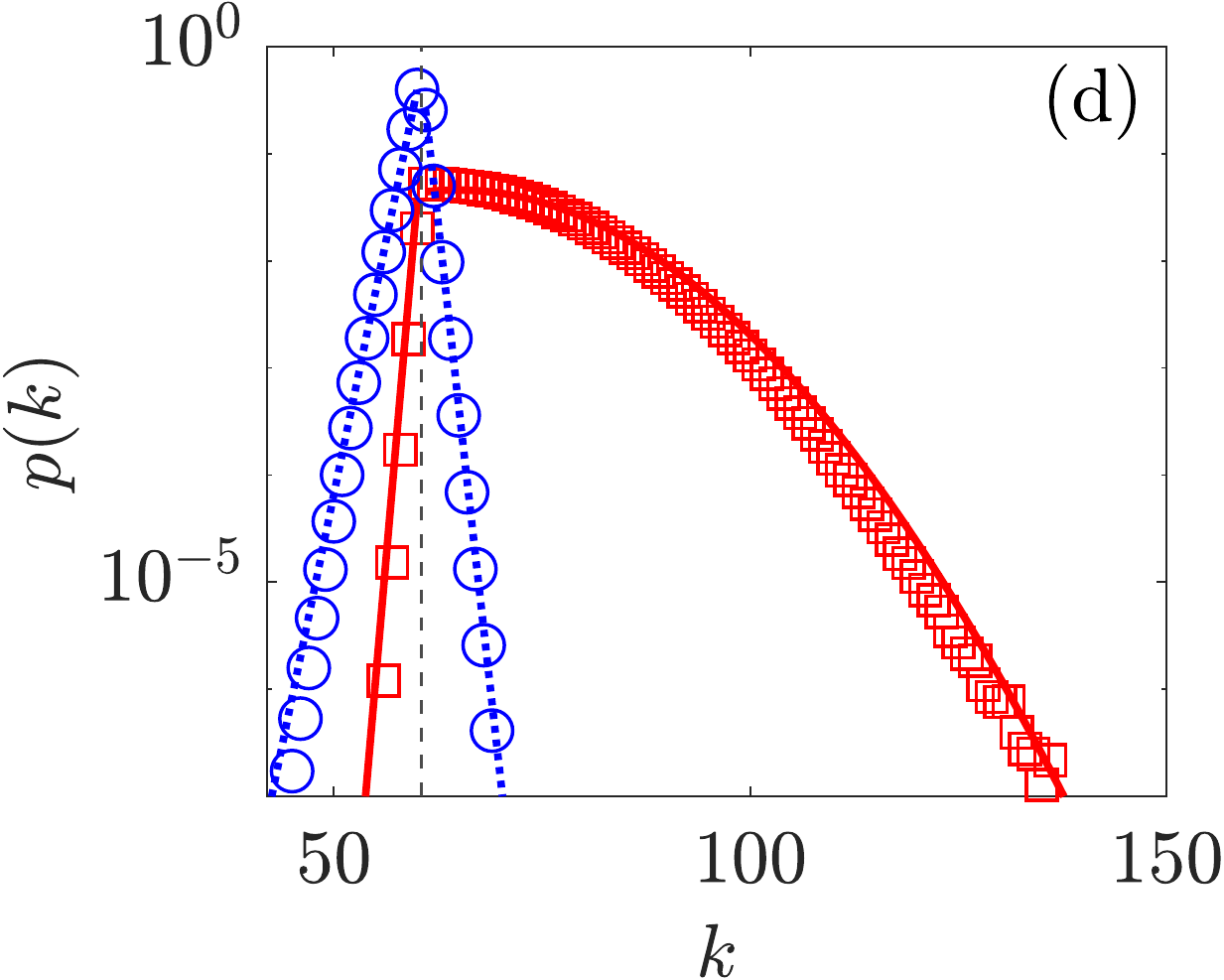}
     \end{subfigure}
     \\
\begin{subfigure}{0.23\textwidth}
         \centering
         \includegraphics[width=\textwidth]{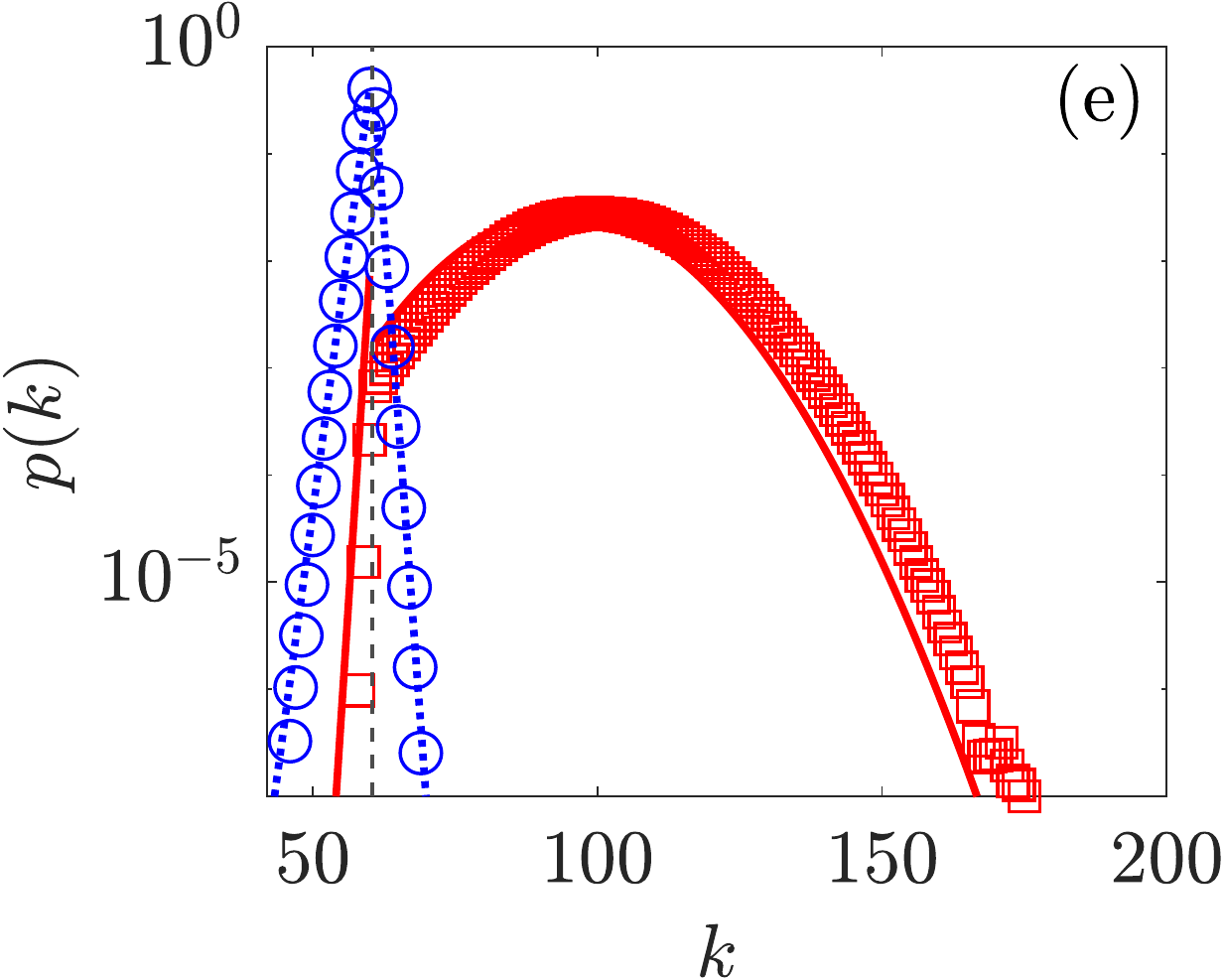}
     \end{subfigure}
     \begin{subfigure}{0.23\textwidth}
         \centering
         \includegraphics[width=\textwidth]{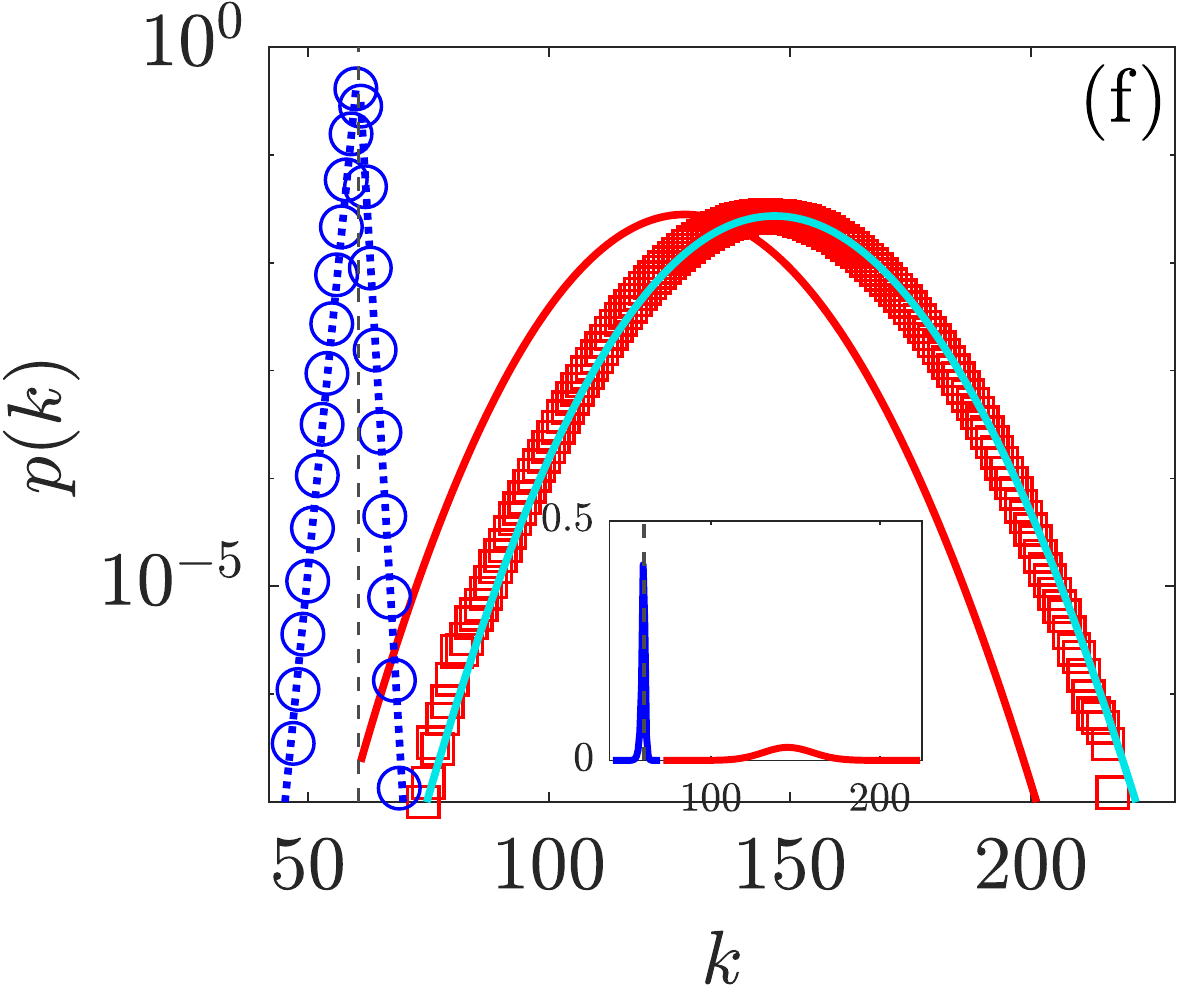}
     \end{subfigure}
\caption{Total degree distributions $p_{\pm}(k)$ for various parameter sets in different regimes with $N=1000$ and $\kappa=60.5$. (a)-(c): $p_{\pm}$ \textit{vs.} $k$ in the ordinary regime;
(a) $(m,J)=(-0.04,-0.5)$; (b) $(m,J)=(-0.6,0.5)$; (c) $(m,J)=(-0.6,-0.3)$.
(d):  $p_{\pm}$ \textit{vs.} $k$ 
$(m,J)=(-0.6,-0.425)$,
at onset of overwhelming transition where 
$p(\ceil{\kappa})\simeq p(\ceil{\kappa}+1)$, see text. 
 (e,f) $p_{\pm}$ \textit{vs.} $k$ in the overwhelming regime;
 (e) $(m,J)=(-0.6,-0.5)$; (f) $(m,J)=(-0.6,-0.6)$.
  Straight blue lines in (a)-(f) are from Eq.~(\ref{DDasy}) given in Sec. V.B; the Gaussian-like curves in (d)-(g) are from Eq.~(\ref{SB}),  derived in Sec.~V.D. Red curves in (d)-(f) are from  Eq.~(\ref{SB}) used together with Eq.~(\ref{MF-QaQc}),  $\mu_-^{a,c}\approx \kappa$ and Eq.~(\ref{alpham}); the cyan curve in (f) is  from Eq.~(\ref{MF-QaQc}) with $\alpha_{-}$ obtained from the simulation data (and $\mu_-^{a,c}\approx \kappa$).
}
  \label{fig:Fig7}
\end{figure}
We now study the total and joint degree distributions. 
At low asymmetry and/or heterophily ($\left\vert m\right\vert \ll 1$ and/or $J>J_{c}$), the DDs $p_{\pm}(k)$ remain approximate exponentials (Laplacian distribution), dropping as $k$ gets further
from $\kappa$. 
As shown in Fig.~\ref{fig:Fig7}(a-c), the Laplacian distributions are not symmetric, as the
slopes on each side (log-linear plot) differ slightly, producing 
$\mu_{\sigma}\neq \kappa$, with the slopes of the DDs in the log-linear plots
and widths being ${\cal O}\left(1\right)$. In our simulations, we found little dependence of the DDs on $\kappa$ and $N_{\pm}$. The slopes of $p_{\pm}$ differ in the opposite directions
for the two communities, corresponding to $\alpha_{\pm}$ deviating from the 
$m=0$ curve in opposite ways.
In Fig.~\ref{fig:Fig7}(c) we see that the right half of the DD of the minority is bent in the log-linear plot, and a Gaussian-like distribution for the cutters in the minority develops gradually as $J$ is decreased further (increased heterophily), with 
$J$ down to $-0.6$ in Fig.~\ref{fig:Fig7}(d-f). These findings illustrate the process of the system transiting from the ordinary 
into the overwhelming regime. To locate the change of phase, the ``overwhelming transition'',
we
assume that the system is in the transition regime when $p(\ceil{\kappa})\simeq p(\ceil{\kappa}+1)$, i.e., the slope of the degree distribution at $\ceil{\kappa}$ of cutters in the minority equals $0$, see  
Fig.~\ref{fig:Fig7}(d).
Within the overwhelming regime, the fraction of adders ($k<\kappa$) decreases substantially, and $p_+(k)$ from an exponential becomes a Gaussian-like distribution. In Fig.~\ref{fig:Fig7}(e,f) we show the DDs deep in the overwhelming phase, where novel behavior emerges: while the
majority keep their DD to be narrowly distributed around $\kappa$, the
dramatic rise of the average degree of the minority agents is accompanied by
significant changes to $p_{+}\left(k\right)$ and a substantial decrease of the adders in the minority.
As illustrated in Fig.~\ref{fig:Fig7}(f) for $m=J=-0.6$, there are no minority nodes with $k<\kappa=60.5$ (i.e. no minority adders), while the distribution is essentially a Gaussian peaking at $k\simeq 145$, well over twice $\kappa$. Unlike the narrow Laplacian, the variance of this Gaussian  is considerably higher,  
$V_{+}\simeq 217$. The inset of Fig.~\ref{fig:Fig7}(f) shows the striking difference between the DDs on linear scale. 

\begin{figure}[th]
\centering
\begin{subfigure}{0.23\textwidth}
         \centering
         \includegraphics[width=\textwidth]{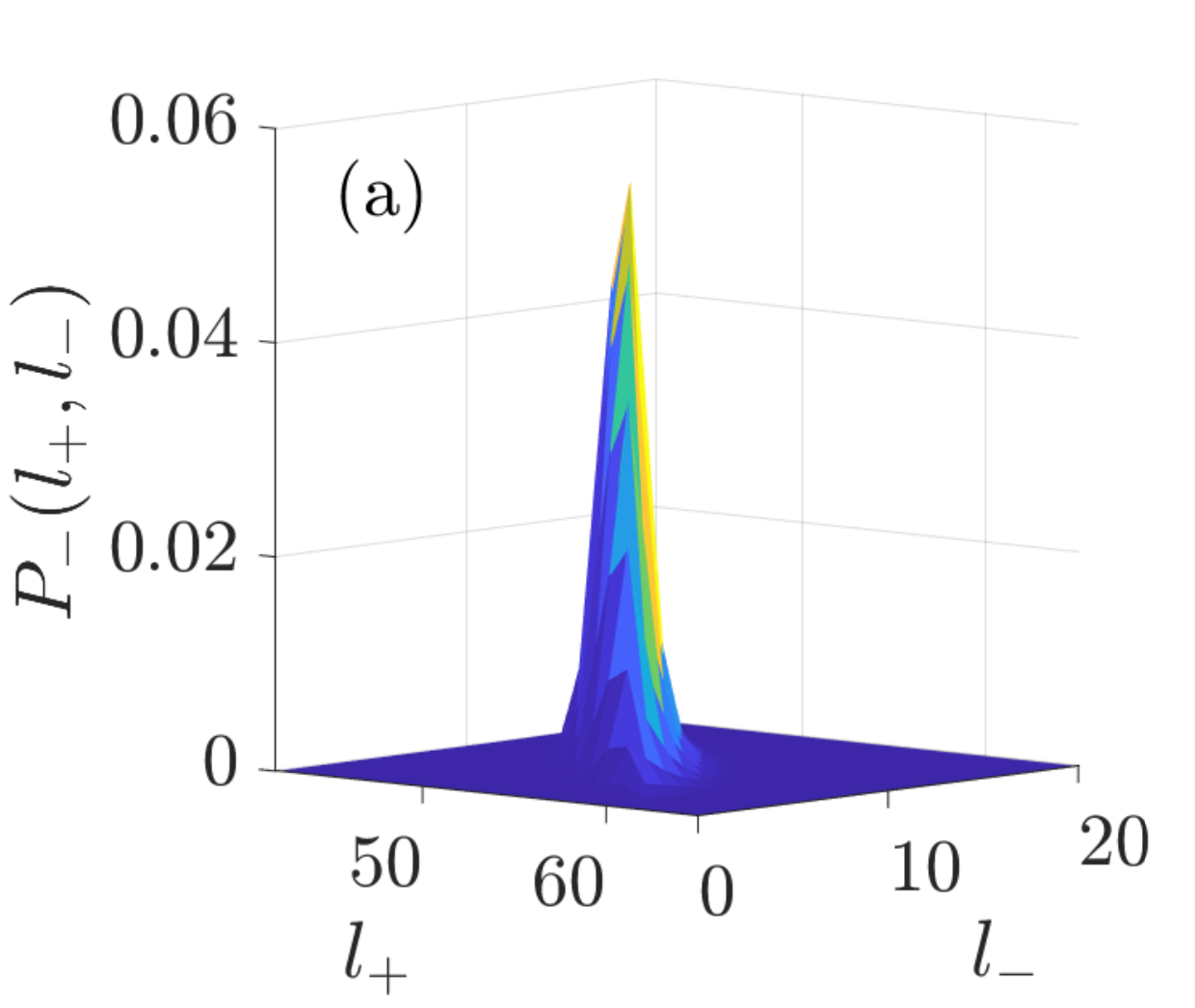}
     \end{subfigure}
\begin{subfigure}{0.23\textwidth}
         \centering
         \includegraphics[width=\textwidth]{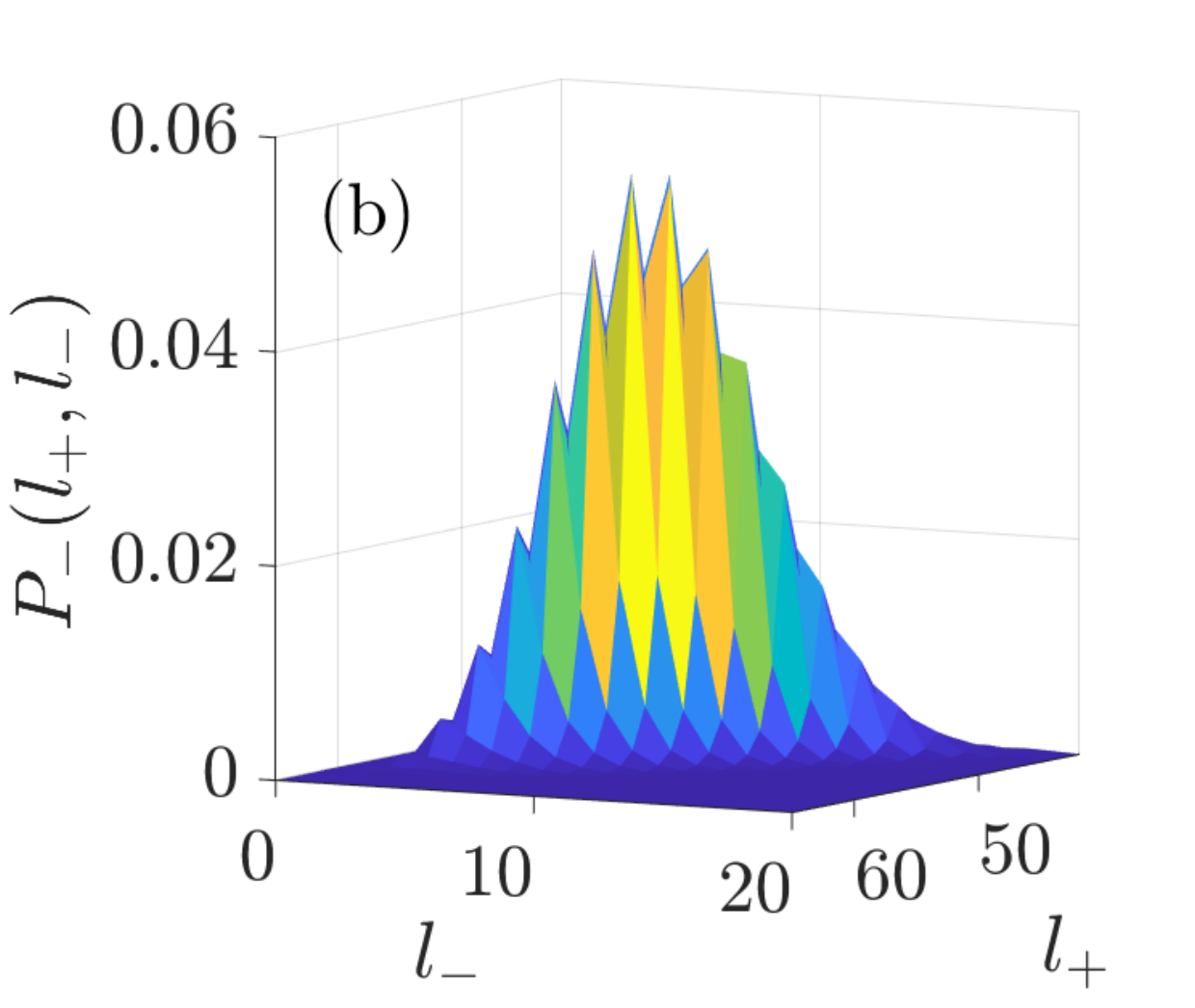}
     \end{subfigure}
\begin{subfigure}{0.23\textwidth}
         \centering
         \includegraphics[width=\textwidth]{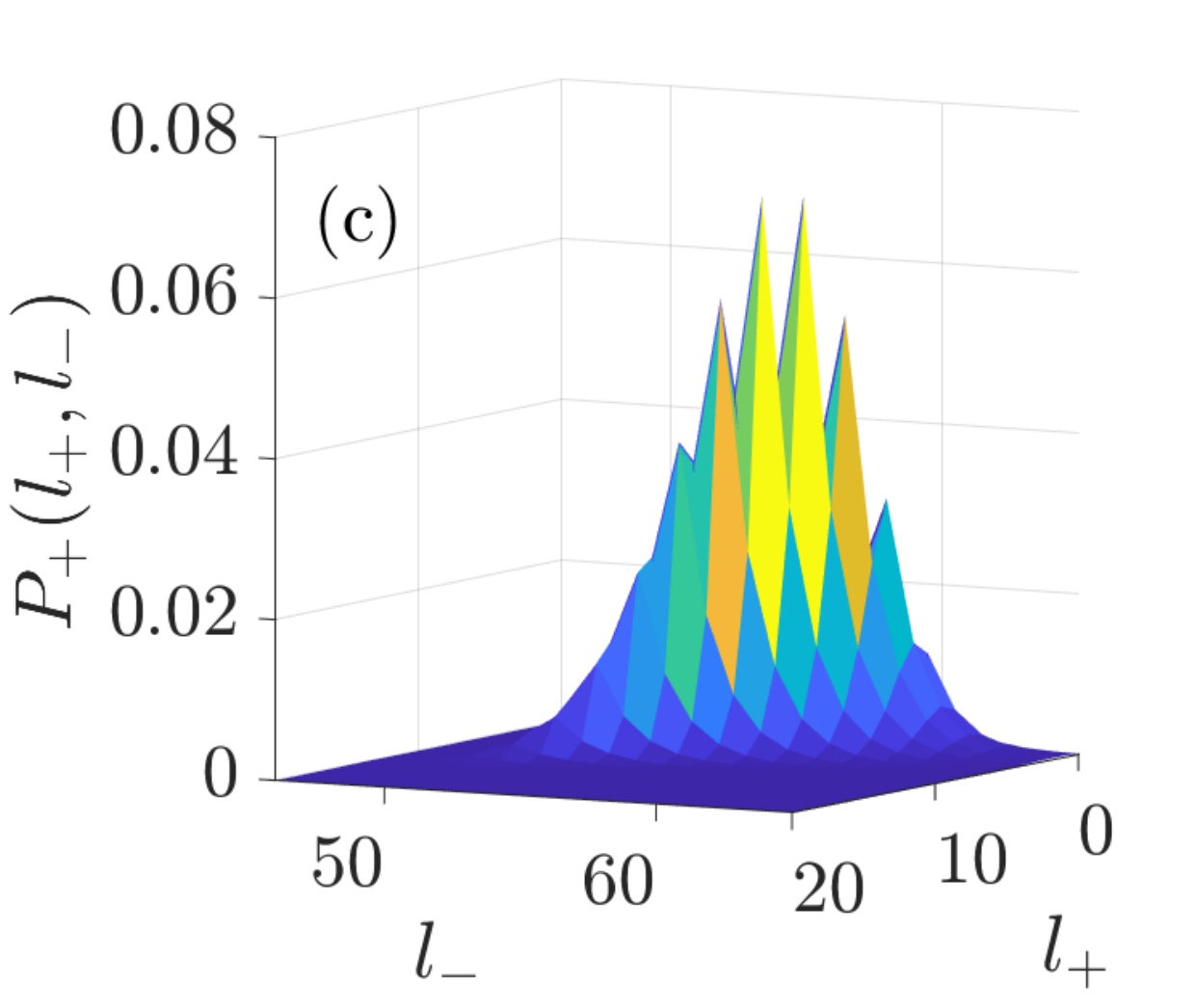}
     \end{subfigure}
\begin{subfigure}{0.23\textwidth}
     \centering
     \includegraphics[width=\textwidth]{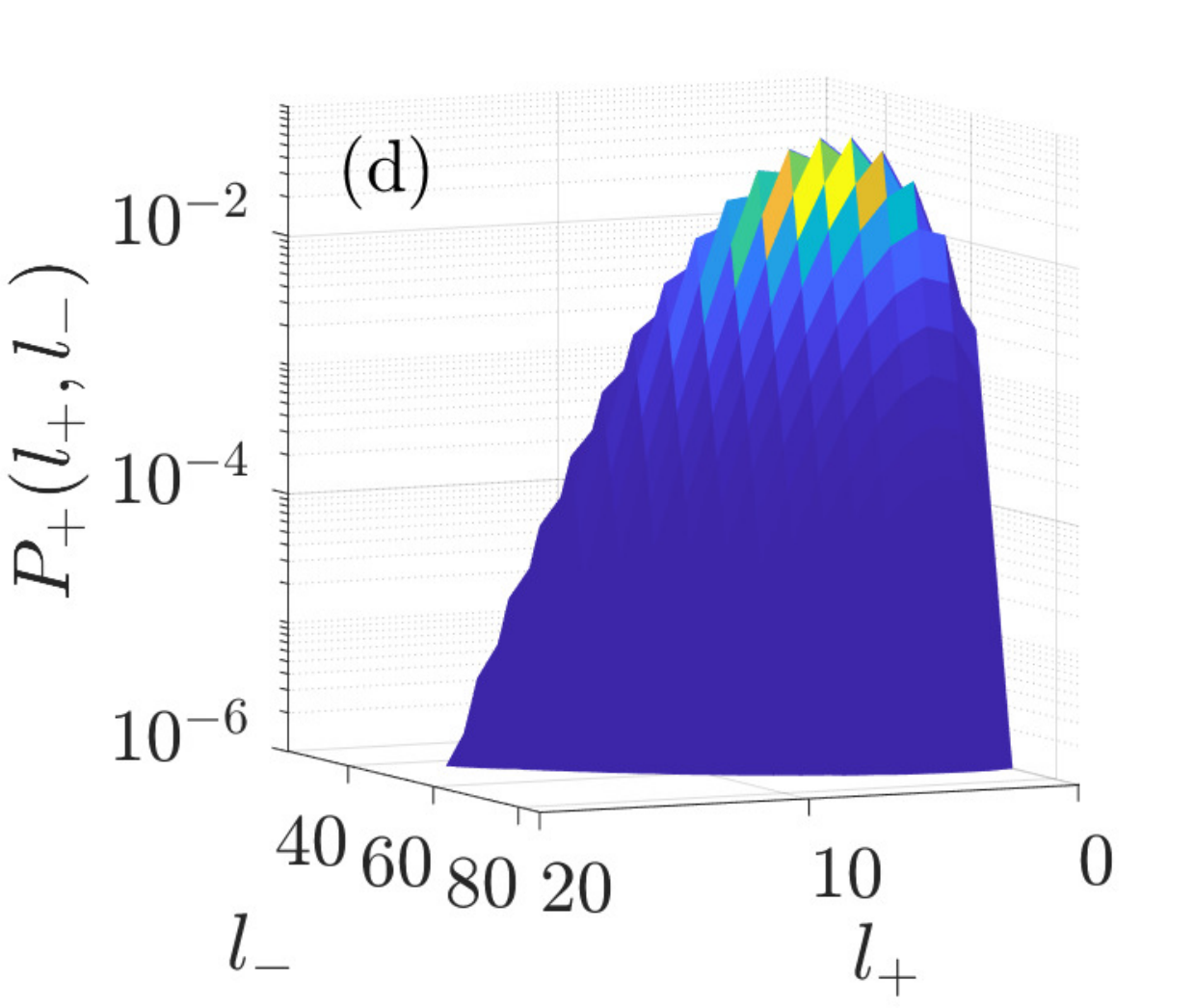}
  \end{subfigure}
  \\
  \begin{subfigure}{0.23\textwidth}
         \centering
         \includegraphics[width=\textwidth]{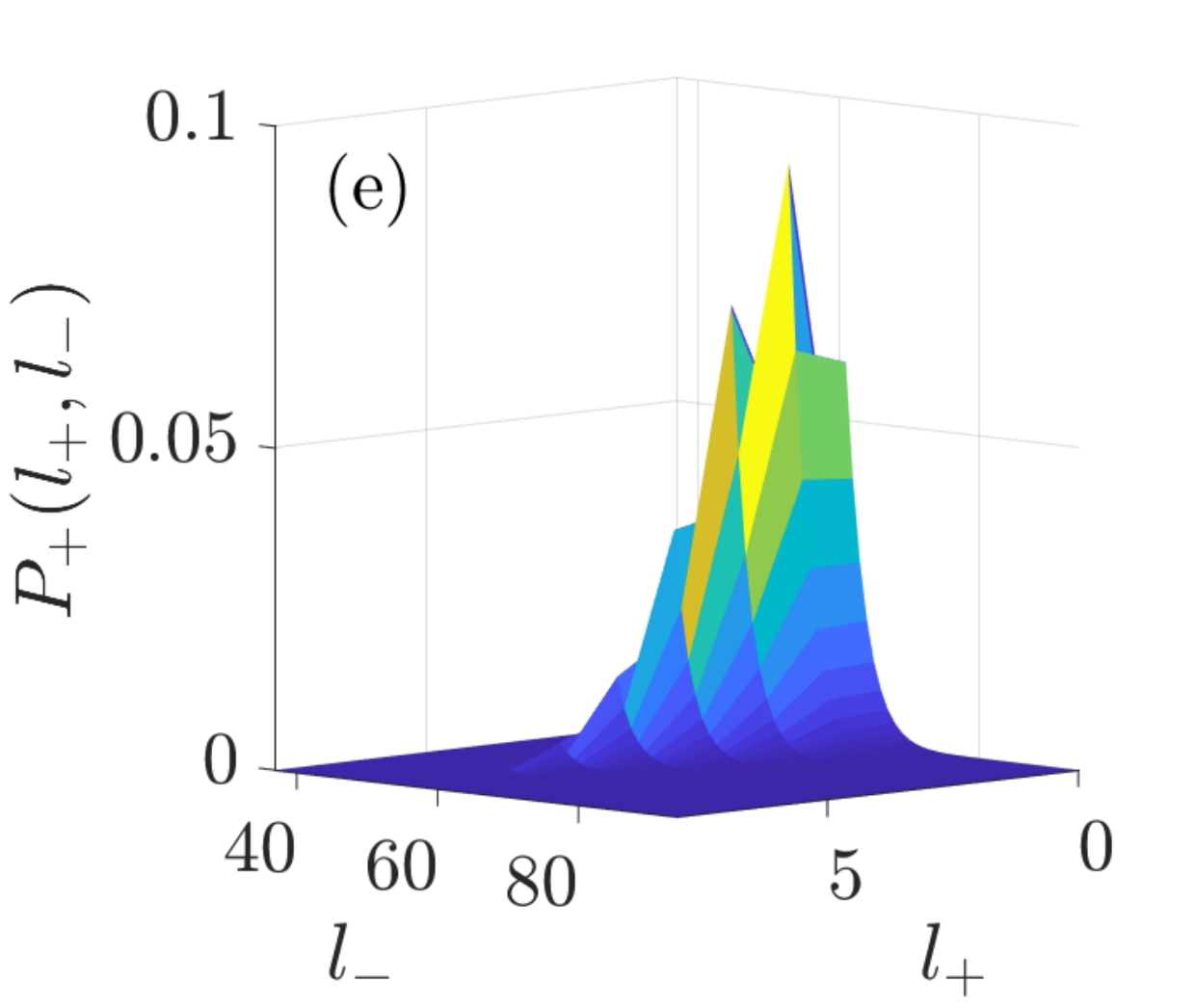}
     \end{subfigure}
\begin{subfigure}{0.23\textwidth}
     \centering
     \includegraphics[width=\textwidth]{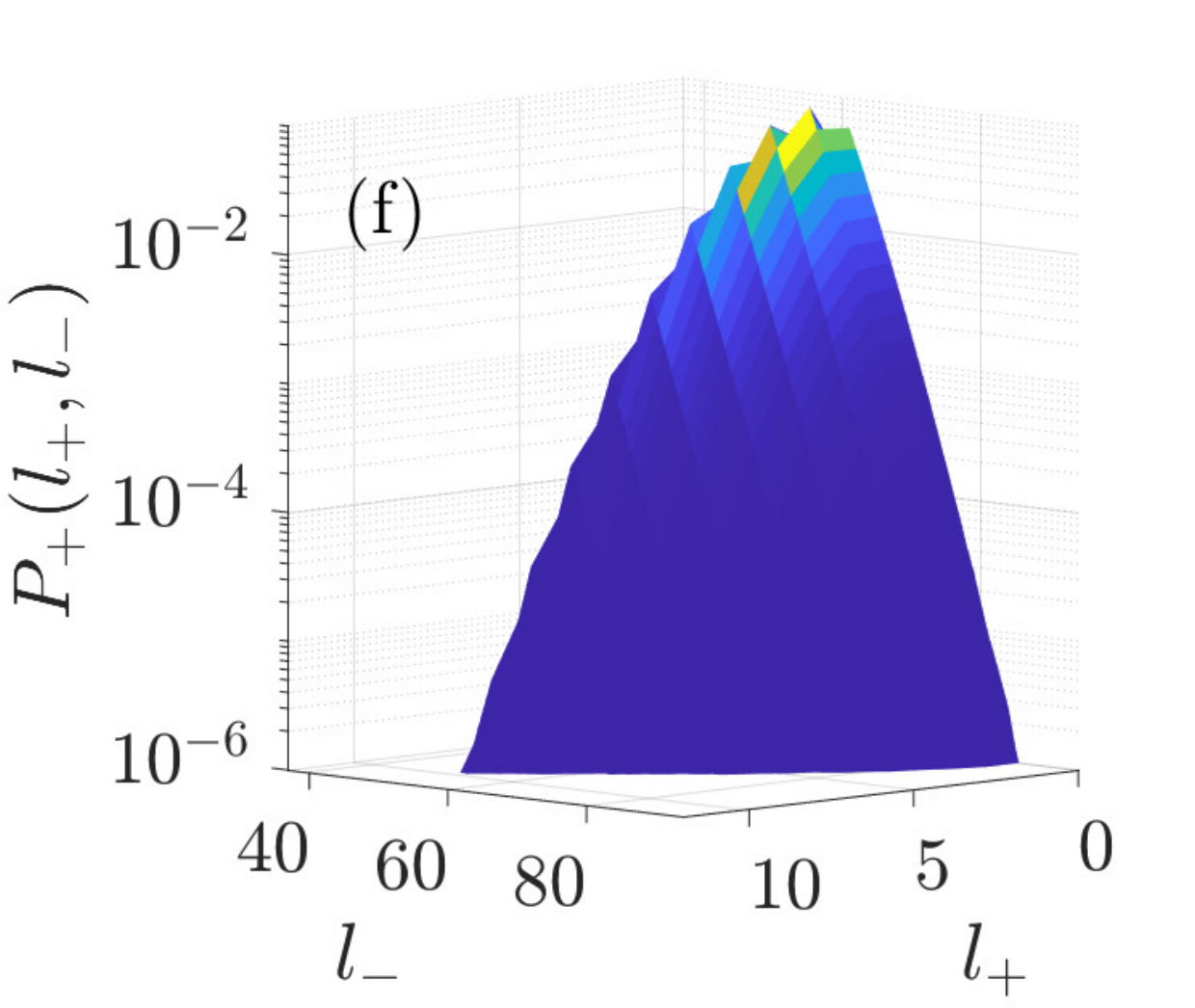}
  \end{subfigure}
\caption{Joint degree distribution $P_{\sigma}(\ell_+,\ell_-)$ with $N=1000$
and $\kappa=60.5$ in the ordinary regime. Parameters are: $(N_+,N_-)=(480,520)$, i.e., $m=-0.04$, and  $J=-0.5$ in (a)-(d); $(N_+,N_-)=(200,800)$, i.e., $m=-0.6$, and $J=-0.3$ in (e)-(f). 
(a,b): Linear plots showing the ``knife-edge'' (a) and broad Gaussian (b) perspectives of $P_{-}(\ell_+,\ell_-)$. 
(c,d): Linear (c) and semi log-scale (d) plots of $P_{+}(\ell_+,\ell_-)$, where the minority JDD  is pressed onto $\ell_+=0$, see text. (e,f): Linear (e) and semi-log (f) scale  plots  of $P_{+}(\ell_+,\ell_-)$
under high asymmetry ($m=-0.6$) and low heterophily.
}
\label{fig:Fig8}
\end{figure}

We also consider the joint degree distributions, 
$P_{\sigma}\left(\ell_{+},\ell_{-}\right)=p_{\sigma}\left(k\right)
q_{\sigma}\left(\ell_{-\sigma}|k\right)$ 
in the ordinary phase, see Eq.~\eqref{eq:q}, where $k=\ell_+ +\ell_-$. In the regime where  $\left\vert mJ\right\vert$ is small,
we can approximate the conditional distribution of cross-links by a 
binomial distribution, i.e., $q_{\sigma}(\ell_{-\sigma}|k)\simeq \binom{k}{\ell_{-\sigma}}\rho_{\sigma}^{\ell_{-\sigma}}\left( 1-\rho_{\sigma} \right) ^{k-\ell_{-\sigma}}$~\cite{pre2021} as in the symmetric case, yielding for the JDDs 
$P_{\sigma}\left(\ell_{+},\ell_{-}\right)\simeq 
\binom{k}{\ell_{-\sigma}}~p_{\sigma}\left( k\right)\rho_{\sigma}^{\ell_{-\sigma}}\left( 1-\rho_{\sigma} \right) ^{k-\ell_{-\sigma}}$, that are of the same form as \eqref{product}. 
%
%
%
Accordingly, the narrow Laplacian and  broad Gaussian are embodied as different perspectives -- in $p_{\sigma}\left(k\right)$ and $q_{\sigma}\left(\ell_{-\sigma}|k\right)$. 
 %
A three-dimensional plot of  $P_{\sigma}\left(\ell_{+},\ell_{-}\right)$  is the best
way to view the ``knife-edge" of the JDD caused by $p_{\sigma}(k)$ following a narrow asymmetric Laplacian
distribution in the ordinary phase. 
 As examples, in Fig.~\ref{fig:Fig8}, we present the JDDs for the case of low asymmetry and intermediate heterophily, and the case of high asymmetry (and low heterophily).

In Fig.~\ref{fig:Fig8}(a,b), the two perspectives for the \textit{majority}
JDD are shown, one along the knife-edge and the other, broadside. Note that $k=\ell_{+}+\ell_{-}$, so that the perspective of the former is aligned
with constant $k$.  The latter clearly gives the
impression of a Gaussian stemming from $q_{\sigma}$. This picture is qualitatively the same for the minority agents.
In Fig.~\ref{fig:Fig8}(c,d), we present the broadside perspective of the minority JDD in linear and semi-logarithmic scales. The latter 
reveals
 the initial stages of the minority being overwhelmed: in Fig.~\ref{fig:Fig8}(d), the right side of the Gaussian is truncated near $\ell_{+}=0$,
and  squeezed towards $\ell_{+}=0$, implying that the probability of a minority node  to have a small
number of ILs (small $\ell_{+}$) is not vanishingly small.
%
%
%
 The graphs of Fig.~\ref{fig:Fig8}(e,f), are similar, but for a low level of heterophily: the effect of heterophily is moderate, with almost half of the Gaussian being squeezed into the $\ell_{+}=0$ plane, with the narrowing of the JDD along $\ell_{+}$ being
accompanied by its broadening along $\ell_{-}$. In this cross-over regime,
the partition of $P_+(\ell_+,\ell_-)$ into the product of a narrow Laplacian and a broad
Gaussian is not valid (see Fig.~\ref{fig:Fig7}(d)). Finally, deep in the overwhelming phase, the
minority JDD collapses entirely onto the $\ell_{+}=0$ plane, and the
product expression is trivially valid: $P_{+}\left( \ell_{+},\ell_{-}\right) =\delta_{\ell_{+},0}~ p_{+}\left(k=\ell_{-}\right)$, where  $p_{+}\left(k\right)$ is a broad Gaussian in the overwhelmed state. Meanwhile, the JDD of the majority agents, $P_{-}\left( \ell_{+},\ell
_{-}\right) $, continues to display the same ``knife-edge" characteristics, as in the ordinary phase. 
In the transition region, the minority JDD cannot be  simply approximated by the
product of the DD and the conditional degree distribution.

\subsection{Measures of polarization}
\begin{figure}[th]
\centering
\begin{subfigure}{0.35\textwidth}
         \centering
         \includegraphics[width=\textwidth]{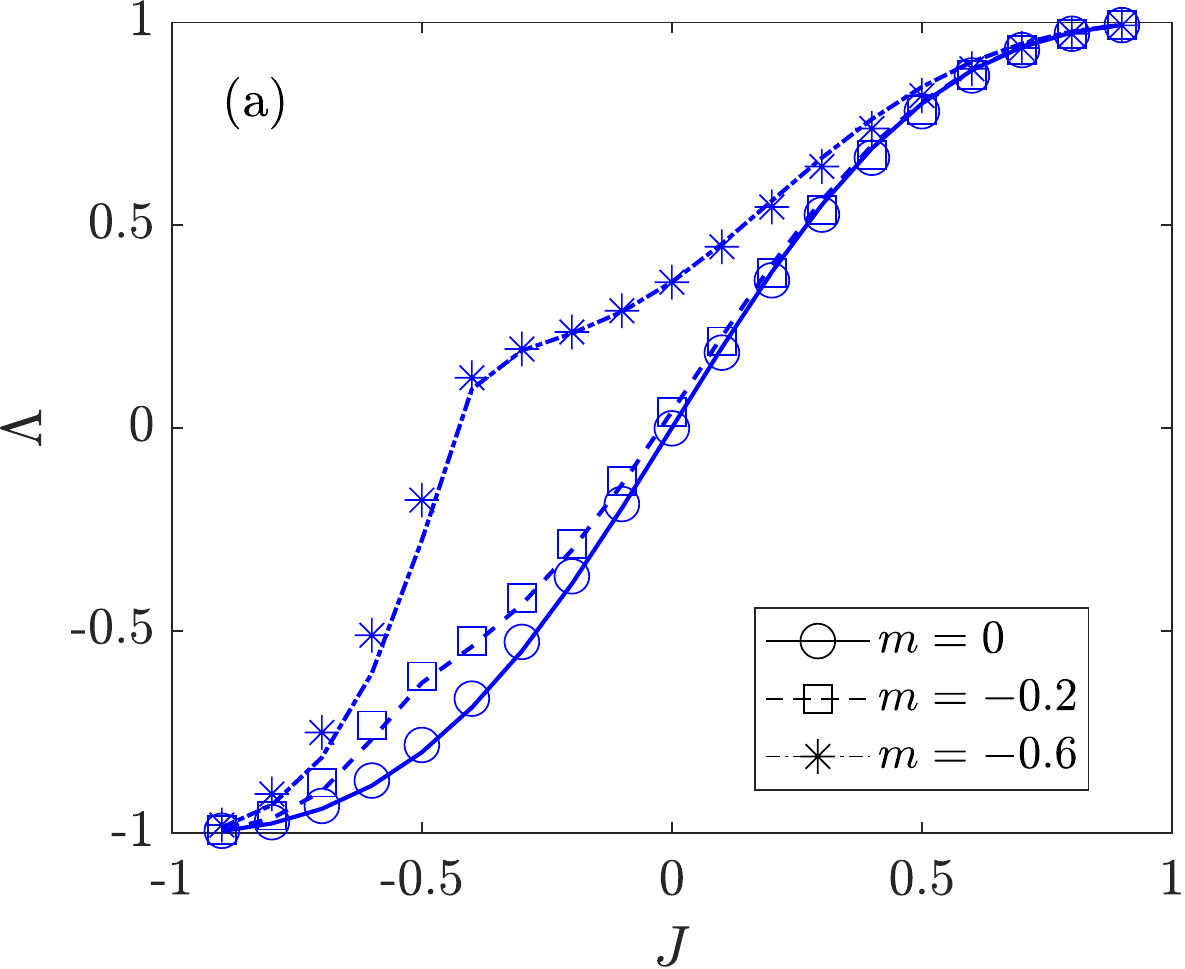}
     \end{subfigure}
     \\
\begin{subfigure}{0.35\textwidth}
         \centering
         \includegraphics[width=\textwidth]{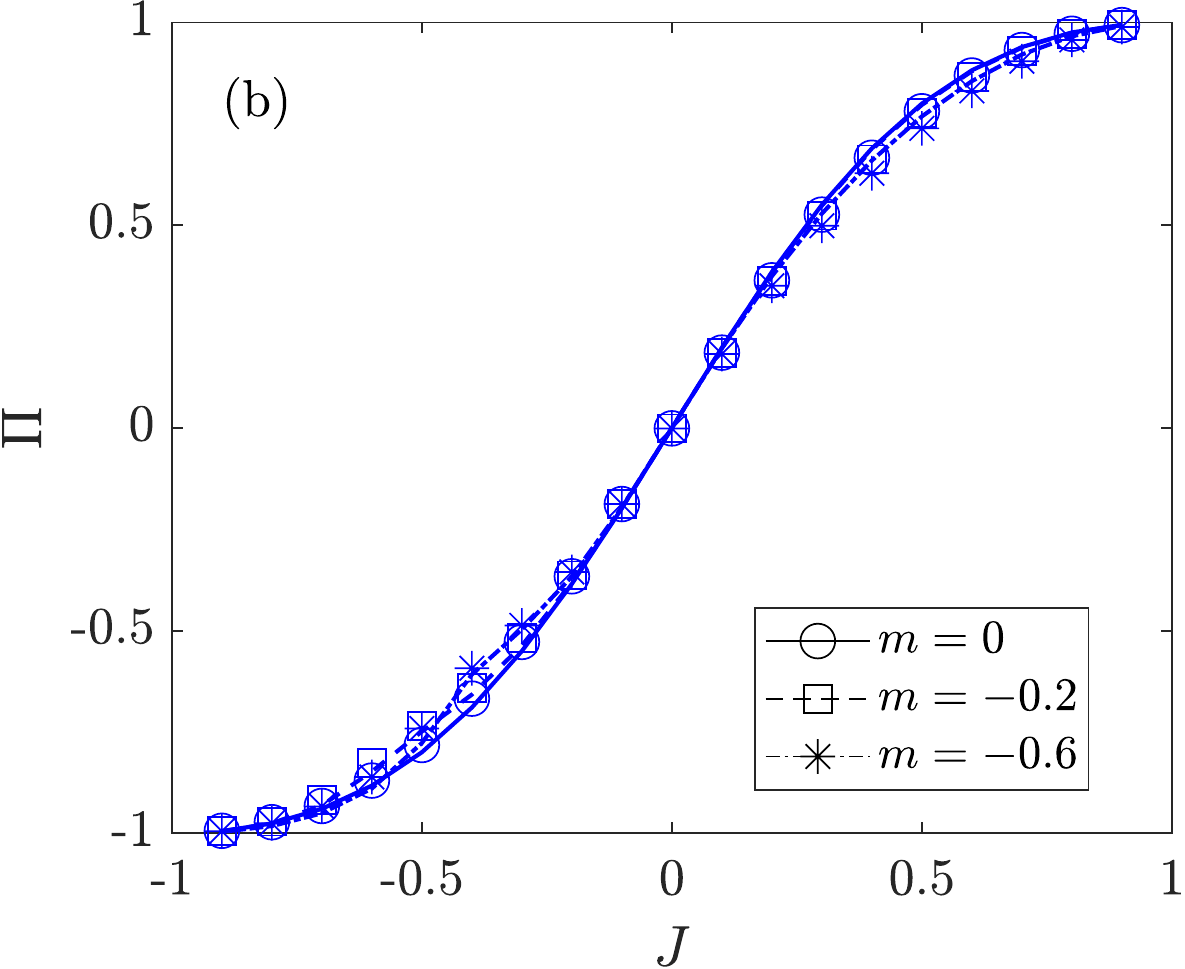}
     \end{subfigure}
\caption{Polarization measures $\Lambda$ and $\Pi$  as functions of $J$ for $m = 0$ (circles, solid), $-0.2$ (squares, dashed) and $-0.6$ (stars, dash-dotted); symbols are from simulation data and lines are mean-field predictions. Here, $N=1000$ and $\kappa=60.5$.
(a) $\Lambda$ \textit{vs.} $J$. 
(b) $\Pi$ \textit{vs.} $J$. 
Lines in (a) are from
Eq.~(S6) and in (b) from Eq.~(S8) of the supplementary material, with 
$\rho_{\pm}$ obtained from  the mean-field theory of Sec.~V.B, see text.
}
\label{fig:Fig9}
\end{figure}

When measuring the network polarization, the quantity $\Lambda$ is commonly used and sometimes referred to as
``average edge homogeneity''~\cite{prasetya2020model,del2016spreading}. See Sec.~S6 in the supplementary material for the derivation of $\Lambda$ in terms of
$\rho_{\pm}$ which are functions of $J$ and $m$ obtained from the mean-field theory of Sec.~V.B.
$\Lambda$ provides a sensible measure of polarization 
in systems with low asymmetry ($m\approx 0$). However, this is generally not the case when $m\ne 0$, especially when $J\approx 0$
and $\Lambda\approx m^2$ fail to predict a vanishing polarization when $J\to 0$, see Fig.~\ref{fig:Fig9}(a) and Ref.~\cite{pre2021}.
This led us to introduce $\Pi$, defined by Eq.~(\ref{Pi-def}), which is constructed to avoid the deficiencies of $\Lambda$ by providing a meaningful measure of polarization for any values of $m$.  
Although $\Pi$ was motivated by the separation of the
CMs of $P_{\sigma}$ (the JDDs of the two groups), we have shown that in the context of the mean-field approximation,
$\Pi$ can be computed from $\rho_{\pm}$ and $m$, see Eq. (S7) in the supplementary material.
While 
$\Pi$ contains the same information displayed in Fig.~\ref{fig:Fig3}(b), its mean-field expression is an instructive  combination of $\rho_{\pm}$  offering a single
meaningful quantity for polarization. We indeed find
that $\Pi \propto 1-\rho_{+}-\rho_{-}$ which vanishes for $J=0$
and any value of $m$ (since $\rho_{\pm}=n_{\mp}$), while it reduces to $\Lambda$ for $m=0$.
When $J=\pm 1$, $\Pi(\pm 1,m)=\Lambda(\pm 1,m)=\pm 1$.
This implies that the sign of $\Pi$ alone signifies if the
system is homophilic or heterophilic, as seen in Fig. \ref{fig:Fig9}(b)
where the mean-field predictions 
are in excellent agreement with simulation data (symbols) for all values of $m$ and $J$.
While $\Pi$ does not appear to be independent of $m$, 
the dependence found in Fig.~\ref{fig:Fig9}(b)
turns out to be weak. 
Finally, we note
that in Fig.~\ref{fig:Fig9}, as expected, $\Lambda$ and $\Pi$ display a signal of the transition from the ordinary
to the overwhelmed state about $J\approx J_c(m)$.

\section{Asymmetric Systems: Theoretical Considerations}
From the phenomena presented in the last section, it is clear that there are two different regimes, the ordinary and overwhelming phases, with quite
distinct properties. This section is devoted to their theoretical characterization in terms of mean-field theories. 
While we believe it is possible to formulate a single
mean-field based theory which would completely describe both regimes, such a theory will be quite complex. 
Instead, here we adopt a simpler approach which has the advantage of being pragmatic and easily comprehensible, see Sec.~V.B,
at the price of being less effective in the overwhelming regime, 
an issue that we circumvent in Sec.~V.C with a refined (complementary)
mean-field theory.


%
\subsection{Framework for a general mean-field analysis}
Our starting point to set a general framework for a mean-field analysis that applies for any choice of $N_{\pm}$, for arbitrary $m$,
is to generalize the mean-field theory devised in Ref.~\cite{pre2021}
for the symmetric case $m=0$, see also Sec.~III. 

For arbitrary $m$,
we now have four unknowns: $\alpha_{\pm}$ (or $n_{\pm}^{a}$) and $\rho_{\pm}$~\footnote{As a reminder, these quantities refer to steady state averages, and every expression involving $\rho_{\sigma}$
has to be interpreted in the context of the mean-field approximation. Any referring
to time dependence will be emphasized by, e.g., $\alpha_{\pm}\left(t\right)$.}.
As in the symmetric case we have two equations and still need two other equations to obtain a self-contained theory.  One such equation is for each $L_{\sigma \sigma}$, while the other is a strict constraint involving $\rho_{\pm}$ in the
identity $L_{+-}=L_{-+}$.

Focusing first on global quantities like $L$, we must
consider the changes for $L_{\sigma \sigma}$, with $\sigma=\pm$. 
Proceeding as in Ref.~\cite{pre2021}, the gain and loss
rates are $n_{\sigma}^{a}n_{\sigma}\hat{J}$
and $n_{\sigma}^{c}\left(1-\rho_{\sigma}\right) \check{J}$,
respectively. For the former, $n_{\sigma}$ means that a node can choose to add a link to \textit{unequal} fraction of partners~\footnote{We have assumed the fraction of connections in the network is small, meaning $\kappa \ll N$, so that an
adder's chance of finding an unconnected node is proportional to $n_{\sigma}$. Discrepancies between theory and data can be generally traced to the
effects of finite size caused by this approximation.}, rather than $1/2$ when $m=0$. Now, there are four ways $L_{\sigma \sigma}$ can arrive at a stationary state. Two correspond to
extreme values of $J$, when one of the rates vanishes and $L_{\sigma \sigma}$ reaches its own bound. Another is when both $n_{\sigma}^{a}$ and $1-\rho_{\sigma}$ vanish (or are vanishingly small), an interesting possibility
we will return to in subsection V.C. Here, we are most
interested in the last scenario, when both rates reach stationary, generic
values. Balancing these, we find for the steady state:~ $\hat{J}n_{\sigma}^{a}n_{\sigma}=\check{J}n_{\sigma}^{c}\left( 1-\rho_{\sigma}\right)$,
which is equivalent to 
\begin{equation}\label{eqn:Na}
\alpha_{\sigma} n_{\sigma}B=(1-\alpha_{\sigma})\left( 1-\rho_{\sigma}\right),
\quad 
\end{equation}
where 
\begin{equation*}\label{B-def}
B\equiv \hat{J}/\check{J}=\frac{1+J}{1-J}
\end{equation*}%
is a convenient way to express the bias due to homophily: $B>1$ for $J>0$.
For CLs, the generalization of the gain/loss rates is slightly
more complicated, yielding
\begin{equation}\label{eqn:CLs}
\left(n_{+}^{a}n_{-}+n_{-}^{a}n_{+}\right) \check{J}=\left(n_{+}^{c}\rho_{+}+n_{-}^{c}\rho_{-}\right) \hat{J}, 
\end{equation}
where we can read the contributions from both groups. 

Another equation comes from the constraint $L_{+-}=L_{-+}$. In terms of the variables here,
this equality (trivially satisfied in the symmetric case) reads
\begin{equation}\label{SecondEqn}
N_{+}\mu_{+}\rho_{+}=N_{-}\mu_{-}\rho_{-}.
\end{equation}
In order to apply Eq.~(\ref{SecondEqn}),
we must first generalize the technique of Ref.~\cite{pre2021} for the
degree distributions $p_{\pm}\left( k\right) $ when $m\neq 0$, from which to extract $%
\mu_{\pm}$ in terms of $\alpha_{\pm}$ and $\rho_{\pm}$. Thus, we
devote the next subsections to studies of the DDs.

\subsection{Systems with low asymmetry or $-J\ll 1$}
As seen in the simulation data, the DDs for asymmetric system with small $m$ and $-J\ll 1$ are qualitatively the same as in the case $m=0$.
Proceeding as in the symmetric case~\cite{pre2021}, we start
from  the balance equation for the addition/deletion of a link at a single node:
\begin{equation}\label{DDbalance}
R^{a}_{\sigma}\left( k\right) p_{\sigma}\left( k\right) =R^{c}_{\sigma}\left(k+1\right) p_{\sigma}\left(k+1\right),
\end{equation}
for which we need expressions for the four $R_{\sigma}^{a,c}$
which are the probabilities in a time unit at which a connection is added ($R_{\sigma}^{a}$) and  cut ($R_{\sigma}^{c}$) in the community $\sigma$. Each $R$ has contributions from both communities, and the
probabilities for adding and cutting links are here associated with the symbols $\eta$ and $\chi$, respectively. We will further distinguish contributions due to the actions ($\eta ,\chi$) of the chosen
node,  or from the rest of the population ($\tilde{\eta}, \tilde{\chi}$)~\cite{pre2021}. In the context of our mean-field theory, the former pair simply reads:
\begin{align}\label{etachi}
\begin{aligned}
\eta_{\sigma} =n_{\sigma}\hat{J}+n_{-\sigma}\check{J},\quad
\chi_{\sigma}=\left( 1-\rho_{\sigma}\right) \check{J}+\rho_{\sigma}
\hat{J}
\end{aligned}  
\end{align}%
as the prefactors of the $J$'s just refer to the chances our node adds/cuts
a link to an agent in its community or otherwise. The latter is similar,
except for the extra factors accounting for the fraction of adders/cutters
in the $\sigma$ group, yielding 
\begin{align}\label{etachit}
\begin{aligned}
\tilde{\eta}_{\sigma}& =\alpha_{\sigma}n_{\sigma}\hat{J}+\alpha
_{-\sigma}n_{-\sigma}\check{J}  
\\
\tilde{\chi}_{\sigma}& =\left( 1-\alpha_{\sigma}\right) \left( 1-\rho
_{\sigma}\right) \check{J}+\left( 1-\alpha_{-\sigma}\right) \rho
_{-\sigma}\left( n_{-\sigma}/n_{\sigma}\right) \hat{J}.  
\end{aligned}
\end{align}
These enter
into the $R$'s as in the case $m=0$~\cite{pre2021}:
\begin{equation}\label{R's}
R_{\sigma}^{a}=\frac{\left[ H\left( \kappa -k\right) \eta_{\sigma}+\tilde{\eta}_{\sigma}\right]}{N};~~R_{\sigma}^{c}=\frac{\left[ H\left( k-\kappa \right) \chi
_{\sigma}+\tilde{\chi}_{\sigma}\right]}{N}.  
\end{equation}%
%
Proceeding as in the symmetric case, solving the recursion relation \eqref{DDbalance}~\cite{pre2021},
we obtain again asymmetric Laplacian distributions (see Sec.~S5 in the supplementary material and \eqref{eqn:finalpk}): 
\begin{align}\label{DDasy}
\begin{aligned}
p_{\sigma}\left( k<\kappa \right) &=p_{\sigma}\left( \left\lfloor \kappa
\right\rfloor \right) \left[ \gamma_{\sigma <}\right] ^{\left\lfloor \kappa
\right\rfloor -k}  
\\
p_{\sigma}\left( k>\kappa \right) &=p_{\sigma}\left( \left\lceil \kappa
\right\rceil \right) \left[ \gamma_{\sigma >}\right] ^{k-\left\lceil \kappa
\right\rceil},
\end{aligned}
\end{align}
where, with \eqref{etachi} and \eqref{etachit},
\begin{equation}\label{gammas}
\gamma_{\sigma <}=\frac{\tilde{\chi}_{\sigma}}{\eta_{\sigma}+\tilde{\eta}
_{\sigma}};~~\gamma_{\sigma >}=\frac{\tilde{\eta}_{\sigma}}{\chi_{\sigma
}+\tilde{\chi}_{\sigma}}  
\end{equation}
are the factors controlling the exponentials.
Finally, by imposing the normalization conditions
\begin{equation}\label{norm}
\frac{p_{\sigma}\left( \left\lfloor \kappa \right\rfloor \right)}{1-\gamma
_{\sigma <}}=\alpha_\pm, \quad \frac{p_{\sigma}\left( \left\lceil \kappa \right\rceil \right) 
}{1-\gamma_{\sigma >}}=1-\alpha_{\pm}, 
\end{equation}%
both $p$'s are uniquely determined in terms of the unknowns $\alpha_{\pm}$
and $\rho_{\pm}$. 

Our goal is to find the averages $\mu_{\sigma}$ of these distributions, which allow us to apply Eq.~(\ref{SecondEqn}) and in turn complete our mean-field theory. From Eqs.~\eqref{DDasy} and~\eqref{norm}, we have
\begin{equation}\label{mu}
\mu_{\sigma}=\left\lfloor \kappa \right\rfloor -\frac{\gamma_{\sigma<}\alpha_{\sigma}}{1-\gamma_{\sigma <}}+\frac{1-\alpha_{\sigma}}{1-\gamma_{\sigma >}}, 
\end{equation}%
where we have neglected terms of order ${\cal O}(\gamma_{\sigma >}^{\left\lceil \kappa \right\rceil},\gamma_{\sigma <}^{\left\lceil \kappa \right\rceil})$ since  $\kappa\gg 1$.
Equation~(\ref{mu}) gives us the expression of $\mu_{\pm}$ in terms of $\alpha_{\pm}$ and $\rho_{\pm}$,  which can be determined by solving Eqs.~(\ref{eqn:Na}), (\ref{eqn:CLs})  and (\ref{SecondEqn}), and in turn allow us to obtain the 
predictions of this theory. Yet, since Eqs.~\eqref{eqn:Na} and (\ref{SecondEqn}) are non-linear, we do not have explicit solutions. Instead we have solved Eqs.~(\ref{eqn:Na}), (\ref{eqn:CLs}),  (\ref{SecondEqn}) and (\ref{mu}) numerically by standard methods. 

\subsubsection{Mean-field predictions of $\alpha_{\sigma},~\rho_{\sigma}, \mu_{\sigma}$ and $p_{\sigma}$}
The theoretical results 
based on Eqs.~\eqref{DDasy}, (\ref{eqn:Na}), (\ref{eqn:CLs}),  (\ref{SecondEqn}) and (\ref{mu})
have been directly used to obtain the mean-field predictions of $\alpha_{\pm}$ and $\rho_{\pm}$, $\Lambda$ and $\Pi$ (see Sec.~S6 in the supplementary material), $\mu_{\pm}$ and ${\cal M}_{\pm}=(\mu_{\pm}/\kappa)-1$, and also the degree distribution $p_\sigma(k)$.

The blue and red lines of Fig.~\ref{fig:Fig3} show the predictions of  
$\alpha_{\sigma}$ and $\rho_{\sigma}$  which are found in good accord with simulation data for all values of $J$, across both the ordinary and overwhelming phases. The agreement is excellent in the ordinary regime, and we note that our mean-field theory   
remarkably captures the ``kinks'' of  $\alpha_{-}$ and $\rho_{-}$ at the onset of the overwhelming transition.
In the overwhelming phase, $|m|={\cal O}(1)$ and $J<J_c<0$ (see Sec.~IV.C), the
degrees of the minority agents are no longer small compared to $N$, which violates a key assumption of our mean-field theory. This leads to  systematic but rather modest deviations between the mean-field predictions and simulation data. 

The comparison of simulation and theoretical results for $\Lambda$ and $\Pi$ in Fig.~\ref{fig:Fig9}
shows a remarkable agreement for different values of $m$ over the entire range of $J$, i.e., across both ordinary and overwhelming phases. In particular, the mean-field theory correctly captures the weak $m$-dependence of $\Pi$, and signals the transition from the ordinary to overwhelming regimes for both $\Lambda$ and $\Pi$.

For $\mu_\pm$ and $\cal{M}_\pm$, as shown in 
Fig.~\ref{fig:Fig4}, mean-field predictions  generally agree well with simulation results, with an agreement that improves as 
$\kappa \ll N$, with $\kappa\gg 1$. Remarkably, the mean-field results give sensible results for $\mu_{\pm}$,  and ${\cal M}_{\pm}$ also in the overwhelming regime, see Sec.~V.C and Sec.~S8 in the supplementary material.

Theoretical predictions of $p_\sigma(k)$
are used to obtain the
red and blue lines in Fig.~\ref{fig:Fig7}, which are generally in good 
 agreement with simulation data, especially 
 in the ordinary regime, i.e., $|mJ|\ll 1$ or  $J>0$, see Fig.~\ref{fig:Fig7}(a,b). As shown in Fig.~\ref{fig:Fig7}(c), for $|m|={\cal O}(1)$ and above a certain level of heterophily ($J<0$ with $|J|={\cal O}(1)$), a Gaussian-like distribution for the cutters in the minority begins to develop, which 
 is not captured by 
 the above mean-field theory. In fact, deep in the overwhelming phase the degree distribution of the minority no longer falls off exponentially as predicted by (\ref{DDasy}), but is 
 a broad Gaussian, see Fig.~\ref{fig:Fig7}(f), characterized in  Sec.~V.C.  We note that the degree distribution of the majority always falls off exponentially and, in both ordinary and overwhelming phases,  is  well described by (\ref{DDasy}), see  blue lines in Fig.~\ref{fig:Fig7}(a-f).

\subsection{A refined mean-field theory for the degree distribution of the overwhelmed minority}
Clearly, the overwhelmed states lie beyond the domain of validity of the above mean-field theory. This chiefly results from the DDs of the minority agents having morphed, see Fig.~\ref{fig:Fig7}(e,f). To get a  reasonable characterization of the quantities in this region, we have to study the degree distribution of the minority agents more carefully. 

For this,  we revise the above theory following the approach of Ref.~\cite{bassler2015networks}, and write 
 the balance equation obeyed by the 
minority DD with a full non-trivial $k$-dependence
of adding/cutting probabilities $R^{a,c}(k)$:
\begin{equation}
\label{eq:recursion_min}
R^{a}_{+}\left( k-1\right) p_{+}\left( k-1\right) =p_{+}\left( k\right)
R^{c}_{+}\left( k\right).  
\end{equation}
Furthermore, 
guided by simulation data, we assume that deep in the overwhelming phase
we have $\alpha_+=0$ (see also at the end of this section).
 To determine $R^{a}\left( k\right)$, we recognize that there are just $\left( N_{-}-k\right)$ majority nodes which
can add a link to our agent of degree $k$,  each of which can be chosen with
probability $1/N$, and only a fraction $\alpha_{-}$ of them would add.
There is also the bias $\check{J}$ 
for adding CLs. Finally, the adder will
choose our agent with probability $1/\left( N-\tilde{k}-1\right) $, where $\tilde{k}$ is the number of links it has, which is a fluctuating
quantity. In the spirit of a mean-field approximation, $\tilde{k}$ is replaced by $\mu_{-}^{a}$ (average degree of  majority adders). Putting everything together, we have
\begin{equation}\label{Rar}
R^{a}_+\left( k-1\right) \simeq \frac{N_{-}-k+1}{N}\frac{\alpha_{-}}{N-1-\mu
_{-}^{a}}\check{J}.  
\end{equation}%
Similar arguments give the expression of $R^{c}_+\left( k\right)$:
\begin{equation}\label{Rcr}
R^{c}_+\left( k\right) \simeq \frac{1}{N}\left( 1+k\frac{1-\alpha_{-}}{\mu
_{-}^{c}}\right) \hat{J}.  
\end{equation}
With \eqref{Rar} and \eqref{Rcr}, the solution of the recursion \eqref{eq:recursion_min} is a ``shifted binomial" (see Sec. S7 in the supplementary material):
\begin{equation}\label{SB}
p_{+}\left( k\right) \propto \frac{\left( Q^{c}/Q^{a}\right) ^{k}\left(
N_{-}\right) !}{\Gamma \left( Q^{c}+k+1\right) \left( N_{-}-k\right) !}, 
\end{equation}
where
\begin{equation}\label{MF-QaQc}
Q^{a}=\left( N-\mu_{-}^{a}-1\right) B/\alpha_{-};~~Q^{c}=\mu_{-}^{c}/\left(
1-\alpha_{-}\right).
\end{equation}%
Since $Q^{c}>0$, this expression is  well-defined as far as $k=0$. Since $\alpha_+\approx 0$ is our assumption in the overwhelming state 
we  expect this theory to be fairly good
deep in the overwhelming regime, but to fail near
the transition line.

For the characterization of the DD  by~(\ref{SB}), we expect $p_+(k)$ to approach a
Gaussian in the limit of large $N,\kappa$ with generic values of $m,J$. Thus, we use the
mode $\hat{k}$ for $\mu_{+}$ and the curvature of $\ln p_{+}$ for $V_{+}$, see also Sec.~S7 in the supplementary material.
\begin{equation}
\label{muplus}
\mu_{+}\simeq \hat{k}\simeq \frac{N_{-}+1-Q^{a}}{1+Q^{a}/Q^{c}},  
\end{equation}%
%
\begin{equation}\label{Vp}
V_+=\frac{N_- Q^c Q^a}{(Q^a+Q^c)^2}+\frac{(Q^a)^2 +2Q^aQ^c-(Q^c)^2+2Q^a(Q^c)^2}{2(Q^a+Q^c)^2}.
\end{equation} 
To determine $Q^a$ and $Q^c$, guided by simulation data showing that the DDs for the majority agents
still follow the asymmetric Laplacian distribution~(\ref{DDasy}), 
we assume that $\mu_-^a\approx \mu_-^c\approx \kappa$.
Also, consistently with our previous assumption, in the overwhelming state,
we set $\alpha_+=0$ in \eqref{eqn:Na} and \eqref{eqn:CLs}, and obtain
\begin{equation}\label{alpham}
\alpha_{-}=\frac{1-J^{2}}{\left( 1-mJ\right) \left(1-m\right)}. 
\end{equation}
These together give us the approximation of $Q^a$ and $Q^c$ in the overwhelming state, and
 thus  the degree distribution $p_+(k)$ of the minority agents (almost all ``cutters'') shown as red curves in Fig.~\ref{fig:Fig7}(d-f).
Since  \eqref{SB} with \eqref{alpham} and 
$\mu_-^{a,c}\approx \kappa$ rely 
on the observed  absence of adders
and ILs in the minority community, it is a ``semi-phenomenological'' refined mean-field theory. A fully self-consistent
refined mean-field theory
is outlined in Sec.~S9 of the supplementary material.
Comparison with simulation results of 
Fig.~\ref{fig:Fig7} shows that this semi-phenomenological  theory gives a good description of the minority DD
not too deep in the overwhelming phase, see Fig.~\ref{fig:Fig7}(d,e).
Yet,  deep in the overwhelming regime of very heterophilic systems ($-J$ close to $1$),
there are  quantitative deviations between the theoretical predictions and simulation data, 
see the red curve in Fig.~\ref{fig:Fig7}(f). These are traced back to the use of  \eqref{alpham}, and a significantly better agreement is found when 
\eqref{SB} is used with  $\alpha_-$ directly obtained from simulations
(and $\mu_-^{a,c}\approx \kappa$), leading to the cyan curve in 
Fig.~\ref{fig:Fig7}(f). The accuracy of the predictions of 
\eqref{SB} with \eqref{alpham}  improves as the system size is increased, i.e. the red and cyan curves of Fig.~\ref{fig:Fig7}(f) will get closer for larger $N$ and $\kappa$.

As a simple  assessment of our refined mean-field theory, we  compare its predictions with the simulation data in the case of $N=1000$ and $\kappa =60.5$, with $m=J=-0.6$. As shown in
Fig.~\ref{fig:Fig7}(f),
$p_{+}\left( k\right)$  in this case study is clearly Gaussian-like, with measured $\mu_+\simeq 146$ and $V_+\simeq 217$. 
These values are compared with the predictions of our  theory based on Eq.~(\ref{MF-QaQc}), \eqref{SB} and \eqref{alpham},   yielding $(\mu_+, V_+)\simeq (128,203)$ from~\eqref{muplus} and~\eqref{Vp}. These results are in reasonable but not perfect agreement with those of simulation.
The data  
can also be compared with \eqref{muplus} and \eqref{Vp} when 
these are used with $\alpha_-$ directly measured from simulations, yielding  
$(\mu_+, V_+)\simeq (146,216)$, which are the approximation of the mean and variance of the cyan curve and
compare remarkably well with those obtained from simulations. 
This agreement gives us confidence that we have devised a
suitable mean-field  description of the DD of the minority agents.

We conclude that our results, illustrated by Figs.~\ref{fig:Fig3}-\ref{fig:Fig9}, show that the ordinary MF approximation gives a sound qualitative and quantitative characterization of all quantities in the ordinary phase, as well as of the global quantities in the overwhelming phase and DDs of the majority phase. Yet, the refined MF is necessary to describe the DD of the minority in the overwhelming phase, see also Sec.~S8 and Fig.~S2 in the supplementary material.

\subsection{Transition line}
In this section, we use the theoretical results of Secs.~V.B and V.C to derive the mean-field prediction of the transition line $(m, J_c(m))$
separating the ordinary
and overwhelming phases (respectively at $J>J_c$ and $J<J_c$, with $m$ fixed), 
see Fig.~\ref{fig:Fig6}. For the sake of concreteness, and without loss of generality, here we consider $m<0$.

To find the point where the transition occurs, we start from the balance equation for a minority node of degree $k=\left\lceil \kappa \right\rceil$ which, from \eqref{DDbalance}, reads 
\begin{equation}\label{eqn:rpa}
R_{+}^a(\ceil{k})~p_{+}(\ceil{k})=p_{+}(\ceil{k}+1)~R_{+}^c(\ceil{k}+1),
\end{equation}
with
\begin{equation}\label{eqn:rpac}
R_{+}^{a}(\ceil{k})=\tilde{\eta}_{+}/N, \quad R_{+}^{c}(\ceil{k})=\left(\chi_{+}+\tilde{\chi}_{+}\right)/N,
\end{equation}
where we have used \eqref{R's} with
$H(\kappa-\ceil{\kappa})=0$ and $H(\ceil{\kappa}-\kappa)=1$.
As discussed in  Sec.~IV.D, we consider that the transition
between the ordinary and overwhelming phases occurs when $p_{+}(\ceil{k})=p_{+}(\ceil{k}+1)$, see Fig.~\ref{fig:Fig7}(d). With  \eqref{eqn:rpa} and \eqref{eqn:rpac}, this readily gives
\begin{equation}
\label{line}
\tilde{\eta}_{+}=\chi_{+}+\tilde{\chi}_{+}.
\end{equation}
Further, we assume that at the onset of the transition, the features of both phases hold: $\alpha_+=0$, $\rho_+=1$ (see Fig.~\ref{fig:Fig3}), and $\mu_+\approx \mu_-$. With these assumptions and Eqs.~\eqref{L=Nkappa}, \eqref{rhosigma}, \eqref{etachi}, \eqref{etachit} and \eqref{line}, we have
\begin{equation}
\label{line2}
\alpha_{-}n_{-}\check{J}=\hat{J}+(1-\alpha_{-})\hat{J},
\end{equation}
where $n_-=(1-m)/2$ and $\alpha_-(J,m)$ can be approximated by Eq.~\eqref{alpham}. The unique physical root of \eqref{line2}
thus gives us the mean-field expression of $J_c(m)$, which explicitly reads (for $m<0$):
\begin{equation}\label{transitionEq}
J_c=\frac{\left( 3m-1-2m^{2}+2\sqrt{m\left( m^{3}-3m^{2}+4m-1\right)}\right)}{1+m}.
\end{equation}
This expression is plotted in Fig.~\ref{fig:Fig6}.
The  predictions of \eqref{transitionEq} are found to generally agree well with
simulation data, with an excellent agreement for $m\lesssim -0.4$ and some noticeable deviations close to the symmetric case ($|m|\ll 1$). These are due to the deterioration of the approximation of $(\alpha_+,\rho_{+}) \approx (0,1)$ that we attribute chiefly to finite size effects,  expected to be important close to the ordinary phase consisting of  a finite fraction of adders and CLs (given by \eqref{alpha*} in both communities). Naturally, this and the limited validity of the crude assumption  $\mu_{+}\approx \mu_{-}$
affect the applicability of \eqref{transitionEq}~\footnote{
In the example of Figs.~\ref{fig:Fig3} and \ref{fig:Fig6}
with  $N=1000$ and $\kappa=60.5$,
in the transition region for $|m|\ll 1$ and $J\lesssim J_c\approx -1$, we have found 
$\alpha_+ \in [0.024, 0.027]$ instead of being strictly equal to zero.
Furthermore, in this example,  $\mu_{-}\approx \kappa$ while $\mu_{+}\gtrsim 70$, hence with $\mu_{-}-\mu_{+}\gg 1$ rather than $\mu_{-}\approx \mu_{+}$.}.


\section{Conclusion and outlook}
We have investigated a dynamic, out-of-equilibrium,  network of individuals that may hold one of two different ``opinions'' in a two-party society. In this work, the opinions of agents are held fixed while inter-party and cross-party links  are endlessly created and deleted in order to satisfy a preferred degree. The evolving network has therefore a fluctuating number of links 
and  is shaped by homophily and heterophily which model forms of social interactions
by which agents tend to establish links  with others having similar or dissimilar opinion, respectively. In our model, homophily/heterophily is modeled by an evolutionary process leading to the continuous ``birth'' and ``death'' of links within and between the communities.
 While the features of the system  where the two opinion groups  are of the
same size (symmetric case) have been studied  elsewhere~\cite{pre2021}, here
we have focused on the generic case of communities of different sizes.
We have thus investigated  how the joint effect of  community size asymmetry and homophily/heterophily influences the network structure in its steady state 
and leads to new phenomena.

The most striking  feature of our model is the transition between distinct phases as the level  
 of homophily/heterophily is varied. As main findings, we unveil the emergence of an ``overwhelming phase'' whose properties are analyzed in detail by a variety of analytical and computational methods presented in Sections IV and V.

 When the level of heterophily is  non-existing or modest, the system is in an ``ordinary phase'' similar to that characterizing the network with communities of equal size. Under intermediate to large heterophily, for sufficient asymmetry in the size of the  communities, the agents of the majority group ``overwhelm'' those of the minority by creating a large number of cross-party links. We refer to this change of regime as 
 the ``overwhelming transition'', and to the regime itself
as the ``overwhelming phase''. In the overwhelming phase, the minority consists  of agents having only cross-party links and large degrees following a broad distribution whose  average can  greatly exceed the preferred degree. 
By means of extensive Monte Carlo simulations and mean-field theories,
 we have determined the transition line separating the ordinary and overwhelming phases, and 
characterized in detail both regimes. In particular, we have studied the 
dependence on
the level of
homophily/heterophily and community size asymmetry
of  the  number of cross-party links,  fraction of agents with fewer links than the preferred degree,  as well as the average degree in each community and the 
level of polarization in the network. In addition to these global quantities, we have 
also determined the total and joint degree distributions of both communities.

We have found that the ordinary phase is characterized by features  similar to those of the symmetric case~\cite{pre2021}. The analysis of these follows from a direct two-community generalization of the  mean-field approach used in the absence of group size asymmetry. The excellent 
agreement between simulation and analytical results has allowed us to show that in the ordinary regime, the network is essentially homogeneous, 
with total degree distribution centred about the preferred degree and falling off exponentially (asymmetric Laplacian distribution), and with a broad distribution of cross-party links resulting in a ``knife-edge'' joint degree distribution.
 
Remarkably, the overwhelming phase displays a number of surprising features: 
generally, 
the agents of the minority, all have a number of edges exceeding greatly the preferred degrees, and all of these are cross-party links. This results in a degree distribution of the minority community that follows a broad Gaussian-like distribution. To characterize the latter, we have 
 devised a nontrivial generalization of the ordinary-phase mean-field analysis which is found to be in good agreement with simulation data.
 Interestingly, the majority community in the overwhelming regime has  essentially the same properties as in the ordinary phase: it forms a homogeneous network whose degree distribution is centred about the preferred degree and that falls off exponentially.
The transition from the  ordinary to the overwhelming phase occurs at finite level of heterophily (when group sizes are asymmetric), and therefore differs from the fragmentation/fission,  
arising in other network models with homophily~\cite{holme2006nonequilibrium,vazquez2008generic,durrett2012graph}. Such a transition, by which the network is split into disconnected communities, is
also found in our model but only under extreme homophily.

It would be interesting to understand whether the existence of an overwhelming transition, the most distinctive features of our simple model, is robust against generalizations of our simple dynamic network model.  As natural further avenues we could consider more than one preferred degree,
or to allow agents to draw their preferred degree from a finite range.
It would also be instructive to investigate other forms of update rules, {\it e.g.}, like networks 
subject to heterophily and growing with preferential attachment~\cite{karimi2018homophily}.
An even more realistic, yet challenging, generalization would be to consider the co-evolutionary dynamics where network varies in response to changes of node states
and the changes of those are coupled  to updates of the network links.
It would be quite relevant to investigate whether an overwhelming phase is a common feature
of all these model extensions, and to what extent our analytical methods can be generalized to tackle the latter. This endeavor, while challenging and likely to unveil  even richer and more complex phenomenology, would allow us to shed further light on the important problem of better understanding the 
general features of dynamic network shaped by social interactions.

\acknowledgments
We are indebted and grateful to Andrew Mellor for substantial input
and helpful discussions. 
The support of a
joint PhD studentship of the Chinese Scholarship Council and University of
Leeds to X.L. is gratefully acknowledged (Grant No.~201803170212). We are also grateful to the London Mathematical Society (Grant No. 41712) and Leeds School of Mathematics for their financial support, and R.K.P.Z. is thankful to the Leeds School of Mathematics for their hospitality at an early stage of this collaboration. This work was undertaken on ARC4, part of the High Performance Computing facilities at the University of Leeds, UK.

\bibliography{refs}
\clearpage


\section*{Supplementary material}

# Effects of homophily and heterophily on preferred-degree networks: mean-field analysis and overwhelming transition

In this Supplementary Material, we provide some further technical details and supplementary information in support of the results discussed in the main text.

In what follows, unless stated otherwise, the notation is the same as in the main text and the sections, equations, figures and references refer to those therein.

## S1  Further details about the model: phenomenology and special cases

This section provides a simple illustration of the phenomenological effect of homophily and the special cases where $J = 0$ and $J = \pm 1$.

### S1.1  Phenomenological effect of homophily

Our system evolves according to stochastic rules and generically settles into a unique stationary state. We are most interested in how homophily/heterophily affects the network topology and the extent of division (polarization) between the opinion groups. In our model, the level of homophily is controlled by the parameter $J \in [-1, 1]$. Positive $J$'s imply that individuals are more likely to establish links with those holding the same opinion [Fig. S1(a,b,d,e)], while negative $J$'s imply that they are more likely to establish links with those of opposite opinions [Fig. S1(c,f)].

Reference [67] gives an account of the model's main features by focusing chiefly on the ("symmetric") case of opinion groups of the same size, see Fig. S1(a-c). Instead, this work aims at a detailed description of the surprisingly rich properties characterizing the generic ("asymmetric") case of communities of different sizes, see Fig. S1(d-f). In particular, as illustrated in Fig. S1(f), the same number of CLs between the two groups leads to a higher average degree in the minority under $J < 0$, a phenomenon due to heterophily favoring the cut/addition of ILs/CLs, as discussed in detail in Secs. IV and V.

In the case of extreme homophily, $J = 1$, the population exhibits "complete polarization": the network breaks up into the disjoint union of the two communities where $\rho = 0$, see Fig. S1(a), a phenomenon often referred to as "fragmentation" or "fission" [27, 30, 32, 33]. In the case of extreme heterophily, $J = -1$, there is "complete anti-polarization": the network reduces to a bipartite graph where each agent is connected only to ones with the opposite opinion where $\rho = 1$, see Fig. S1(c). In the following, we will give a more clear illustration of these special cases.

At last, by the symmetry of opinions of both communities [e.g., compare panels of Fig. S1(d,e)], without loss of generality, we can only study systems with $N_+ < N_-$, i.e., $m < 0$.

### S1.2  Special cases $J = 0$, $J = \pm 1$

Regarding our network evolution model, some properties of certain special cases are easily understandable, and the investigation of these cases may shed light on the properties of the general system.

(i) The simplest case is $J = 0$, when the actions of the nodes are not opinion dependent. As there is no real distinction between the two communities, the model reduces to a homogeneous system [65, 70]. However, since the probability for taking action is $1/2$ regardless of the opinion of the partner node (in contrast to a link being always added/cut in the original PDN), the details of the evolution are slightly different. Nevertheless, we do not expect the time-independent properties of the stationary state to be different. As a result, regardless of $m$, the degree distribution $p_\sigma(k) = p(k)$ is the same Laplacian distribution. Furthermore, the partition into the two communities being essentially nominal, the related quantities are purely "random", and we therefore expect $\alpha = 1/2$, $\rho = 2N_+N_-/N(N-1) \approx (1 - m^2)/2$, $n_\sigma^\beta = n_\sigma/2$, and $(\rho_\sigma, 1 - \rho_\sigma) = (n_{-\sigma}, n_\sigma)$.

(ii) The cases $J = \pm 1$ of extreme homophily/heterophily, leading to complete polarization/anti-polarization, are particularly interesting. In these cases, a node can only add ILs or cut CLs for $J = 1$, and create CLs and delete ILs for $J = -1$. The final network is static and eventually all the agents will have more links than $\kappa$. To see this behavior,

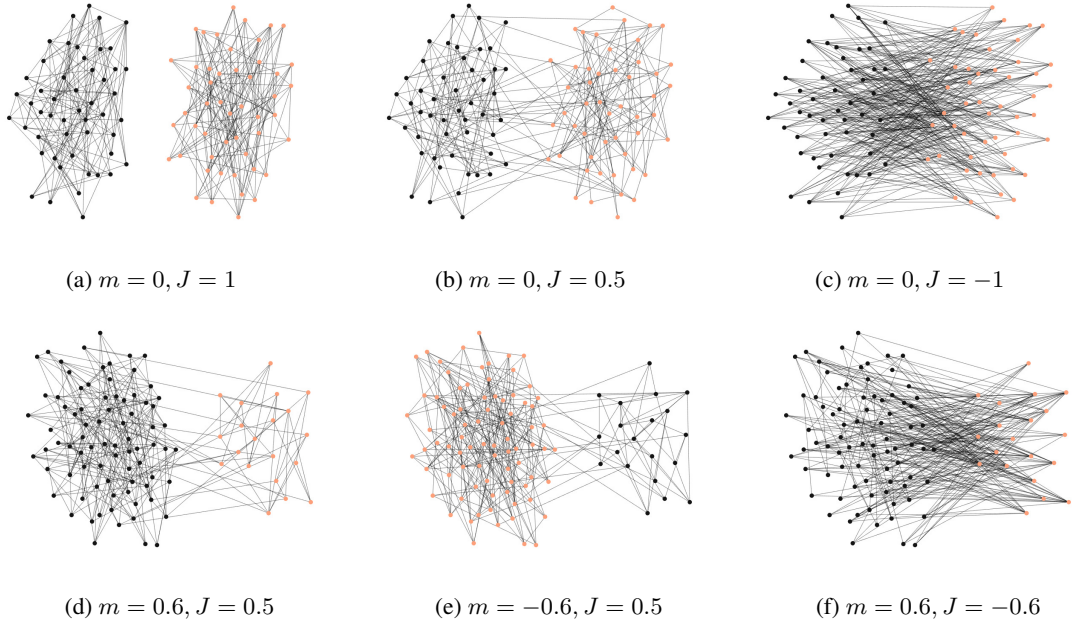

Figure S1: Typical different polarization scenarios for various levels of homophily/heterophily, in a network of total size $N = 100$ and $\kappa = 4.5$, after 100 Monte Carlo steps (MCS, 1 MCS = $N$ update steps) in opinion groups of the same size ("symmetric case") (a)-(c), and in opinion groups of different sizes ("asymmetric case") (d)-(f). Light and dark dots are respectively $i$ and $j$ voters/nodes holding opinion $-1$ ($\sigma_i = -1$) and $+1$ ($\sigma_j = +1$). Top row, symmetric case: same number of light and dark dots ($m = 0$, $N_+ = N_- = 50$, see text). Bottom row, asymmetric case: (d,f) the number of dark dots is four times that of light dots ($m = 0.6$, $N_+ = 80$, $N_- = 20$); (e) there are four times more $-1$ light nodes than $+1$ dark nodes ($m = 0.6$, $N_+ = 20$, $N_- = 80$). The level of homophily is (a) $J = 1$ (complete polarization); (b,d,e) $J = 0.5$, (c) $J = -1$ (complete anti-polarization), and (f) $J = -0.6$, see text. In (f), agents of the minority group have almost only links to those of the majority community and are hence "overwhelmed" by the latter (that have links across both groups), see Secs. IV.D and V.C.

let us consider $J = 1$: since each node adds only ILs and cuts CLs, any initial network will quickly evolve into one with no CLs. Meanwhile, as soon as a node exceeds $\kappa$ links, it becomes passive. However, as others in the community with fewer than $\kappa$ links can choose to link to it, its degree can only increase [*]. While we always consider an initial network without any links, we can here consider the most atypical example consisting of an initial complete network, with $\kappa \ll N_{\pm}$. Then, in every time step, an agent will cut a link — CL or IL — depending on $J = \pm 1$, until attaining a steady state consisting (regardless of $m$) of either two disjoint complete networks or a single bipartite graph as in Fig. S1(a,c).

(iii) Another singular limit of this model is extreme asymmetry: $|m| \to 1$. With only one set of agents, the population is homogeneous and the update rules reduce to adding (cutting) links with probability $\hat{J}\,(\check{J})$. Thus, balancing $\alpha \hat{J}$ with $(1 - \alpha)\,\check{J}$, we conclude $\alpha = (1 - J)/2$. This result is consistent with those above: there are no adders in the final state when $J = 1$. The opposite limit, $J = -1$, is more interesting as the naive conclusion would be that there are no cutters $(1 - \alpha = 0)$ in the steady state. But this appears to contradict (ii) above. The resolution lies in the singular nature of the points $J = -1, |m| = 1$, so that the final states depend on which direction these points are approached. If $J$ is taken to $-1$ first, then the final state will indeed have no adders. Such a singularity affects $\alpha\,(m, J)$ before these points are reached, namely, there is a non-uniform convergence towards $\alpha \to 0$ when $J \to -1$. As shown in Sec. V, as the asymmetry becomes more pronounced (i.e. as $|m|$ approaches 1), the fraction of adders $\alpha\,(m, J)$ in the majority tends to $(1 - J)/2$ for larger portions of $J \in [-1, 1]$, before dropping steeply towards $\alpha = 0$ near $J = -1$. These turning points are significant as they signal transitions to very different final states. These discussions can also shed light on the properties of the general system, e.g., the properties of $\alpha_{\pm}$ displayed in Fig. 2(a). First, we see that when $J$ is not close to $-1$ and $m = -0.6$, the fraction of majority adders, $\alpha_{-}(J)$, can be approximated by $(1 - J)/2$. Meanwhile, much of the minorities' response is governed by the action of the majority on the CLs. With $J > 0$, CLs are preferentially cut. As there are more majority cutters $(\alpha_{-} < 1/2)$, $L_{\times}$ is likely to suffer more losses, which can be compensated by having the minority agents add more links. Thus, we find more minority adders $(\alpha_{+} > 1/2)$. Of course, when homophily is increased further, the level of polarization is higher. With the dominance by the majority lessened, the minority agents can revert to the behavior of a single population with $J \simeq 1$, i.e., $\alpha_{+} \to 0$.

## S2 Master equation and detailed balance

The complete description of our system is given by the adjacency matrix $\mathbb{A}$ with elements $A_{ij}$ being 1 or 0, if a link between nodes $i$ and $j$ is present or not. Clearly, $\mathbb{A}$ is symmetric and we define $A_{ii} \equiv 0$. To be specific, we will order the nodes so that all those with opinion $+1$ comes first: $\sigma_i = 1$ for $i = 1, \cdots, N_+$. It is then convenient to use the notation

$$\mathbb{A} = \left( \begin{array}{cc} \mathbb{A}_{++} & \mathbb{A}_{+-} \\ \mathbb{A}_{-+} & \mathbb{A}_{--} \end{array} \right)$$

where the blocks represent ILs and CLs. Obviously, $\mathbb{A}_{-+} = \mathbb{A}_{+-}^{T}$ is rectangular in general. Starting with some initial network $\mathbb{A}_{ini}$ and generic $J$, the system's stochastic time evolution obeys a master equation for the distribution $\mathcal{P}\,(\mathbb{A}, t)$ — an abbreviation of the conditional probability $\mathcal{P}\,(\mathbb{A}, t | \mathbb{A}_{ini}, 0)$ — and arrive at a stationary $\mathcal{P}^{*}\,(\mathbb{A})$ as $t \to \infty$. We will show that detailed balance is violated by the dynamics of this model that is therefore genuinely out of thermal equilibrium. As a consequence, determining the full evolution of $\mathcal{P}$ or even an explicit form for $\mathcal{P}^{*}$ are beyond reach. Thus, in the main text we focus our attention mainly on investigating the interesting *steady state* properties of our model.

The stochastic aspects of the system is embodied in $\mathcal{P}\,(\mathbb{A}, t)$, the probability of having $\mathbb{A}$ at discrete times $t = 0, 1, \cdots$. The evolution of $\mathcal{P}$ is governed by a Master equation:

$$\mathcal{P}\,(\mathbb{A}, t + 1) = \sum_{\{\mathbb{A}'\}} R\,[\mathbb{A} \leftarrow \mathbb{A}']\,\mathcal{P}\,(\mathbb{A}', t) \tag{S1}$$

where, $R\,[\mathbb{A} \leftarrow \mathbb{A}']$ stands for transition probabilities from $\mathbb{A}'$ to $\mathbb{A}$. From one $t$ to the next, an update consists of choosing a node at random, deciding if it adds or cuts a link (based on $\kappa$), choosing another node at random for this

---

[*]Although the process appears Poisson-like, it is much more complex, as the number of adders *not already* connected to our node depends on $k$ and dwindles with time. Even the case with the lowest $\kappa$ (0.5) is non-trivial, though the $N \to \infty$ limit may be analytically accessible (H. Hilhorst, private communication (2020))

III

action, and then applying the rules of homophily (controlled by $J \in [-1, 1]$). Given in words in Sec. II, these rules are presented concisely by

$$R\left[\mathbb{A} \leftarrow \mathbb{A}'\right] = \frac{1}{N} \sum_i \frac{H\left(\kappa - k_i'\right)}{N - 1 - k_i'} \sum_j a_{ij} \left(1 - a_{ij}'\right) h_{ij} \Delta_{ij}$$

$$+ \frac{1}{N} \sum_i \frac{H\left(k_i' - \kappa\right)}{k_i'} \sum_j \left(1 - a_{ij}\right) a_{ij}' h_{ij} \Delta_{ij}, \qquad \text{(when } \mathbb{A}' \neq \mathbb{A}) \tag{S2}$$

and

$$R\left[\mathbb{A} \leftarrow \mathbb{A}\right] = 1 - \sum_{\mathbb{A}' \neq \mathbb{A}} R\left[\mathbb{A}' \leftarrow \mathbb{A}\right], \tag{S3}$$

where

$$k_i' = \sum_j a_{ij}' \text{ and } h_{ij} \equiv \frac{a_{ij}(1 + J\sigma_i\sigma_j)}{2} + \frac{a_{ij}'(1 - J\sigma_i\sigma_j)}{2} \tag{S4}$$

accounts for the effects of homophily, and

$$\Delta_{ij} \equiv \prod_{\substack{(k,l) \neq (i,j) \\ \text{and } (l,k) \neq (i,j)}} \delta\left(a_{kl} - a_{kl}'\right) \tag{S5}$$

ensures that only $a_{ij}$ and $a_{ji}$ are changed. In Eq. (S2), $H$ is the Heaviside step function, while the denominator accounts for the total number of choices node $i$ has to act (i.e., add/cut the link to node $j \neq i$).

It is clear that, for $J = 0$, the opinions are irrelevant so that the system reduces to the case of "a homogeneous population" as in Ref. [64, 76]. Since detailed balance was shown to be violated in that model, we can expect that to be generically true here. For example, let $\kappa = 10.5$ and nodes $i$, $j$, and $\ell$ with $(k_i, k_j, k_\ell) = (9, 15, 6)$. Now, consider the loop $(9, 15, 6) \rightarrow (10, 15, 7) \rightarrow (11, 16, 7) \rightarrow (10, 16, 6) \rightarrow (9, 15, 6)$ with non-zero rates at every step (by $i$ adding link to $k$, $i$ adding to $j$, then $i$ cutting to $k$, and then $j$ cutting to $i$). But the reverse has $R\left[(9, 15, 6) \leftarrow (10, 15, 7)\right] = 0$, since $k_i$ and $k_\ell$ are both $< \kappa$. Thus, the product of $R$'s for the forward loop cannot be the same as for the reverse (regardless of the opinions of the nodes) and, applying the Kolmogorov criterion [77], detailed balance is violated. Since $\kappa$ is picked so that a chosen node always adds or cuts a link, it is easy to verify that the system is ergodic (any configuration $\mathbb{A}$ can be reached from any other). Thus, the Frobenius theorem guarantees $\mathcal{P}$ settles in a stationary state $\mathcal{P}^*\left(\mathbb{A}\right)$, which is the right eigenvector of $R$ with unit eigenvalue. Since detailed balance is violated, $\mathcal{P}^*\left(\mathbb{A}\right)$ should be regarded as a *non-equilibrium* steady state. In particular, unlike typical equilibrium states associated with the Boltzmann-Gibbs ensemble, there are non-trivial probability currents in this steady state [78].

## S3 Quantities of interest: table & exact expressions for quantities of interest

We also provide explicit expressions for various quantities defined in Sec. II, summarized in Table 1. To be specific, we choose $\sigma_i = 1$ for $i = 1, \cdots, N_+$ and $-1$ for $i = N_+ + 1, \cdots, N$. Since $\sigma$'s are not variables, there is no need to express them explicitly in $\mathbb{A}$.

The average of an observable $\mathcal{O}$ in the steady state is

$$\langle \mathcal{O} \rangle \equiv \sum_{\{\mathbb{A}\}} \mathcal{O} \mathcal{P}^*\left(\mathbb{A}\right).$$

Some relevant observables and averages are listed below.
— Degree for node $i$:

$$k_i = \sum_{j=1}^N A_{ij}$$

— Number of links to node of opinion $\tau$:

$$k_{i,\tau} = \sum_{j=1}^N A_{ij} \delta\left(\tau - \sigma_j\right).$$

Table 1: Summary of notation. The quantities $J, \kappa, m$ and $N$ are fixed control parameters.

| | |
|---|---|
| $N$ | Total number of nodes |
| $J \in [-1, 1]$ | Degree of homophily/heterophily |
| $\kappa$ | Preferred degree of all the nodes |
| $\sigma_i = \pm 1$ | State/opinion of node $i$ |
| $m \in [-1, 1]$ | Average of opinions |
| $k_i$ | Degree of node $i$ |
| $\hat{J}, \check{J}$ | $\hat{J} = (1 + J)/2, \check{J} = (1 - J)/2$ |
| $L_{\sigma\tau}$ | Number of links from agents with opinion $\sigma$ to those with $\tau$; $\sigma, \tau \in \{+, -\}$ |
| $L_\times, L_\odot$ | Total number of CL and IL links, $L_\times = L_{+-} = L_{-+}, L_\odot = L_{++} + L_{--}$ |
| $L$ | Total number of links: $L = L_\times + L_\odot$ |
| $\alpha \in [0, 1]$ | Fraction of adders, $\alpha \equiv n^a = N^a/N$ |
| $p_\sigma(k)$ | Degree distribution of the community of opinion $\sigma$ |
| $P_\sigma(\ell_+, \ell_-)$ | Joint degree distribution of the community of opinion $\sigma$ |
| $(\bar{\ell}_\pm)_\sigma$ | The average fraction of neighbours of opinion $\pm$ for an agent of opinion $\sigma$, see Eq. (1) |
| $\alpha_\sigma \in [0, 1]$ | Fractions of adders within each opinion group, $\alpha_\sigma \equiv N^a_\sigma/N_\sigma = n^a_\sigma/n_\sigma$ |
| $\rho \in [0, 1]$ | Fraction of CLs, $\rho \equiv L_\times/L$ |
| $\rho_\sigma \in [0, 1]$ | Fractions of CLs within each opinion group, see Eq. (6) |
| $\mu_\sigma$ | Average degree of the community of opinion $\sigma$ |
| $q_\sigma$ | Conditional degree distribution, see Eq. (2) |
| $N_+, N_-$ | Number of nodes with opinion $+1$ and $-1$, $N = N_+ + N_-$ |
| $n_+, n_-$ | Fraction of nodes with opinion $+1$ and $-1$ |
| $N^a, N^c$ | Number of adders and cutters, $N = N^a + N^c$ |
| $N^{a,c}_\sigma$ | Number of adders and cutters of opinion $\sigma$, $N_\sigma = N^a_\sigma + N^c_\sigma$ |
| $n^{a,c}_\sigma$ | Fraction of adders and cutters of opinion $\sigma$, $n_\sigma = n^a_\sigma + n^c_\sigma$ |
| $\mathbb{A} = (a_{ij})$ | Adjacency matrix of the network |
| $\Lambda, \Pi \in [-1, 1]$ | Measures of polarization, see Eq. (5) and (7) |

— Total number of links:

$$L = \frac{1}{2} \sum_{i,j} A_{ij} = \|\mathbb{A}\| / 2,$$

where $\|\cdot\|$ is the 1-norm of a matrix.

— Number of ILs and CLs:

$$L_{\sigma\tau} = \frac{1}{1 + \delta(\sigma - \tau)} \sum_{i,j} A_{ij} \delta(\sigma - \sigma_i) \delta(\tau - \sigma_j)$$

$$= \frac{\|\mathbb{A}_{\sigma\tau}\|}{1 + \delta(\sigma - \tau)}.$$

— Number of adders and cutters holding opinion $\sigma$:

$$N^a_\sigma = \sum_i H(\kappa - k_i)\delta(\sigma - \sigma_i),$$

$$N^c_\sigma = \sum_i H(k_i - \kappa)\delta(\sigma - \sigma_i).$$

— Average degree partitions:

$$\langle k_{\sigma\tau} \rangle = \sum_{i,j} A_{ij} \delta(\sigma - \sigma_i) \delta(\tau - \sigma_j) / N_\sigma.$$

— Degree distribution of nodes with opinion $\sigma$:

$$p_\sigma(k) = \frac{1}{N_\sigma} \left\langle \sum_i \delta(k - \sum_j A_{ij})\delta(\sigma - \sigma_i) \right\rangle,$$

so that the moments are

$$\left\langle k^\ell \right\rangle_\sigma \equiv \sum_k k^\ell p_\sigma\left(k\right),$$

with $\mu_\sigma \equiv \langle k \rangle_\sigma = \left(\bar{\ell}_+\right)_\sigma + \left(\bar{\ell}_-\right)_\sigma$.

— Joint distribution of nodes with opinion $\sigma$:

$$\tilde{p}_\sigma\left(k_{\sigma\sigma}, k_{\sigma,-\sigma}\right) = \left\langle \sum_i \left[ \delta(k_{\sigma\sigma} - \sum_j A_{ij}\delta\left(\sigma - \sigma_j\right))\delta\left(\sigma - \sigma_i\right) \right] \right\rangle,$$

so that $p_\sigma\left(k\right) = \sum_{k_{\sigma+}, k_{\sigma-}} \delta(k_{\sigma+} + k_{\sigma-} - k)\tilde{p}_\sigma\left(k_{\sigma+}, k_{\sigma-}\right).$

## S4 Discrete derivatives

Since we have taken measurements at unevenly spaced parameter values, we provide some details on how we arrive at the plots of first and second derivatives of $\alpha$ and $\rho$. Given $f\left(x_i\right)$, where the points $x_i$ $(i = 1, 2, \cdots)$ are not necessarily evenly spaced, we define discrete derivatives by

$$\Delta f\left(\frac{x_i + x_{i+1}}{2}\right) \equiv \frac{f\left(x_{i+1}\right) - f\left(x_i\right)}{x_{i+1} - x_i},$$

and

$$\Delta^2 f\left(x_i\right) \equiv \frac{\Delta f\left(\frac{x_i + x_{i+1}}{2}\right) - \Delta f\left(\frac{x_i + x_{i-1}}{2}\right)}{\frac{x_{i+1} - x_{i-1}}{2}}.$$

If we want to be consistent, then we should assign this $\Delta^2 f$ value to

$$\tilde{x}_i = \frac{x_{i-1} + 2x_i + x_{i+1}}{4},$$

which of course reduces to $x_i$ if the spacings are equal. It is interesting to note that the difference, $\tilde{x}_i - x_i$, looks like the ordinary curvature $(x_{i-1} - 2x_i + x_{i+1})$.

## S5 Average and variance of an asymmetric Laplacian distribution (ALD)

The degree distributions considered in this paper are well approximated by ALDs, see *e.g.* Eq. (23) in Sec. V.B:

$$Q\left(\ell\right) = \left\{ \begin{array}{l} Q\left(1\right)\gamma_>^\ell \ \text{ for } \ell > 0 \\ Q\left(0\right)\gamma_<^{-\ell} \ \text{ for } \ell \leq 0 \end{array} \right.$$

for $\ell > 0$ or $\leq 0$ respectively. Along with the $\gamma$'s (both $< 1$), the discontinuity (given by the ratio $\Gamma$)

$$\Gamma \equiv Q\left(0\right)/Q\left(1\right),$$

is needed to specify the distribution uniquely. Normalization implies

$$\left[\frac{\Gamma}{1 - \gamma_<} + \frac{\gamma_>}{1 - \gamma_>}\right]Q\left(1\right) = 1.$$

Thus, the fraction with $\ell \leq 0$ is

$$\alpha \equiv \frac{Q\left(0\right)}{1 - \gamma_<} = \frac{\Gamma Q\left(1\right)}{1 - \gamma_<} = \frac{\Gamma\left(1 - \gamma_>\right)}{\gamma_>\left(1 - \gamma_<\right) + \Gamma\left(1 - \gamma_>\right)},$$

and, of course, the fraction with $\ell > 0$ is

$$1 - \alpha = \frac{\gamma_> \left(1 - \gamma_<\right)}{\gamma_> \left(1 - \gamma_<\right) + \Gamma \left(1 - \gamma_>\right)}.$$

The average can be computed by "pieces", starting with the result

$$\langle \ell \rangle = (1 - x) \sum_{\ell \geq 0} \ell x^\ell = \frac{x}{1 - x}.$$

In particular,

$$\langle \ell \rangle_< = \frac{-\gamma_<}{1 - \gamma_<} = 1 - \frac{1}{1 - \gamma_<},$$

$$\langle \ell \rangle_> = \frac{1}{1 - \gamma_>},$$

while

$$\langle \ell \rangle = \langle \ell \rangle_< \alpha + \langle \ell \rangle_> \left(1 - \alpha\right)$$
$$= \frac{-\gamma_<}{1 - \gamma_<} \alpha + \frac{1}{1 - \gamma_>} \left(1 - \alpha\right)$$
$$= \frac{1}{1 - \gamma_>} - \alpha \left[\frac{1}{1 - \gamma_>} + \frac{\gamma_<}{1 - \gamma_<}\right].$$

## S6  Computing $\Lambda$ and $\Pi$ from $\rho_\pm$ and $m$

The quantity $\Lambda$ is commonly used and sometimes referred to as "average edge homogeneity" [47, 72]. From its definition, in the realm of the mean-field approximation, we have

$$\Lambda(J, m) \equiv 1 - 2\rho = 1 - \frac{4\rho_+ \rho_-}{\rho_+ + \rho_-}, \tag{S6}$$

where we have used $1/\rho_+ + 1/\rho_- = 2/\rho$, based on Eq. (6).

A salient feature of $\Pi$ is its simple mean-field expression in terms of the fraction of connections *within each of the three sectors*: $f_\pm \equiv 2L_{\pm\pm}/N_\pm^2$ and $f_\times \equiv L_\times/N_+N_-$. Then, Eq. (3) implies $\left(\bar{\ell}_\sigma\right)_\sigma = f_\sigma N_\sigma$ and $\left(\bar{\ell}_\sigma\right)_{-\sigma} = f_\times N_\sigma$, so that by plugging these into Eq. (7), i.e., our definition of $\Pi$ is equivalent to,

$$\Pi = \frac{f_+ f_- - f_\times^2}{\left(f_+ + f_\times\right)\left(f_- + f_\times\right)}. \tag{S7}$$

Again, it clearly reduces to $\pm 1$ for $J = \pm 1$, since $f_\pm$ or $f_\times$ vanishes. For $J = 0$, every nodes has exactly the same average degree, $\kappa$, so that the same fraction, $f = \kappa/\left(N - 1\right)$, prevails in each sector and $\Pi = 0$. For calculating $\Pi$ by Eq. (S7), we recognize that we only need the two ratios $f_\pm/f_\times$ that will be again treated in the spirit of the mean-field approximation, see below. Using Eq. (6), we find

$$\rho_\sigma = \frac{f_\times N_{-\sigma}}{f_\times N_{-\sigma} + f_\sigma N_\sigma} = \frac{1}{1 + \frac{f_\sigma N_\sigma}{f_\times N_{-\sigma}}},$$

and so

$$\frac{f_\sigma}{f_\times} = \left(\frac{1 - \rho_\sigma}{\rho_\sigma}\right) \left(\frac{1 - \sigma m}{1 + \sigma m}\right).$$

This leads immediately to the numerator of $\Pi$ being

$$\frac{f_+ f_-}{f_\times^2} - 1 = \frac{1 - \rho_+ - \rho_-}{\rho_+ \rho_-},$$

and

$$\Pi \propto 1 - \rho_+ - \rho_-,$$

which vanishes at $J = 0$ (where $\rho_\sigma = n_{-\sigma}$) regardless of asymmetry. We obtain

$$\Pi = \frac{(1 - \rho_+ - \rho_-)\left(1 - m^2\right)}{(1 + 2\rho_+ m)\left(1 - 2\rho_- m\right) + (2\rho_+ + 2\rho_- - 1)m^2}, \tag{S8}$$

which is an expression of $\Pi$ in terms of $\rho_\pm$ and $m$ (obtained from the mean-field theory of Sec. V.B). It is worth emphasizing that (S7) is equivalent to (7), but the above expression of (S8) of $\Pi$ has again to be interpreted in a mean-field sense. This is because $f_\pm$ and $f_\times$ are fluctuating quantities and in the above derivation of (S8), we have approximated the average of their ratio by the ratio of their average.

## S7   A "shifted" binomial distribution

By introducing a refined theory to better characterize the behavior of the minority agents in the overwhelmed state, as discussed in Sec. V.C, we obtain a "shifted" binomial distribution for the minority agents. In this section, we show several properties of the distribution. In particular, here we provide a procedure of approximating the mean and variance of the distribution.

The well-known binomial distribution, normalized by unity at $n = 0$, is

$$\binom{N}{n} z^n, \ \ n \in [1, N]$$

with the overall normalization factor $(1 + z)^N$. This can be cast in terms of the $\Gamma$ function, or more compactly by using the Pochhammer symbol, $(x)_n = x\,(x + 1) \cdots (x + n - 1)$ [79]:

$$\frac{(N + 1 - n)_n}{(1)_n} z^n.$$

In Sec. V.C, we encounter what might be termed a "shifted binomial" distribution [†]

$$S\,(n; N, z, a, b) \equiv \frac{(a + 1 - n)_n}{(b)_n} z^n, \ \ n \in [1, N] \tag{S9}$$

where we have again normalized $S$ by unity at $n = 0$. Of course, we could have also written $(N + a - n)$ for the numerator (so that $a = b = 1$ is the ordinary binomial). As it is, the explicit expression is

$$S\,(n) = \frac{a\,(a - 1) \cdots (a + 1 - n)}{b\,(b + 1) \cdots (b + n - 1)} z^n$$

and so, we must specify the range of validity for $n$ — chosen to be $[0, N]$ here. Clearly, $S$ satisfies the recursion relation

$$z\,(a - n)\,S\,(n) = (b + n)\,S\,(n + 1)$$

and requires a normalization constant

$$\mathcal{Z}\,(N, z, a, b) \equiv 1 + \sum_{n=1}^{N} S\,(n; N, z, a, b)$$

for $\mathcal{S} \equiv S / \mathcal{Z}$ to be a proper distribution.

In one "thermodynamic limit" — $N, a \to \infty; z, b = O\,(1)$, $S$ peaks at $\hat{n}$, given by

$$z\,(a - \hat{n}) \cong (b + \hat{n}) \implies \hat{n} \cong \frac{za - b}{z + 1} \to \frac{za}{z + 1}. \tag{S10}$$

---

[†] A neater form is $z^n a^{(n)} / (b)_n$, where $a^{(n)} \equiv a\,(a - 1) \cdots (a - n + 1)$ is the "falling factorial." But that notation can cause much confusion.

If $\hat{n}$ lies within $[0, N]$, then $S$ is well approximated by a Gaussian around it, with variance given by

$$1/V \cong -\ln\left[S\left(\hat{n}+1\right)S\left(\hat{n}-1\right)/S\left(\hat{n}\right)^2\right].$$

But

$$\ln\left(\frac{b+\hat{n}-1}{z\left(a-\hat{n}+1\right)}\right) \simeq \ln\left(\frac{\hat{n}+b-1}{\hat{n}+b+z}\right) \simeq \frac{-1-z}{\hat{n}},$$

so that

$$V \cong \frac{za}{\left(1+z\right)^2}. \tag{S11}$$

For example, for the standard distribution of $N$ tosses of an uneven coin (with probabilities $p$ and $1-p$), $z = p/\left(1-p\right)$ so that $\hat{n} \cong Np$ and $V \cong Np\left(1-p\right)$.

Another "thermodynamic limit" is $N, a \to \infty; \zeta \equiv za, b = O\left(1\right)$. Then, to leading order,

$$S \propto \zeta^n / \left(b\right)_n$$

is what might be termed the "shifted" Poisson distribution (with $b = 1$ being the standard Poisson).

Another familiar distribution — the geometric $S \propto \zeta^n$ — can be obtained from the limit $a, b \to \infty; \zeta \equiv za/b = O\left(1\right)$. Obviously, if $\zeta < 1$, $S$ peaks at $n = 0$ and we can consider $N \to \infty$ as well. Otherwise, the distribution peaks at $n = N$, and has the same characteristic after normalization.

Base on Eq. (S10) and (S11), for the shifted binomial degree distribution of the minority agents as shown in Sec. V.C, we can give an approximation of $\mu_+$ and $V_+$ by letting $z = Q^c/Q^a, a = N_-$, as given by (32) and (33) (but for (33) in Sec. V.C, we kept another order of approximation).

## S8    Mean-field prediction for the average degree of the minority group

In Secs. V.A and V.B, we provide an ordinary mean-field theory to characterize the network's topology in the steady-state. Although we do not expect it to be effective across both phases, it surprisingly gives us good approximations of several quantities even in the overwhelming regime, e.g., see Fig. 3(c,d). As a result, the mean-field predictions of the average degree of the minority group may be directly obtained by solving Eqs. (16-18) and (26), which is shown as the dotted line in Fig. S2, which generally has good agreement with the simulation data.

As discussed in Sec. V.C, in the overwhelming phase, the degree distribution of the minority is de facto morphed from exponentials to Gaussian-like (see Fig. 6(d-f)), which encourages us to introduce a refined theory to capture the behaviour of those agents better. In terms of their average degree, using the refined theory, we have Eq. (32), where we still need to determine the values of $Q^a$ and $Q^c$ by approximating $\alpha_-$. One way to determine $\alpha_-$ is to use the semi-phenomenological assumption that $\alpha_+ = 0$. In this way, we have Eq. (34), and hence Eq. (32) gives us the solid line shown in Fig. S2. Comparison with the simulation data gives us the impression that it agrees well with the simulation data in the transition region. However, deviations become increasingly visible as $-J$ gets larger, i.e., deep in the overwhelming regime. These can be traced back to the use of Eq. (34), which has slight discrepancies with the simulation data of $\alpha_-$ deep in the overwhelming regime, see Fig. 2(a). Once we circumvent this issue, i.e., obtaining $\alpha_-$ directly from the simulation data, we have excellent agreement shown as the dashed line in Fig. S2. This makes us believe that the refined mean-field approach is quite successful at predicting the observed average degree of the minority agents.

## S9    About a self-consistent refined mean-field theory for the degree distributions

In Sec. V.C, we obtain predictions for the degree distribution of the minority community in the overwhelming phase based on (30) that for practical purposes, and guided by simulation data, we have used together with the crude approximations (34) and $\mu_-^{a,c} \approx \kappa$, yielding a semi-phenomenological refined mean-field theory.

As noted in the main text, a fully self-consistent refined mean-field theory that would not rely on the above assumptions, in the spirit of Ref. [66], is in principle feasible but would require a re-analysis of the majority degree distribution. In fact, this program would involve finding appropriate expressions for $R_\sigma^{a,c}$ for majority community in terms of a full set of unknowns like $n_\sigma^{a,c}$ and $\rho_\sigma$, setting up a set of "self-consistent" equations and solving for them numerically as in Ref. [66]. Such an intensive program is beyond the scope of this work. Here, we merely remark

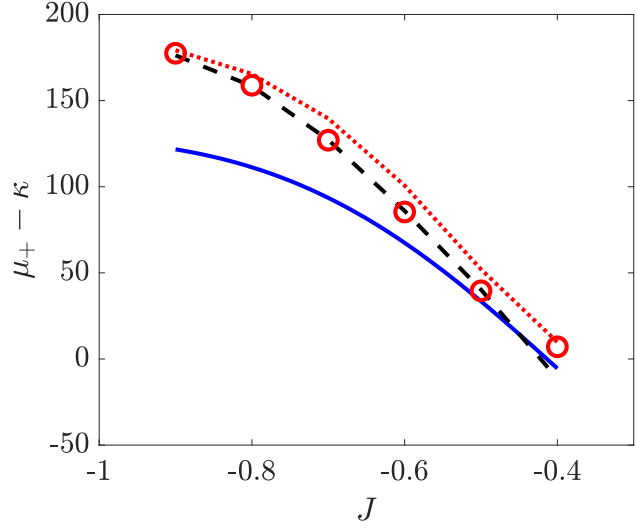

Figure S2: $\mu_+ - \kappa$ *vs* $J$ for $N = 1000, \kappa = 60.5$ and $m = -0.6$ in the overwhelming phase ($J \leq -0.4$). Markers are results from simulation; The solid and dashed lines are from (32) with $\alpha_-$ determined by (34) and the simulation data, respectively, and $\mu_-^{a,c} = \kappa$; the dotted line is from the predictions of (26).

that $p_-(k)$, the degree distribution of the majority nodes, and $p_+(k < \kappa)$, the degree distribution of the adders of the minority nodes, are again asymmetric Laplacian distributions (see Fig. 6). Hence, the mean-field theory in Sec. V.B, yielding exponential fall off of the degree distribution of both communities according to (30), still provides a good approximation for these nodes, as indicated by the red and blue straight lines in Fig. 6(d-f).


\end{document}